\documentstyle[epsf,11pt]{article}
\def\m@thcombine#1#2{%
  \setbox0=\hbox{$#1$}
  \setbox1=\hbox{$#2$}
  \ifdim\wd0>\wd1
    \setbox0=\hbox to\wd1{\hss\box0\hss}
  \else
    \setbox1=\hbox to\wd0{\hss\box1\hss}
  \fi
  \mathop{\vcenter{
    \offinterlineskip\box0\box1}}}
\def\lesim{\m@thcombine<\sim}
\def\gesim{\m@thcombine>\sim}
\def\lessgtr{\m@thcombine<>}
\def\gtrless{\m@thcombine><}

\newcommand{\ket}[1]{\left| #1 \right\rangle}

\newcommand{\beq}{\begin{equation}}
\newcommand{\beqa}{\begin{eqnarray}}
\newcommand{\eeq}{\end{equation}}
\newcommand{\eeqa}{\end{eqnarray}}

\def\mGamma{{\mit\Gamma}}
\def\mDelta{{\mit\Delta}}
\def\mDeltan{{\mit\Delta}_{\rm n}}
\def\mDeltap{{\mit\Delta}_{\rm p}}
\def\Nout{{N_{\rm out}}}
\def\Nband{{N_{\rm band}}}
\def\Eout{{E_{\rm out}}}
\def\Iout{{I_{\rm out}}}

\def\Edamp{{E_{\rm damp}}}
\def\nbranch{{n_{\rm branch}}}
\def\Gammat{{\mGamma_{\rm t}}}
\def\Gammas{{\mGamma_{\rm s}}}
\def\Gamman{{\mGamma_{\rm n}}}
\def\rhos{{\rho_{\rm s}}}
\def\rhon{{\rho_{\rm n}}}
\def\Ds{{D_{\rm s}}}
\def\Dn{{D_{\rm n}}}
\def\Qs{{Q_{\rm s}}}
\def\Qn{{Q_{\rm n}}}
\def\Jmoms{{{\cal J}_{\rm s}}}
\def\Jmomn{{{\cal J}_{\rm n}}}
\def\Dtot{{D_{\rm tot}}}
\def\Eone{{\rm E1}}
\def\Etwo{{\rm E2}}
\def\ND{{\rm ND}}
\def\SD{{\rm SD}}
\def\NDyr{{\rm NDyr}}
\def\SDyr{{\rm SDyr}}
\def\FG{{\rm FG}}
\def\FGn{{\rm FG,n}}
\def\mass{{\rm mass}}
\def\omegas{{\omega_{\rm s}}}

\def\omegab{{\omega_{\rm b}}}
\def\Vb{{V_{\rm b}}}
\def\GDR{{\rm GDR}}
\def\hop{{\rm hop}}
\def\eff{{\rm eff}}
\def\ex{{\rm ex}}

\begin{document}
\begin{center}
{\large\bf Barrier penetration and rotational damping
of thermally excited superdeformed nuclei}
\end{center}
\begin{center}
K. Yoshida$^{\rm a}$, M. Matsuo$^{\rm b}$  and Y.R. Shimizu$^{\rm c}$
\vspace{3mm}
\end{center}
\begin{center}
{\it $^{\rm a}$ Institute of Natural Science, Nara University, 
Nara 631-8502, Japan} \\
{\it $^{\rm b}$ Graduate School of Science and Technology,
Niigata University, Niigata 950-2181, Japan}\\
{\it $^{\rm c}$ Department of Physics, Kyushu
University, Fukuoka 812-8581, Japan}
\end{center}

\begin{abstract}
We construct a microscopic model of 
thermally excited superdeformed states that describes both
the barrier penetration mechanism, leading to the decay-out 
transitions to normal deformed states, and the rotational
damping causing fragmentation of rotational E2 transitions.
We describe the barrier penetration by means of a tunneling path 
in the two-dimensional deformation energy surface, 
which is calculated with 
the cranked Nilsson-Strutinsky model. The individual excited superdeformed 
states and associated E2 transition strengths are calculated
by the shell model diagonalization of the
many-particle many-hole excitations interacting with 
the delta-type residual two-body force. 
The effect of the decay-out
on the excited superdeformed states are discussed in detail for
$^{152}$Dy, $^{143}$Eu and $^{192}$Hg. 
The model predicts that the decay-out brings about a characteristic decrease
in the effective number of excited superdeformed rotational bands.

\end{abstract}

\section{Introduction}\label{Sec:Intro}

Quasi-continuum gamma-ray spectra observed in heavy ion
fusion reactions have attracted much attention recently
as they carry information on structure of
rapidly rotating and thermally excited nuclei.
Indeed studies of the ridge and valley structures
and the fluctuations in the double coincident spectra 
revealed occurrence of the rotational damping. Namely  
the collective rotation of a deformed nucleus
becomes a damped motion when the thermal excitation energy is
provided to the nucleus~\cite{Lauritzen,FAM}. This is in contrast to 
the rotational band structures known for the levels
near the yrast line.
The rotational damping is intimately related to a basic
feature of  highly excited compound states that
the wave functions of excited levels become 
complex mixture of many-particle many-hole or
many quasiparticle configurations due to the residual two-body
nuclear interaction~\cite{Lauritzen,FAM,Aberg,Matsuo,Bracco-simul}.

The quasi-continuum gamma-rays are also produced in the reactions
forming superdeformed (SD)
nuclei~\cite{Dy-cont,Khoo,Hg192-cont-decay,Leoni-Npath,Leoni-decay}. 
However, their
properties are much more complex than in normal deformed (ND) nuclei.
Recent experiments~\cite{Leoni-decay} reveal that
the continuum gamma-rays in superdeformed nuclei contain
more than one components. This complexity arises from a
characteristic feature that the superdeformed  and the normal deformed states
coexist in the same region of spin and excitation energy.
It is known that the observed SD rotational bands keep their
identity remarkably well even though the SD bands are embedded
in a sea of compound levels having normal deformation.
This is because the SD and ND states are separated by
a potential barrier in the deformation space.
On the other hand, observed SD rotational bands 
terminate suddenly at spin value around $10-30\hbar$ by decaying to
ND states. The sudden decay-out of the SD rotational bands are
interpreted as a barrier penetration
phenomenon~\cite{Vigezzi,ShimizuA,ShimizuB,%
Schiffer,Exp-decayout,Khoo,Leoni-decay}.
As thermally excited superdeformed states relevant
for the quasi-continuum gamma-ray spectra are concerned,
the barrier penetration is expected to be more effective.
Therefore, it is important to incorporate
not only the rotational damping effects (or the complex configuration
mixing) but also the barrier penetration
in order to describe 
the thermally excited superdeformed states.

It is interesting to study the thermally excited SD bands
because we can learn about the collective rotational motion
and the collective motion in the shape degrees of freedom at the same time.
The rotational damping phenomenon is related to
the study of quantum chaos and is crucial to investigate
the very existence of rotation bands
in nuclei at thermally excited states.
The decay-out of the SD bands as a barrier penetration problem
tells us information on how nuclear shape evolves as a function of
the excitation energy and the angular momentum.
The potential energy surface in deformation coordinates
is often used in the nuclear structure problems, but
the reliability of the energy surface has been tested
only near the minimum points in most cases.
The barrier penetration problem
gives a rare chance to explore the portions of the energy surface
far from the minimum points, i.e. it reflects the effect of
the large amplitude shape dynamics
extending from one minimum to the other through a barrier region.
The significance of studying the thermally excited SD bands
is similar to that of spontaneous fissions,
but the present problem is unique in the sense that it gives possibility
to examine such shape dynamics
under the influences of the thermal motion
and collective rotational motion.

In the present paper, we attempt to construct
a microscopic nuclear structure model including
both the rotational damping and decay-out effects. 
Theoretical treatment of the barrier penetration mechanism
relevant to the decay-out of the SD bands 
was formulated in a consistent way 
firstly by Vigezzi et al.~\cite{Vigezzi}. The validity of the
theory have been checked by more firm theoretical
considerations~\cite{Gu-Weiden},
and also by analysis of the experimental data~\cite{Exp-decayout}.
The calculation based on this theory combined with a
microscopic potential energy surface of the cranked Nilsson-Strutinsky
type and the collective mass parameter of the pair hopping model
gives good account
of the sudden decay-out of the SD bands
in $A \approx$ 150 nuclei~\cite{ShimizuB}.
On the other hand, a microscopic model of the rotational damping in
thermally excited rotating nuclei has been
formulated also on the basis of the cranked Nilsson-Strutinsky 
mean-field~\cite{Aberg,Matsuo}.
This  model treats not only the mean-field but also
the shell model configuration mixing by incorporating 
the many-particle many-hole excitations built on the cranked
Nilsson potential and by performing a shell model diagonalization of the
residual effective interactions.
Quantitative success of the model has been demonstrated recently for
normal deformed nuclei in the rare-earth region~\cite{Matsuo,Bracco-simul}
and also in $A \approx$ 110 deformed nuclei~\cite{Te-cont}. 
It was also applied to the rotational damping 
in the superdeformed nuclei~\cite{Yoshida1,Yoshida2,Leoni-Npath}.
However, so far the model does not take into account
decay-out caused by the barrier penetration.
In this paper, we extend this approach by combining with the 
barrier penetration model~\cite{Vigezzi,ShimizuA,ShimizuB}.
Previously Monte-Carlo statistical simulation models that combine
the two effects were proposed~\cite{Schiffer,Khoo}. However,
these models are designed in phenomenological ways
so that they can be used to analyze the 
experimental data with  parameter fitting or with use of
deduced parameter values.
It is important for further quantitative study to construct a 
theoretical model on the basis of the microscopic 
description of the rotational damping and the decay-out.
The purpose of the present paper is to provide with such a microscopic model.

In the next section we describe details of the model formulation.
After recapitulating briefly the cranked shell model description
of the thermally excited superdeformed states,
we discuss in detail the barrier penetration mechanism
relevant for the thermally excited superdeformed states.
Here we extend the barrier penetration model of 
Refs.~\cite{Vigezzi,ShimizuA,ShimizuB} developed
for the yrast SD bands in order to describe the decay-out of
the thermally excited SD states. 
In \S3, we present main results and discuss 
their implication on the properties of the thermally
excited SD states and associated gamma-ray transitions.
The calculations
are done for typical superdeformed nuclei,
$^{152}$Dy, $^{143}$Eu and $^{192}$Hg
in the $A \approx$ 150 and 190 mass regions.
In \S4 we 
summarize the points of this paper.

\section{The Model}\label{Sec:Model}

\subsection{Configuration mixing and E2 transitions among SD states}
\label{Sec:csm}

Excited superdeformed states lying highly above the yrast states
may consist of many-particle many-hole configurations built on
the superdeformed mean-field, but they hardly have pure 
mean-field configurations due to the two kinds of mixing effects. One is the 
mixing among SD states caused by the residual two-body interaction.
This brings about the rotational damping, which leads to fragmentation
of the rotational E2 transitions among SD states.  The other is 
the mixing between SD and ND states caused by the barrier penetration.
This leads to decay-out transitions to ND states. 
In this subsection we briefly recapitulate a microscopic
formulation of the first kind of configuration mixing
among the SD states~\cite{Matsuo,Yoshida1,Yoshida2}.

We start with the cranked Nilsson single-particle
Hamiltonian $h_\omega = h_{\rm Nilsson} - \omega J_x$
whose quadrupole and hexadecapole deformation parameters
are determined by the minimization of the potential
energy surface calculated by the Strutinsky method. 
The cranked Nilsson Hamiltonian $h_\omega$ 
defines the single-particle orbits associated
with the rotating superdeformed potential
as a function of the rotational frequency $\omega$.
Many-body shell model basis configurations are then defined in terms of the
cranked Nilsson single-particle orbits. We employ all
many-particle many-hole configurations but with
a truncation  with an upper limit on the excitation energy. 
Changing the variable from
the rotational frequency to the rotational spin $I$,
where $I$ means the expectation value of $J_x$
(angular momentum along the rotation axis),
the shell model basis states are constructed 
for each spin and parity,
as usually prescribed by the cranking model. 
This shell model basis for the superdeformed states
is denoted $\{\ket{\mu(I\pi)}\}$.

In the shell model Hamiltonian describing the mixing
among  $\{\ket{\mu(I\pi)}\}$, we include the volume-type
delta force $v(1,2) = v_{\tau}\delta ({\vec{x}}_1 - {\vec{x}}_2)$
as the residual two-body interaction. The force
strength $v_\tau$ is taken the same as Refs.~\cite{Bush,Yoshida1,Yoshida2}.
Diagonalizing the 
shell model Hamiltonian, we obtain a set of 
eigen solutions for each spin $I$ and parity $\pi$. We denote their energy
$E_\alpha(I\pi)$, and state vectors 
$\ket{\alpha(I\pi)}=\sum_\mu X_\mu^\alpha(I\pi)\ket{\mu(I\pi)}$.
The $B$(E2) strength of stretched rotational E2 transitions from an
eigenstate $\alpha$ at spin $I$ to states $\beta$ at $I-2$ are
calculated by using the obtained wavefunctions. 
It is convenient to normalize the $B$(E2) strength so that
the sum over final states for a fixed initial state becomes the unity.
We denote the normalized strength $S_{\alpha I, \beta I-2}$, which 
approximately corresponds to the branching ratio of the E2 transitions
from the state $\ket{\alpha(I\pi)}$. 
See Refs.~\cite{Yoshida1,Yoshida2,Matsuo} for details
of the cranked shell model calculation. 
The calculation of $\ket{\alpha(I\pi)}, E_\alpha(I\pi)$ and
$S_{\alpha I, \beta I-2}$ in the present paper
is the same as those in Refs.~\cite{Yoshida1,Yoshida2}.

\subsection{Barrier penetration and decay-out transitions to ND states} 
\label{Sec:tunnel}

Because of the barrier penetration effect, 
the superdeformed states $\ket{\alpha(I\pi)}$ obtained in the
cranked shell model will couple further to normally
deformed (ND) compound 
states which lie energetically near $E_\alpha(I\pi)$. 
We follow the theory proposed by Vigezzi et al.~\cite{Vigezzi} 
to describe this coupling between SD and ND states. 

It is assumed that ND compound states are distributed 
randomly around
a SD state $\ket{\alpha(I\pi)}$, and that  the 
coupling matrix elements between  $\ket{\alpha(I\pi)}$ and the
ND states have an average size of $v$. 
The Vigezzi model describes this coupling 
in terms of a quantum tunneling taking place
through the potential energy barrier in the deformation space.
This leads to the relation between $v$ and the tunneling width $\Gammat$ as,
\begin{equation}
  \Gammat = \frac{2\pi v^2}{\Dn} = 2\pi v^2 \rhon,
\label{EQ:GamTun}
\end{equation}
where $\rhon$ is the level density of the ND states and $\Dn=1/\rhon$
is the mean level spacing.

Due to the coupling, some components of normal deformed
states are mixed in the superdeformed state $\ket{\alpha(I\pi)}$.
Since the normal deformed states have also electromagnetic transition
probabilities, the SD states thus mixed have possibility to decay not only 
with the rotational E2 gamma-rays feeding other SD states 
but also with the gamma-rays feeding to normal deformed states.
The latter is the origin of the decay-out transitions.
For the electromagnetic transitions associated with ND states,
we consider the statistical E1 and the rotational E2 transitions.
A key quantity that reflects the effect of decay-out
is the branching ratio of the decay-out transitions,
$\Nout$; it depends on the transition probabilities
of the competing electromagnetic transitions, i.e. 
the E2 transition width $\Gammas$ among superdeformed states,
and the E1 and E2 transition width $\Gamman$ for the
normal deformed states, and also depends on the amplitudes
of the normal deformed states mixed in the superdeformed states.
The mixing amplitudes may fluctuate strongly depending on the
relative positions between the SD and ND states. In addition 
statistical fluctuation in coupling matrix elements is caused
by the compound nature of the ND states. We can evaluate
the average value of $\Nout$ assuming 
random matrix model description of the ND states and 
the Gaussian orthogonal ensemble (GOE)~\cite{Vigezzi}. 
The resultant average value $\left<\Nout\right>$  of the
decay-out branching ratio 
is a function of the tunneling width $\Gammat$ or the coupling
matrix element $v$, 
the decay width $\Gamman$ of the ND transitions,
the decay width $\Gammas$ of the SD transitions, and the average 
level spacing $\Dn$ of the ND states.
More specifically, $\left<\Nout\right>$ 
is a function of the two ratios $\Gammat/\Dn$ 
and $\Gammas/\Gamman$ in the model of Ref.~\cite{Vigezzi},
and we have calculated its functional form by numerically diagonalizing 
GOE matrices of dimension four hundreds.
Recently theoretical models to calculate $\left<\Nout\right>$ 
have been proposed~\cite{Weidenmuller,Barrett,Gu-Weiden}.
Approximate treatment of Ref.~\cite{Weidenmuller} gives a simple result,
$\left<\Nout\right> = \Gammat/(\Gammat+\Gammas)$,
but it was found that the approximation is not valid in the realistic
situations of decay-out.
In Ref.~\cite{Gu-Weiden} the statistical theory
using the supersymmetry technique has been rigorously solved
in general conditions, and a simple two-level model has been
investigated in~\cite{Barrett},
which is appropriate in the weak coupling limit.
We examined the result of the Vigezzi model~\cite{Vigezzi}
by comparing with that of~\cite{Gu-Weiden}, and we found
that both theories give the same results within the relevant
range of parameters $\Gammas,\Gamman,\Gammat$ and $\Dn$
for the decay-out of SD bands in both the $A \approx$ 150 and 190 regions.

If the tunneling width $\Gammat$, the level spacing
$\Dn$ of the ND states,  and the electromagnetic decay rates 
$\Gammas$ and $\Gamman$ are given, one can calculate
the average decay-out probability $\left<\Nout(E,I)\right>$ 
for any specified values
of the spin and the excitation energy of the superdeformed states.
It is then possible to include the effect of the decay-out on 
the excited superdeformed states  $\ket{\alpha(I\pi)}$ calculated
by the shell model configuration mixing. 
Now the state  $\ket{\alpha(I\pi)}$ with energy $E_\alpha$ has an average 
branching ratio $1-\left<\Nout(E_\alpha,I)\right>$ 
for the rotational E2 transitions feeding
to other superdeformed states at $I-2$, and the ratio 
$\left<\Nout(E_\alpha,I)\right>$ 
for the decay-out transitions. Thus E2 transition probabilities 
$S_{\alpha I,\beta I-2}$ calculated by the cranked shell model
(see \S\ref{Sec:csm}) are renormalized as
\begin{equation} 
\tilde{S}_{\alpha I,\beta I-2}=
 \Bigr(1-\left<\Nout(E_\alpha,I)\right>\Bigl)\,S_{\alpha I,\beta I-2}. 
\label{EQ:E2-renorm}
\end{equation}
Hereafter we denote the average decay-out
probability  $\Nout$ without the symbol of average 
for simplicity of notation.

In the next section, we analyze properties of rotational 
E2 gamma-rays from the
excited SD states in terms of this renormalized E2
transition probabilities. 
In the rest of this section
let us describe more details of the model constituents.

\subsection{Collective potential and mass tensor} 
\label{Sec:PotMass}

In order to specify the tunneling width $\Gammat$,
we assume that the barrier penetration takes place as 
the nucleus changes its shape from the superdeformation 
to the normal deformation.
We consider the two quadrupole deformation parameters
$(\epsilon_2,\gamma)$ of the cranked Nilsson model in the Lund convention
as the dynamical variables relevant for the deformation change.
The collective Hamiltonian for the deformation variables is given by
\begin{equation}
  H=\frac{1}{2}\,\sum_{ij=1}^2 m_{ij}(q_1,q_2)\,\dot{q_i}\dot{q_j}+V(q_1,q_2) 
\end{equation}
where the collective coordinates $q_1$ and $q_2$ are defined by
\begin{equation}
  q_1=\epsilon_2 \cos{(\gamma + 30^\circ)}, \quad
  q_2=\epsilon_2 \sin{(\gamma + 30^\circ)}.
\end{equation}

The collective potential energy  $V(q_1,q_2)$ is calculated for each signature
and parity by means of the cranked Nilsson-Strutinsky method
with the monopole pairing residual interaction
as a function of the deformation parameters ($\epsilon_2, \gamma$),
or equivalently $(q_1,q_2)$.
Both the static and dynamic pairing correlations are taken into account
in terms of the cranked Hartree-Bogoliubov method and
the random phase approximation (RPA) as has been done in Ref.~\cite{RMP89}.
Namely the response function technique is used with small
imaginary part of frequency, $\delta={\rm Im}(\omega)=200$ keV$/\hbar$,
which is slightly larger value but saves computation time considerably.
The cut-off parameter for calculating the effective RPA pairing
gap~\cite{RMP89,ShimBrog}, which is used in Eq.~(\ref{EQ:MassTens}),
is taken as $\omega_{\rm cut}=400$ keV$/\hbar$.
Note that exchange contributions are excluded in both
the RPA correlation energy and RPA pairing gap
($\delta {\tilde E'}_{\rm RPA}$,
${\tilde \mDelta}_{\rm RPA}$ in \cite{ShimBrog}).
The angular momentum is conserved in the tunneling problem so that
the potential energy surface should be calculated with fixed spin values.
Spin-interpolation is used for this purpose.
The hexadecapole deformation ($\epsilon_4$ in the Nilsson potential) 
is determined so as to minimize the potential energy for each point of
$(\epsilon_2,\gamma)$ deformation parameters and the spin values.
The grid of calculation for the deformation parameters
($q_1, q_2, \epsilon_4$) are
$-0.18 \le q_1 \le 0.60$, $-0.12 \le q_2 \le 0.42$ and
$-0.10 \le \epsilon_4 \le 0.14$ with an interval of 0.06.
The grid for rotational frequency, which is used for the spin-interpolation,
is determined by $\mDelta\omega = \mDelta I/{\cal J}_{\rm str}$ with
$\mDelta I=3$, where ${\cal J}_{\rm str}$ is Strutinsky-smoothed
moment of inertia calculated for each set of deformation parameters
($\epsilon_2,\gamma, \epsilon_4$),
and the maximum frequency is extended
to give calculated spin value up to 60 $\hbar$.
The $ls$ and $l^2$ parameters of the Nilsson potential
is taken from Ref.~\cite{BenRag}.
The remaining parameter for the potential energy calculation
is the monopole pairing force strength, which are determined by
the smoothed gap method~\cite{Brack}
for each set of deformation parameters with
the model-space energy cut-off of $1.2\hbar\omega_0$
below and above the Fermi surface
($\hbar\omega_0 =41.0/A^{1/3}$ MeV is the oscillator frequency).
Note, however, that in Ref.~\cite{ShimizuB} 
the minimization of energy with respect to
the $\epsilon_4$ deformation has been done
only approximately, namely the value determined
without pairing correlations has been used, and
the strength of pairing correlations has been determined
by the standard smoothed pairing gap ${\tilde \mDelta} =12 /\sqrt{A}$ MeV.
We have found that the full minimization against $\epsilon_4$
including the static and dynamic pairing correlations
leads considerably smaller height of barrier compared with
the previous calculations~\cite{ShimizuB},
which makes the tunneling probability too large.
We, therefore, employed smaller pairing force strength determined by
${\tilde \mDelta} =10 /\sqrt{A}$ MeV in the present calculations.
This value is chosen to obtain barrier height which gives
similar tunneling probability to that of the previous calculations.
The pairing force strength thus determined is smaller by about 6\%
for nuclei in both the Dy and Hg region.

As for the collective mass tenser $m_{ij}(q_1,q_2)$,
we use the one based on the pair hopping model~\cite{Bertsch,Barranco},
where the semiclassical (Fermi-gas) estimate
for counting the number of pair-level crossing
has been extended to the two dimensional case
of the volume-conserving ellipsoidal deformation~\cite{ShimizuA,MassTensor}.
The mass tensor is given explicitly as\footnote{
  There is a misprint of missing factor (3/2) in
  Eqs.~(22) and (26) of Ref.~\cite{MassTensor}.
  It is corrected in Eq.~(\ref{EQ:HopMass}). }
\begin{equation}
 m_{ij}(q,I) = \frac{2 \mDelta_0^2}
  { \Bigr[\mDeltan^{(\eff)}(q,I)\Bigl]^2
  + \Bigr[\mDeltap^{(\eff)}(q,I)\Bigl]^2} \, m_{ij}^{(\hop)}(q),
\label{EQ:MassTens}
\end{equation}
where $\mDeltan^{(\eff)}$ and $\mDeltap^{(\eff)}$ are the effective RPA
pairing gap~\cite{RMP89,ShimBrog} for neutron and proton calculated at
each deformation $(q)$ and spin $I$,
and $m_{ij}^{(\hop)}(q)$ is written analytically as 
\begin{equation}
 m_{ij}^{(\hop)} (q) = m_0^{(\hop)} \,\frac{3}{2}
 \biggr[ \,\frac{1}{a^2} \,
 \frac{\partial a }{\partial q_i} \frac{\partial a }{\partial q_j}
 + \frac{1}{b^2} \,
 \frac{\partial b }{\partial q_i} \frac{\partial b }{\partial q_j}
 + \frac{1}{c^2} \,
 \frac{\partial c }{\partial q_i} \frac{\partial c }{\partial q_j}
 \,\biggl],
\label{EQ:HopMass}
\end{equation}
with
\begin{equation}
 m_0^{(\hop)} = \frac{\hbar^2 G_0}{\mDelta_0^2} \, \frac{A^2}{27}.
\label{EQ:sphericalHopMass}
\end{equation}
Here the three axis lengths, $a$, $b$, and $c$, for the volume-conserving
ellipsoidal shape can be specified
by the deformation parameter ($\epsilon_2, \gamma$),
and $m_{ij}^{(\hop)}(q=0)=m_0^{(\hop)}\delta_{ij} $,
i.e.  $m_0^{(\hop)}$ is the hopping mass at the spherical shape.
The empirical values for the constant parameters,
$G_0=25/A$ MeV and $\mDelta_0=12/\sqrt{A}$, are employed~\cite{Barranco}.
The same tunneling model consisting of the collective potential
and the mass tensor explained above has been used for the problem of decay of
high-$K$ isomers, see Ref.~\cite{Kisomer} for details.

\subsection{Tunneling decay width} 
\label{Sec:TunWidth}

If the decay-out of the yrast SD band is concerned, the tunneling width
of this state can be evaluated semiclassically as 
\begin{equation}
   \Gammat = \frac{\hbar\omegas}{2\pi}(1+\exp{2S})^{-1},
\label{EQ:WKB1}
\end{equation}
\noindent
where $T=(1+\exp{2S})^{-1}$ is the tunneling transmission coefficient and
$S$ is the (imaginary time) action integral along a classical tunneling path, 
\begin{equation}
 S(E)=\int_{\rm path}ds\sqrt{2M_0(V(q(s))-E)},
\label{EQ:action}
\end{equation}
where $s$ is a parameter along the classical tunneling path,
and is defined according to
\begin{equation}
 M_0 ds^2 = \sum_{ij}m_{ij}(q)dq_i dq_j.
\label{EQ:s}
\end{equation}
Here $M_0$ is a mass unit and we take $M_0=m_0^{(\hop)}$.
See \S\ref{Sec:path} for determination of the path.
When we apply to the yrast SD band, 
the energy $E$ of the classical tunneling path
is chosen as the 
zero-point energy of the shape motion around the SD minimum. The factor
$\hbar\omegas/{2\pi}$ in Eq.~(\ref{EQ:WKB1}) 
is the knocking probability where
$\omegas$ is the vibrational frequency of the collective
motion along the tunneling path near the SD potential minimum.
Eq.~(\ref{EQ:WKB1}) has been used for 
description of the decay-out
of SD rotational bands~\cite{Vigezzi,ShimizuA,ShimizuB,Khoo,Exp-decayout}.

The above expression, however, may not be directly 
applied to the tunneling of the excited superdeformed states
with high thermal excitation energy since it assumes that 
the internal nucleon configuration is kept adiabatically along
the tunneling process, considering 
only the collective shape dynamics. If the highly excited
SD states are concerned, the collective motion associated with the
tunneling will be coupled with
the complex internal excitations, which is a basic feature of
highly excited compound states caused by the residual two-body
interaction. 
Following Bj{\o}rnholm-Lynn~\cite{Bjornholm-Lynn}, 
which discusses the tunneling of fission isomer states, 
we consider that collective vibrational excited states
form doorway states of tunneling, and the strength
of the vibrational states are spread over many compound states.
If we assume that the coupling is sufficiently strong and
the strength is distributed uniformly, 
the average strength of  doorway states in a compound energy
level is evaluated as $\approx \Ds/\hbar\omegas$. 
Here $\Ds=1/\rhos$ is the average level spacing
of the superdeformed states at a given excitation energy 
($\rhos$ is the level density of SD states).
The knocking 
probability for the tunneling of the compound level is then reduced as
$\approx (\hbar\omegas/2\pi)(\Ds/\hbar\omegas)=\Ds/2\pi$  in average.
The tunneling width in this situation is given by
\begin{equation}
 \Gammat = {\Ds \over 2\pi}(1+\exp{2S})^{-1}.
\label{EQ:WKB2}
\end{equation}
Here the action $S$ of the classical tunneling path is given
by Eq.~(\ref{EQ:action}) whereas the energy $E$ of the path 
is chosen as the energy of the excited SD state under consideration.
The knocking probability $\Ds/2\pi$ 
has been used also in the previous simulation model of
Ref.~\cite{Schiffer,Leoni-decay}.

An important feature of Eq.~(\ref{EQ:WKB2}) is that
it can describe the limit of statistical mixing when applied to
the excited states located above the barrier height.  
Since the 
SD and ND states above the barrier are expected to mix completely 
with the statistical weight
($\rhon$ vs. $\rhos$) and to form the compound states,
an average strength of a SD state $\ket{\SD}$
in the compound states $\ket{i}$ is given by
$|\left<i|\SD\right>|^2 \approx \rhos/(\rhon + \rhos) = \Dtot/\Ds$,
where $\Dtot$ denotes the average spacing of the
statistically mixed compound states.
On the other hand, if 
we assume $S=0$ above the barrier, the tunneling
width given by Eq.~(\ref{EQ:WKB2})
reads $\Gammat = \Ds/4\pi$. Using this expression, the average strength 
of the superdeformed state can be evaluated as
$|\left<i|\SD\right>|^2 \approx \Dtot/\Gammat\approx 4\pi \Dtot/\Ds $,
which agrees with the value for the statistical mixing 
except a numerical factor.
If the factor $1/2\pi$ in Eq.~(\ref{EQ:WKB2})
were removed, better agreement with the statistical limit would be achieved.
Since the argument of Ref.~\cite{Bjornholm-Lynn} for the derivation
of Eq.~(\ref{EQ:WKB2}) concerns an accuracy of an order of magnitude,
such possibility may also be allowed. As shown later, however,
the difference in the numerical factor has only minor effect.
It is also interesting to see that an 
expression of the tunneling width similar to Eq.~(\ref{EQ:WKB2}) is
derived  for 
the tunneling of a chaotic system with two degrees of freedom on the
basis of the semiclassical theory~\cite{Creagh}. 
There the tunneling width is proportional
to the average level spacing of the
initial states and the tunneling transmission coefficients, 
as in Eq.~(\ref{EQ:WKB2}), but it also has 
an additional prefactor related to a second 
order dynamical correction~\cite{Creagh}.  Eq.~(\ref{EQ:WKB2})
does not include such a dynamical factor, which is, however, not known
for many-body fermion systems.

We evaluate the tunneling width $\Gammat$ of the excited SD states
by means of  Eq.~(\ref{EQ:WKB2}) since it describes the
average value of the tunneling width $\Gammat$, assuming the
uniform distribution of doorway strength.
However, if the strength distribution of vibrational states is
not uniform and concentrated to some excited SD states, the 
tunneling width for those states may be larger than that given by
Eq.~(\ref{EQ:WKB2}), and some other states may have smaller 
tunneling width. Although it is not easy to evaluate such fluctuations,
we can evaluate an upper limit of $\Gammat$. It is
given by Eq.~(\ref{EQ:WKB1}) and calculated at the excitation energy
under consideration since this equation corresponds to
a special situation where a excited SD state contains large amount 
of the vibrational strength.  We may also 
use Eq.~(\ref{EQ:WKB1}) for the excited SD states in the following analysis
in order to estimate an upper limit of $\Gammat$. We adopt
Eq.~(\ref{EQ:WKB1}) also to describe the decay-out of yrast SD states.
It is important to note that the two evaluations, 
Eqs.(\ref{EQ:WKB1}) and (\ref{EQ:WKB2}), 
give  different results.
Since Eq.~(\ref{EQ:WKB2}) contains
the knocking probability $\Ds/2\pi$ that
decreases steeply with increasing excitation energy,
it gives significantly smaller value of the tunneling width than
Eq.~(\ref{EQ:WKB1}) at high excitation energy.
As far as the the SD states near the yrast
states are concerned, on the other hand, 
the two expression gives more or less similar
value since the level density $\Ds$ near the yrast line may be
order of several hundred keV, which is approximately
the same as the value of  $\hbar\omegas$ (See Table~\ref{Tab:TUN}).
In the following section, we will discuss the difference by using 
numerical results.

We remark that it is also possible~\cite{Aberg-Chaos} for
the tunneling width $\Gammat$ to be reduced from the average
value given by Eq.~(\ref{EQ:WKB2}), if the strength of doorway
states is not uniformly distributed but concentrated in between
the excited SD states and so the coupling between them is minimal.
However, we do not
evaluate such reduction since a precise modeling of the doorway 
states is required and it is beyond the scope of the present paper.

\subsection{Tunneling path and classical action}
\label{Sec:path}

\begin{figure} 
\centerline{
\epsfxsize=120mm\epsffile{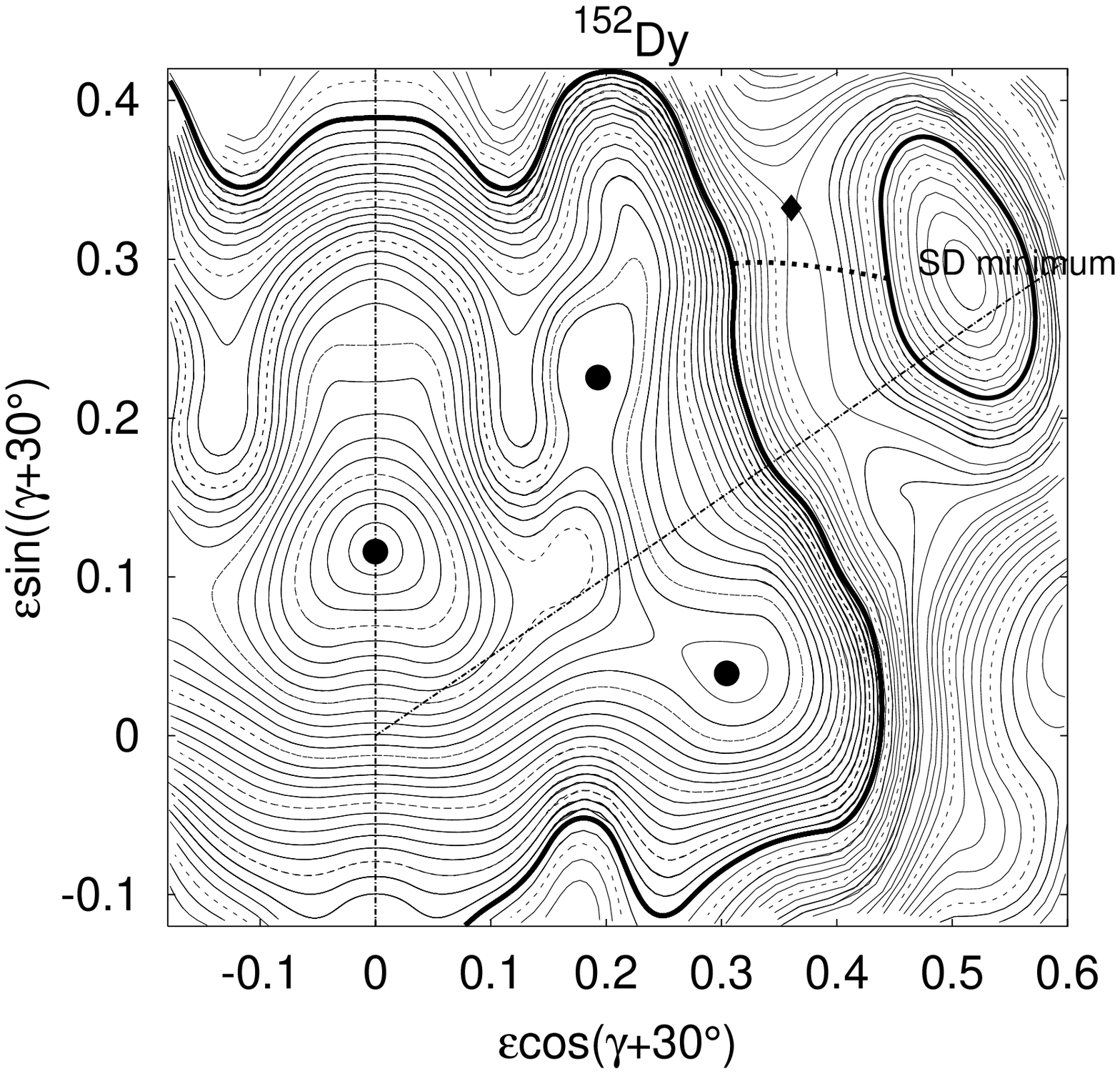}}
\centerline{
\epsfxsize=60mm\epsffile{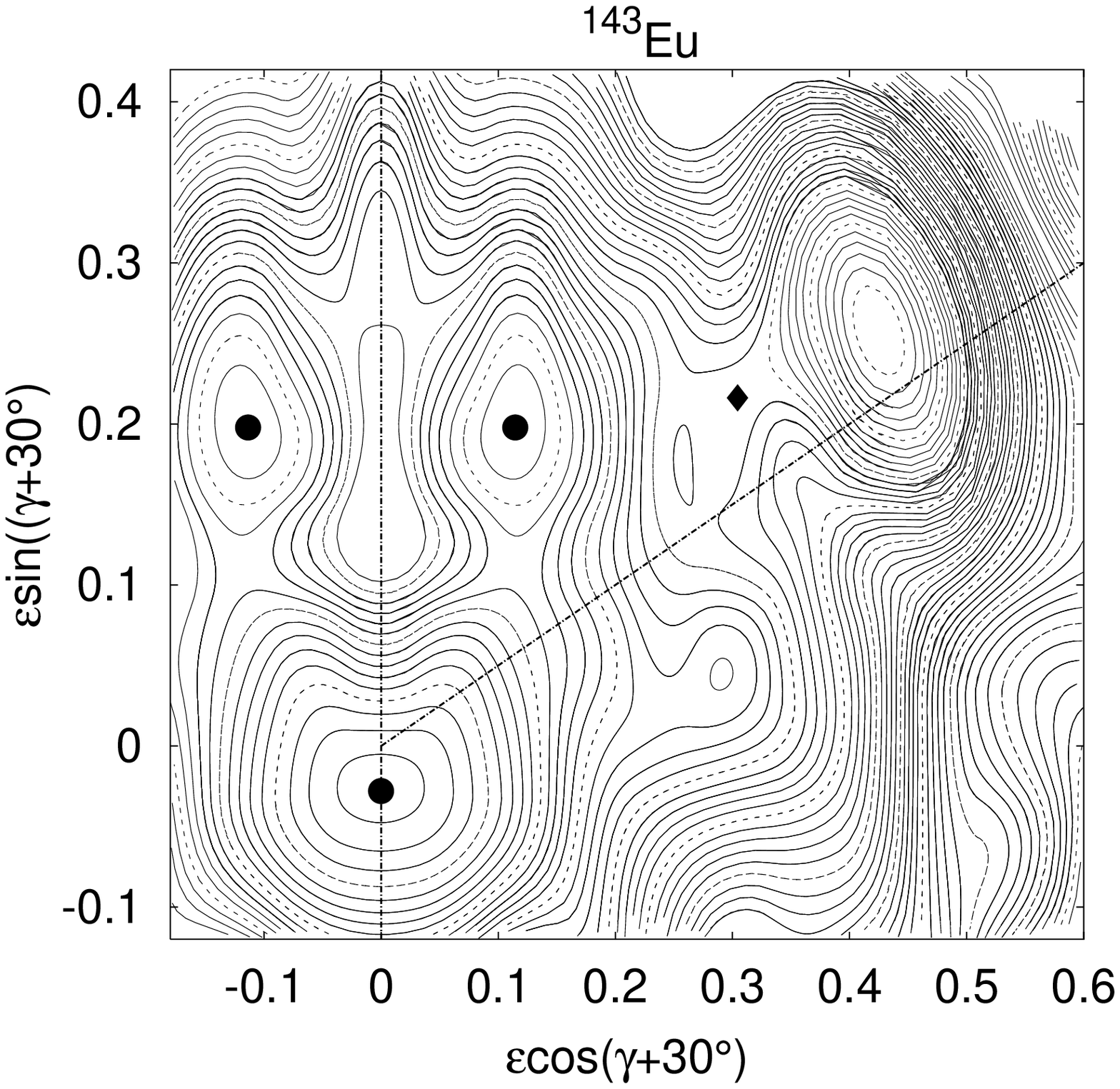}
\epsfxsize=60mm\epsffile{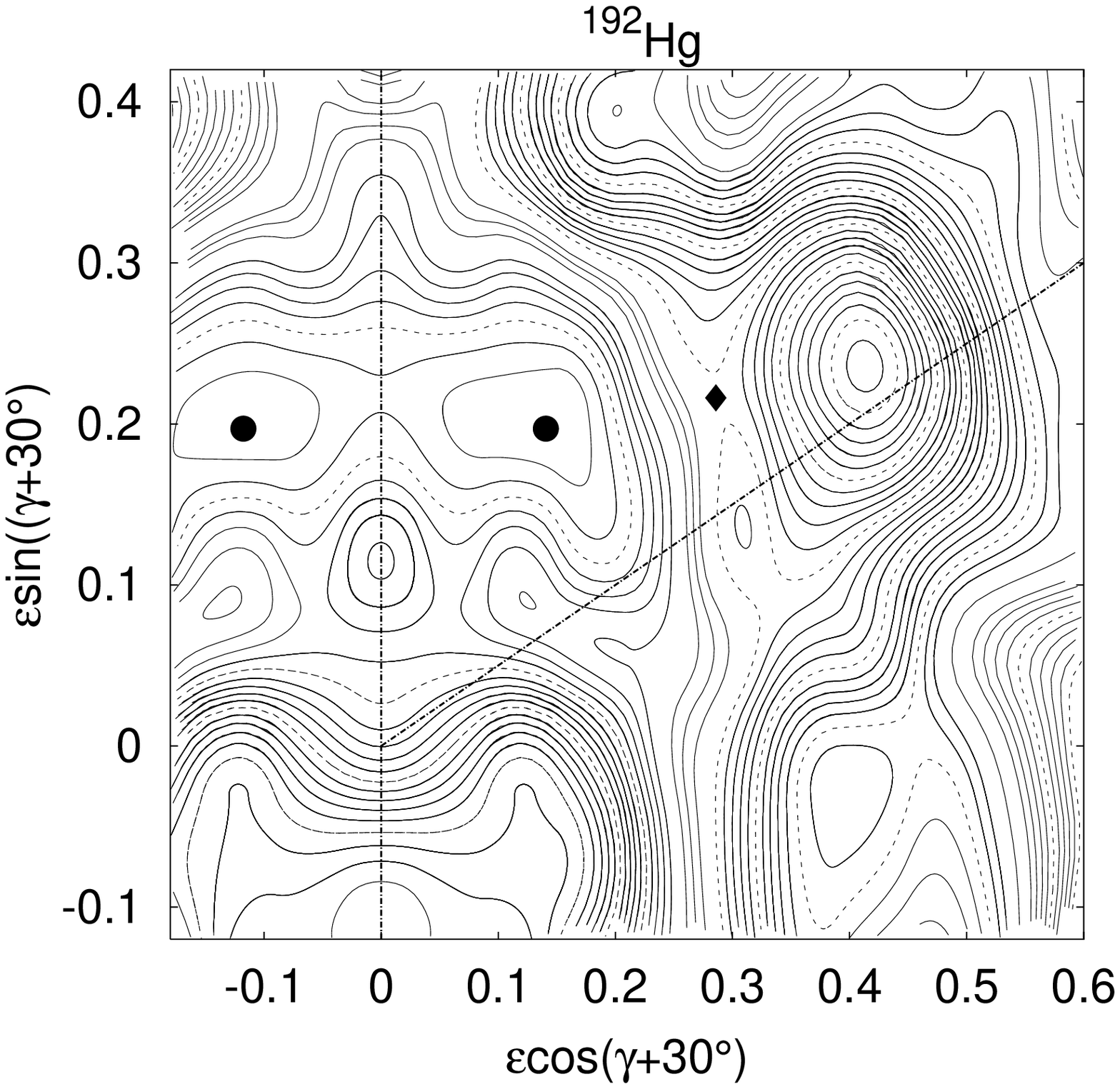}}
\caption{\label{Fig:PES}
The calculated potential energy surfaces $V(q_1,q_2)$ for 
$^{152}$Dy ($I=40^{+}$), $^{143}$Eu ($I=61/2^{+}$) and
$^{192}$Hg ($I=30^{+}$) shown in the top, the left bottom, and 
the right bottom panels, respectively. 
Interval of energy contours is 0.2 MeV and $\gamma=0$ is 
indicated by an oblique line.
The SD minima are located around $\epsilon_2=\sqrt{q_1^2+q_2^2}=0.45\sim0.6$,
$\gamma \approx 0$.
In the top panel, the energy contour with $E=E_\SD^0+1.5$ MeV
is shown with thick solid curves, and 
the least action path at this energy with a thick dashed curve. The
solid circles indicate local minima of the ND side whereas the saddle-point is
also shown with a diamond. }
\end{figure}

Examples of the calculated Nilsson-Strutinsky
potential energy surface $V(q_1,q_2)$  are shown in
Fig.~\ref{Fig:PES}. The local minimum at large quadrupole deformation 
$\epsilon_2 \approx 0.4-0.6$, $\gamma \approx 0$ corresponds to the 
superdeformed state. The minimum with smaller quadrupole deformation
corresponds to normal deformation. The potential barrier 
and the saddle-point separating
the two classes of states are also seen in the figure. 
The quantum tunneling takes place through the barrier 
from the excited superdeformed states situated at the superdeformed minimum 
to the normal deformation.

The tunneling path and the associated classical action integral is
calculated in the following way. For a given energy $E$, 
the deformation space $(q_1,q_2)$ is separated into classically
allowed and forbidden regions, whose borders are given by energy
contours with energy $E$. There are two allowed regions,
one surrounding the SD minimum and the other corresponding to normal
deformation. The tunneling path is the one that starts at a
point on the border of the allowed region at the SD side, travels
through the barrier region, and reaches a point on the border of the other 
allowed region at the ND side (See Fig.~\ref{Fig:PES}). 

The tunneling path should minimize the action $S(E)$, Eq.~(\ref{EQ:action}),
under a constraint of fixed energy $E$. The minimal action path
obeys a classical equation of motion
\begin{equation}
  \sum_{j} m_{ij} \frac{d^2q_j}{d\tau^2}
  +\sum_{jk} \frac{\partial m_{ij}}{\partial q_k}
             \frac{dq_j}{d\tau} \frac{dq_k}{d\tau}
  -\frac{1}{2} \sum_{jk} \frac{\partial m_{jk}}{\partial q_i}
             \frac{dq_j}{d\tau} \frac{dq_k}{d\tau}
   - \frac{\partial V}{\partial q_i} = 0,
\label{EQ:eqmotion}
\end{equation}
where $\tau=it$ is the imaginary time and it is related to
the path parameter $s$, Eq.~(\ref{EQ:s}), through
\begin{equation}
   \frac{d\tau}{ds} = \sqrt{\frac{M_0}{2\,(V(q)-E)}}.
\label{EQ:taus}
\end{equation}
Namely, the tunneling motion corresponds to the classical motion
in the inverted potential energy, $-(V(q)-E)$.
Solving the classical equation
of motion with an initial condition starting at a point on the border 
at the SD side, a classical trajectory is obtained. 
The equation is of the second order and the initial condition is
given by $q_i=a_i$ and $dq_i/d\tau=0$ at $\tau=0$, where
the point $P^{(0)}=(a_1, a_2)$ is on the energy contour at the SD side.
The zero initial velocity (in imaginary time) is the consequence
of the energy conservation.
An arbitrary classical trajectory does not 
necessarily reach the energy contour at the ND side,
but if we choose a proper initial point $P^{(0)}$, 
the classical path reaches the ND energy contour.
This path, called the escape path, gives the minimum action.
We search such a solution by changing the initial point $P^{(0)}$
of the classical path on the energy contour.
This method to find the minimal action path is parallel
to that of the Schmid's method~\cite{Schmid}, 
which deals with the path starting at the potential minimum.
Since in this method the initial point $P^{(0)}$ is placed at
a potential minimum and
the force ${\partial V}/{\partial q_i}=0$, 
an infinitesimal initial velocity should be imposed in some direction,
which should be chosen properly to obtain the escape path
reaching to the contour of the ND side.
The search of direction of the infinitesimal initial velocity
in the Schmid's method just corresponds to the search of the initial point
on the contour in our case of the finite excitation energy.
The numerical calculation is relatively easy in our case
since it is a one dimensional search
in the two dimensional collective coordinate.
Depending on the shape of the barrier and the mass tensor, more than
one solutions may exist as local minima of the action.
In such cases, we adopt the solution which gives the smallest action.

An example of the obtained tunneling path is shown 
in Fig.~\ref{Fig:PES} for the excitation energy 
$E_\ex=1.5$ MeV measured from the SD potential minimum.
For energy below the saddle-point,
the tunneling path and the action $S$
is obtained by the procedure just described.
Above the saddle-point energy we set $S=0$.
When the excitation energy under consideration becomes lower than the ND
yrast, there should be no tunneling, i.e., $\Gammat =0$.
In the following, we define the barrier energy $\Vb$ 
as a excitation energy of
the saddle-point, i.e. $\Vb = E_{\rm saddle}$
since the action vanishes when the excitation energy exceeds
the barrier energy, i.e.  $S=0$ if $E \ge \Vb$.
Note, on the other hand, that 
the least action path 
does not necessarily go through the saddle point,
see e.g. Ref.~\cite{Kisomer} and also Fig.~\ref{Fig:PES}.

We take into account of the zero-point oscillation of the
shape dynamics for the description of the tunneling. This can be
achieved just by assuming that 
the yrast superdeformed state has 
an zero-point oscillation energy $\hbar\omegas/2$ above the 
SD minimum (with energy $E_\SD^0$) of the potential energy surface
$V(q_1,q_2)$. Here
$\omegas$ is the frequency of the oscillation at the SD potential minimum 
given by a small amplitude approximation along the tunneling path
at the energy $E \approx E_\SD^0$ (see Table~~\ref{Tab:TUN}).
Namely, the tunneling path and the action $S(E)$ 
for the yrast superdeformed states are calculated with
the energy $E=E_\SDyr \equiv E_\SD^0+ \hbar\omegas/2$.
For the excited states having the excitation energy $U$ (measured from
the SD yrast state), we calculate the tunneling path with 
the energy  $E=E_\SD^0 + \hbar\omegas/2 + U$.
In the following, we use $U$ to denote the physical excitation energy 
measured from the SD yrast energy 
$E_\SDyr=E_\SD^0 + \hbar\omegas/2$ while we use
another notation $E_\ex=E-E_\SD^0$ to represent the energy relative to
the potential energy minimum $E_\SD^0$. 

In Ref.~\cite{ShimizuB} a brute force method has been used in order to
obtain the minimum action path, i.e. the path is divided
into finite segments and the positions of all segments are varied
in such a way to realize the minimum action.
In Ref.~\cite{Kisomer} the Schmid's method has been used to obtain
the minimum action path at $E_\ex=0$, along which the action is
calculated with using $E_\ex=\hbar\omega_s/2$ in Eq.~(\ref{EQ:action}).
This treatment can be justified
if $E$ is much smaller than the energy of the saddle point.
In the present paper, on the other hand, we
solve the least action path explicitly, starting from the finite energy
contour as is explained above, since the excitation energy
ranges up to the saddle-point energy.

\begin{figure} 
\centerline{
\epsfxsize=50mm\epsffile{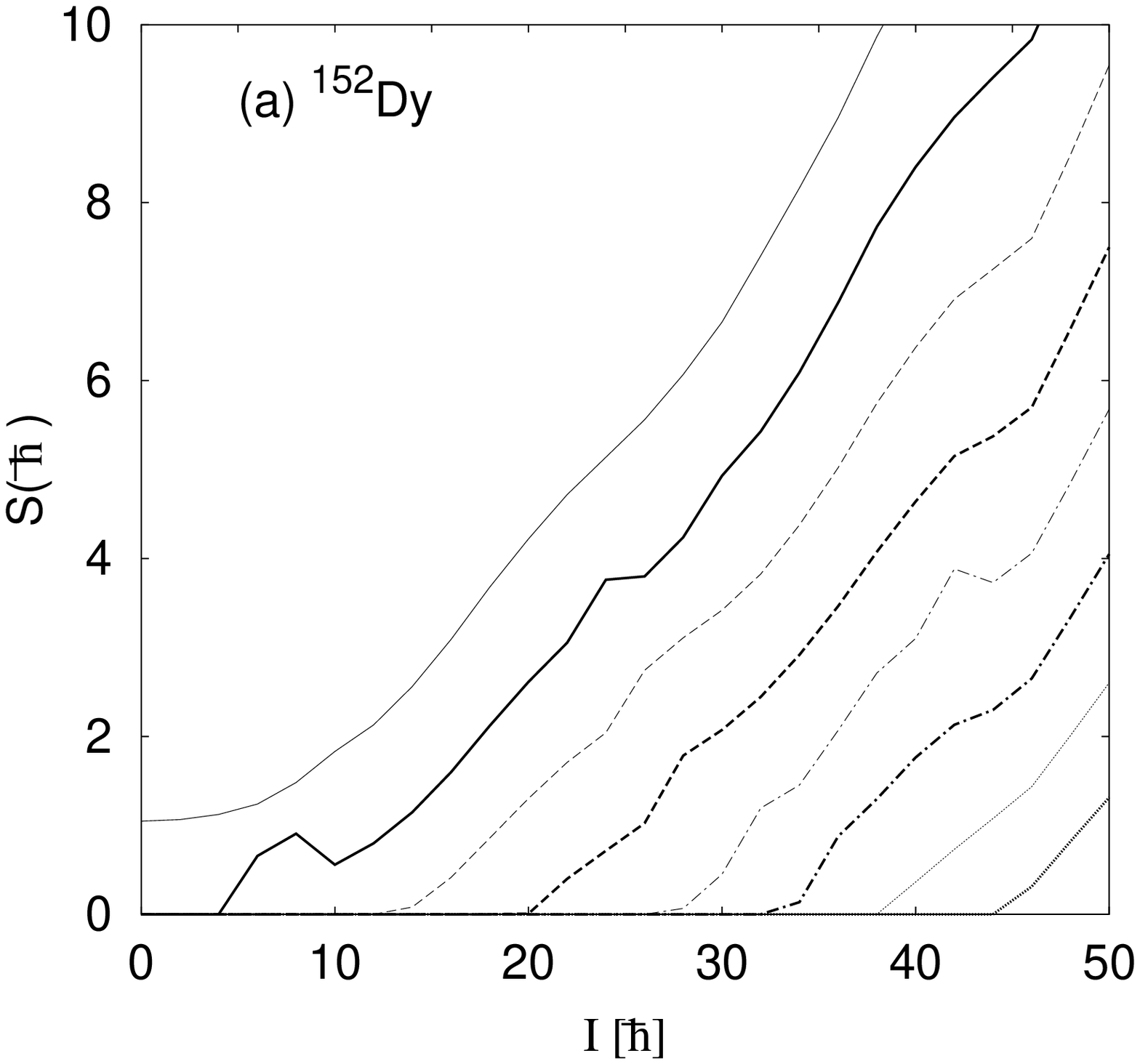}
\epsfxsize=50mm\epsffile{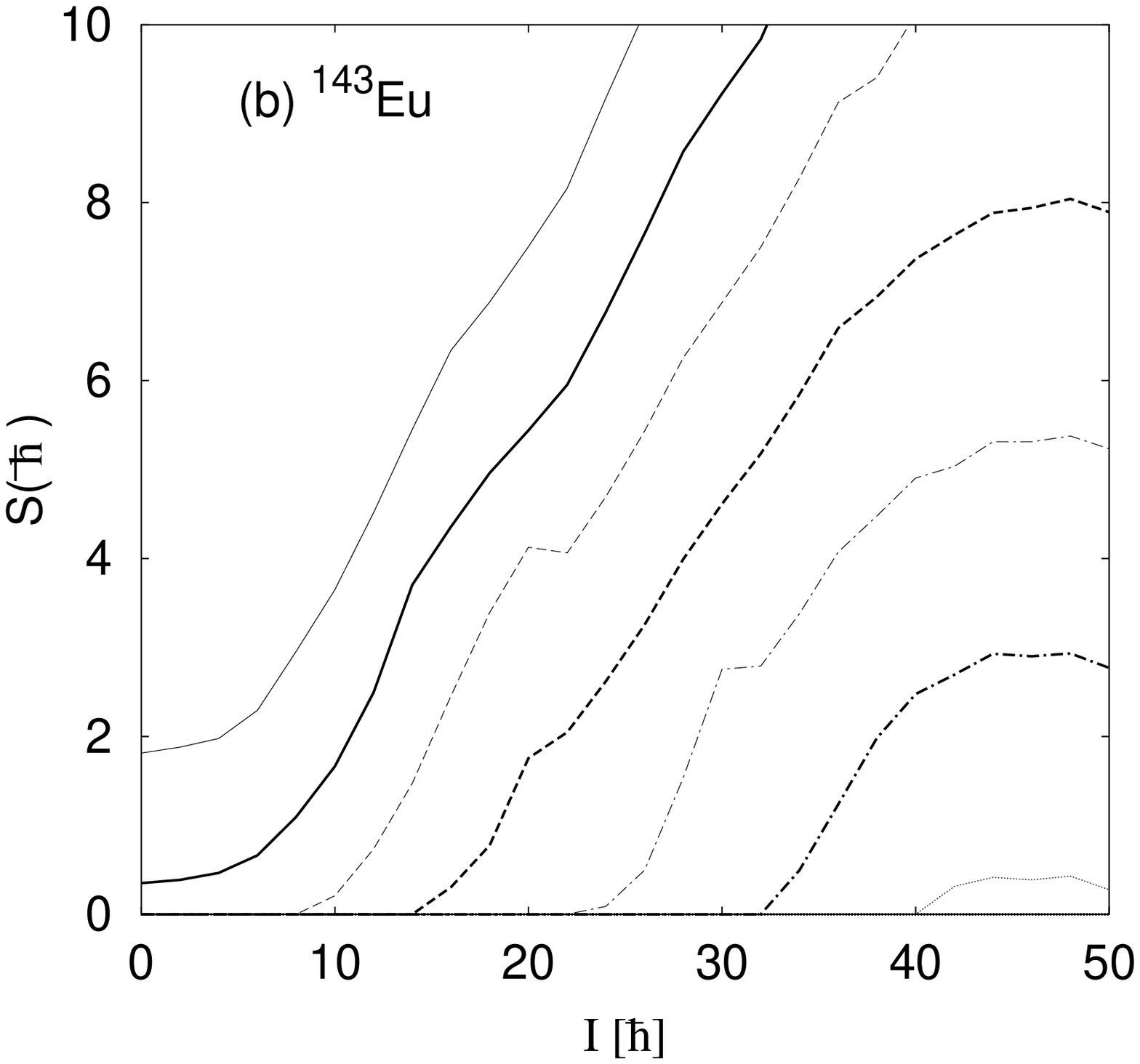}
\epsfxsize=50mm\epsffile{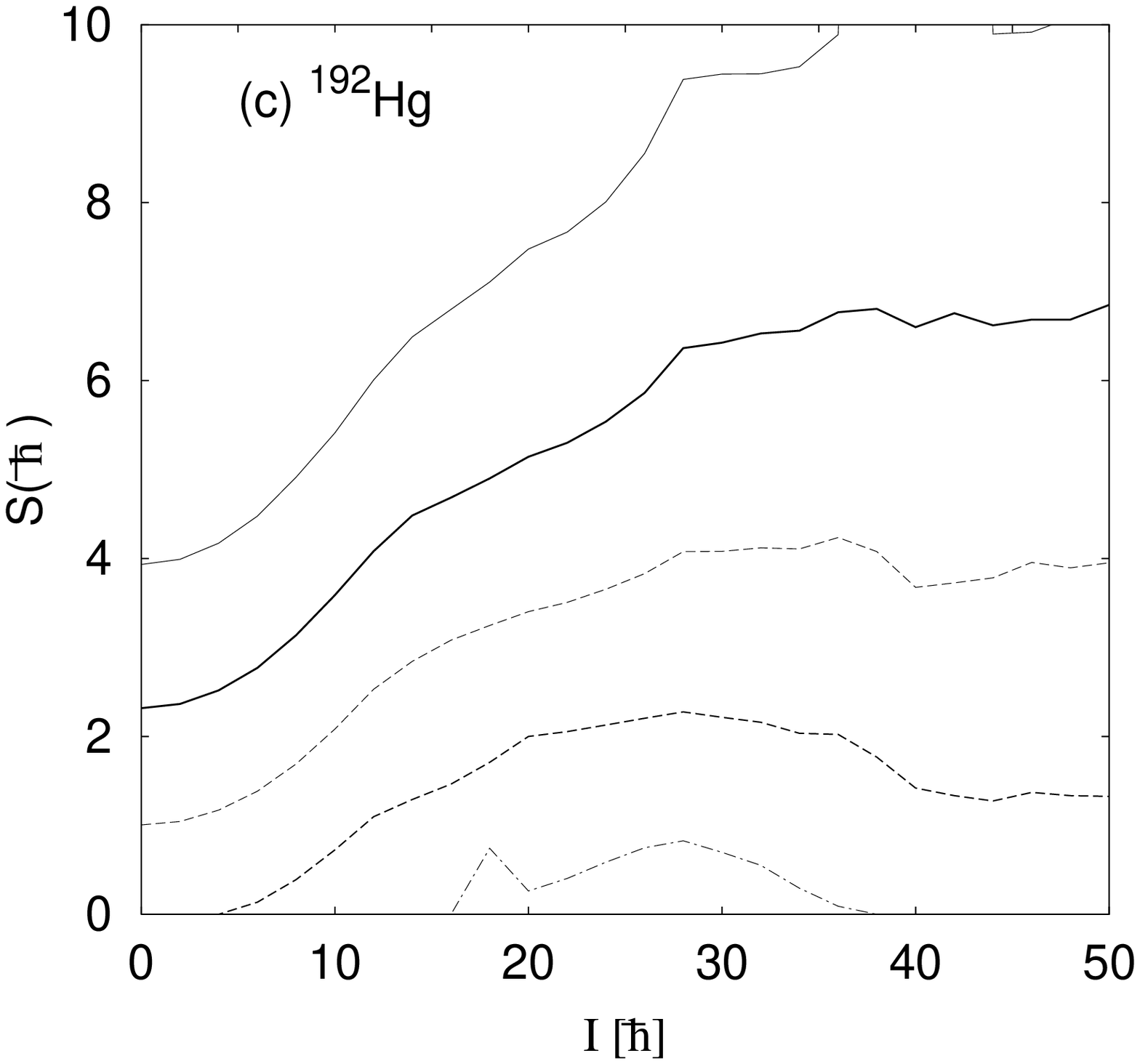}}
\caption{\label{Fig:ACT}
The tunneling action $S(E,I)$ plotted as a function of spin 
for different excitation energies for $^{152}$Dy $(\alpha,\pi)=(0,+)$,
$^{143}$Eu $(1/2,+)$ and
$^{192}$Hg $(0,+)$.
The signature and parity  $(\alpha,\pi)$ for each nuclei 
is taken the same as the yrast SD configuration,
as in Fig.~\ref{Fig:PES}.
The thin solid curve with the largest value of action 
corresponds to the zero excitation energy
$E_\ex=0.0$ (the energy of the SD potential minimum) while
the others are for $E_\ex=0.5, 1.0, 1.5, \cdots$ MeV. 
}
\end{figure}

In Fig.~\ref{Fig:ACT} we plot the action value of the 
classical tunneling path 
as a function of angular momentum for different values of
excitation energies for the same parity and signature quantum
numbers as those in Fig.~\ref{Fig:PES}.
The thick solid line corresponds approximately to the
action for the yrast SD band. (This is because the zero-point energy
is about 0.5 MeV.) 
The increase in the action  with spin is caused by
increasing barrier height, $\Vb-E_\SDyr$,
compared to the SD yrast and by increase
in the collective mass due to the quenching of pairing.
The action $S(E)$ decreases with excitation energy at a fixed spin
since length of the tunneling path becomes shorter. 
It becomes zero as the excitation energy $U$ reaches the barrier height.

\begin{table}
\begin{center}
\begin{tabular}{|c|c|c|c|c|c|c|c|c|c|}
\hline
&\multicolumn{3}{c|}{$^{152}$Dy}&\multicolumn{3}{c|}{$^{143}$Eu}&
\multicolumn{3}{c|}{$^{192}$Hg}\\
\hline
$I_0(\hbar)$
&$\hbar\omegab$&$\Vb-E_\SDyr$&$\hbar\omegas$
&$\hbar\omegab$&$\Vb-E_\SDyr$&$\hbar\omegas$
&$\hbar\omegab$&$\Vb-E_\SDyr$&$\hbar\omegas$\\
\hline
10&1.38&0.29&1.33&0.94&0.53&1.49&1.04&1.24&1.12\\
20&1.12&0.82&1.25&0.77&1.20&1.32&0.87&1.60&0.95\\
30&0.97&1.57&1.10&0.62&1.76&1.26&0.75&1.83&0.86\\
40&1.03&2.67&0.85&0.62&2.47&1.07&0.65&1.48&0.79\\
50&0.85&3.57&0.80&0.50&2.59&0.93&0.52&1.43&0.67\\
\hline
\end{tabular}
\end{center}
\caption{\label{Tab:TUN}
The calculated barrier parameters, $\hbar\omegab$, $\Vb-E_\SDyr$
and $\hbar\omegas$, for several spin values.
The signature and parity is taken the same as in
Fig.~\ref{Fig:ACT}. Here $I_0=I-\alpha$ in the first column
is a combination of the spin $I$ and the signature $\alpha$.
The unit for the barrier parameters is MeV.
}
\end{table}

The decrease in $S(E)$ with excitation energy is almost linear 
as seen in Fig.~\ref{Fig:ACT}.
If the potential barrier were one-dimensional and had a 
inverted quadratic shape, the action integral would have a
form $S(E)=\pi \,(\Vb - E) / \hbar\omegab$ where $\Vb$ is the
barrier energy and $\omegab$ is an oscillation frequency characterizing
the barrier. Using the linear energy-dependence and the formula of
the harmonic approximation, we can characterize 
the excitation energy dependence of
the microscopically calculated action
with the two parameters $\Vb$ and $\omegab$. 
In practice, the barrier energy $\Vb$ is evaluated as
the saddle-point energy, and the barrier frequency $\omegab$ is
evaluated as $\hbar\omegab=\pi \,(\Vb-E)/S(E)$ at
the energy $E=E_\SDyr$ of the SD yrast state 
(or at the energy of the ND yrast state if the latter is higher).  
Table~\ref{Tab:TUN} shows the calculated values of 
the barrier frequency $\hbar\omegab$ and the relative barrier height 
$\Vb-E_\SDyr$  for several spin values, 
together with the zero-point oscillation frequency $\omegas$ 
around the SD minimum.
The barrier height has  rather strong spin dependence which is
common to other calculations of potential energy 
surfaces~\cite{WernerDudek,Satula}.
Note that the barrier frequency $\omegab$ also shows sizable 
spin dependence. 
This is related to increase of the collective mass caused by 
decrease of pairing correlation at higher spins.

\begin{figure} 
\centerline{
\epsfxsize=50mm\epsffile{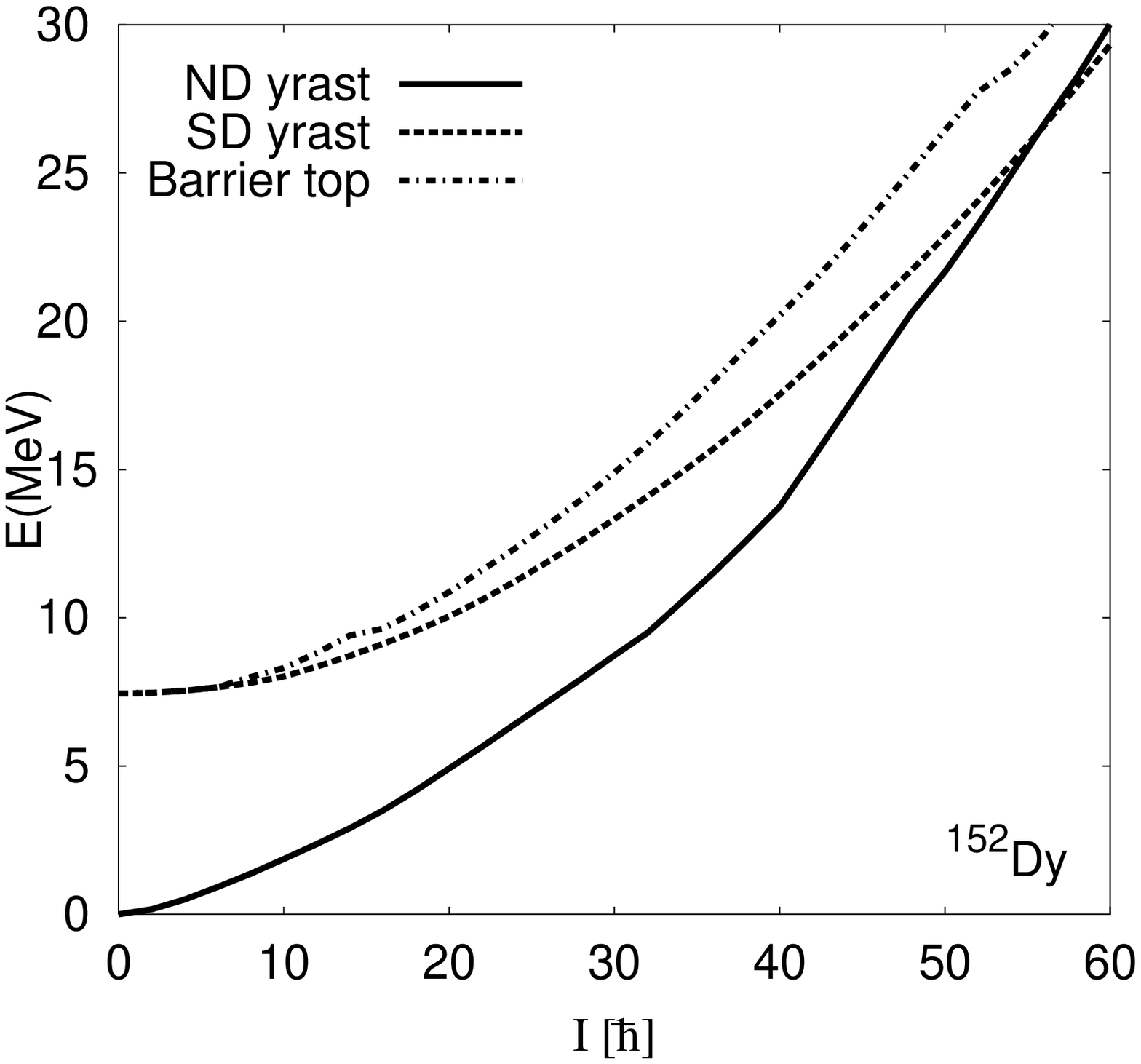}
\epsfxsize=50mm\epsffile{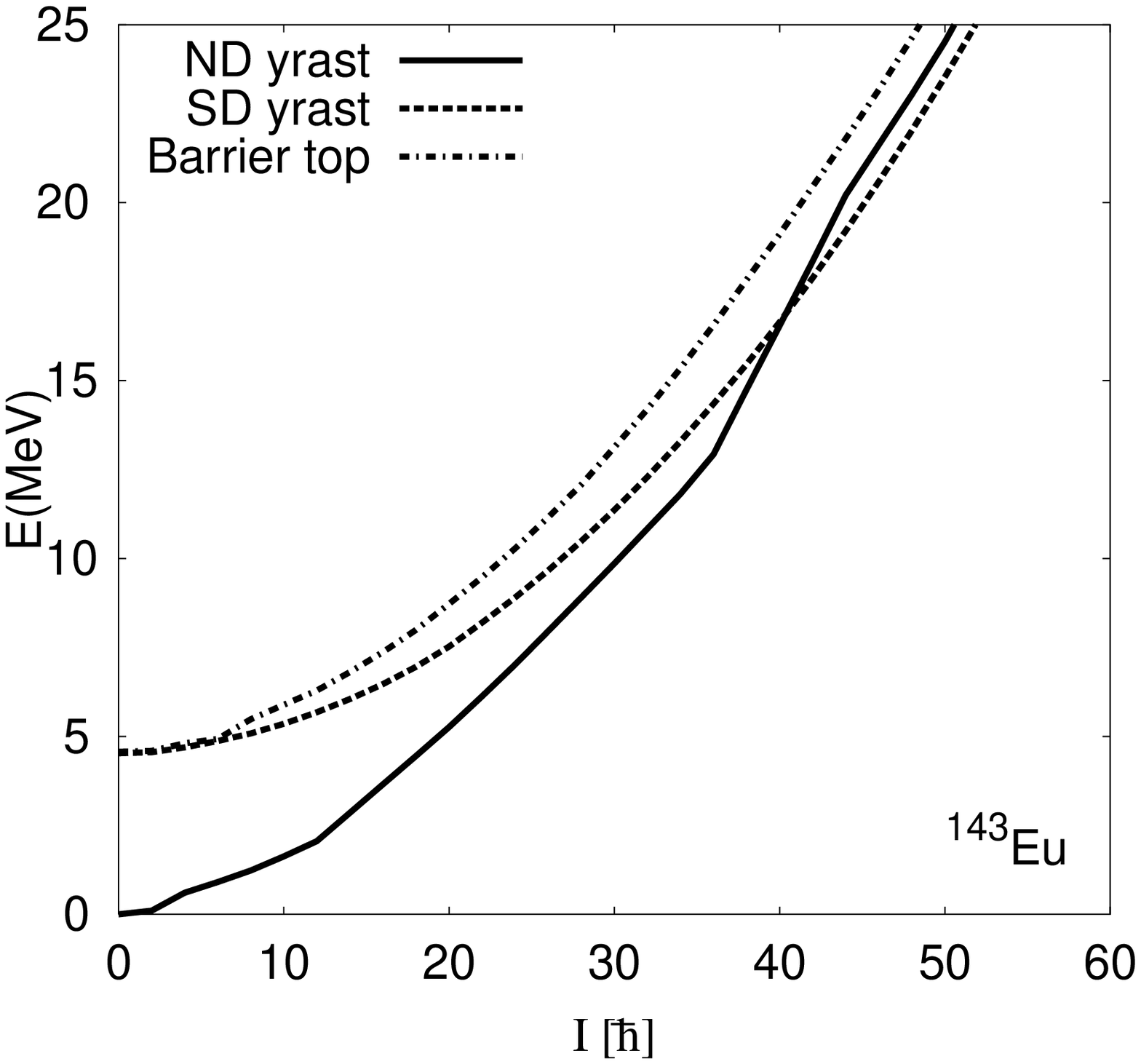}
\epsfxsize=50mm\epsffile{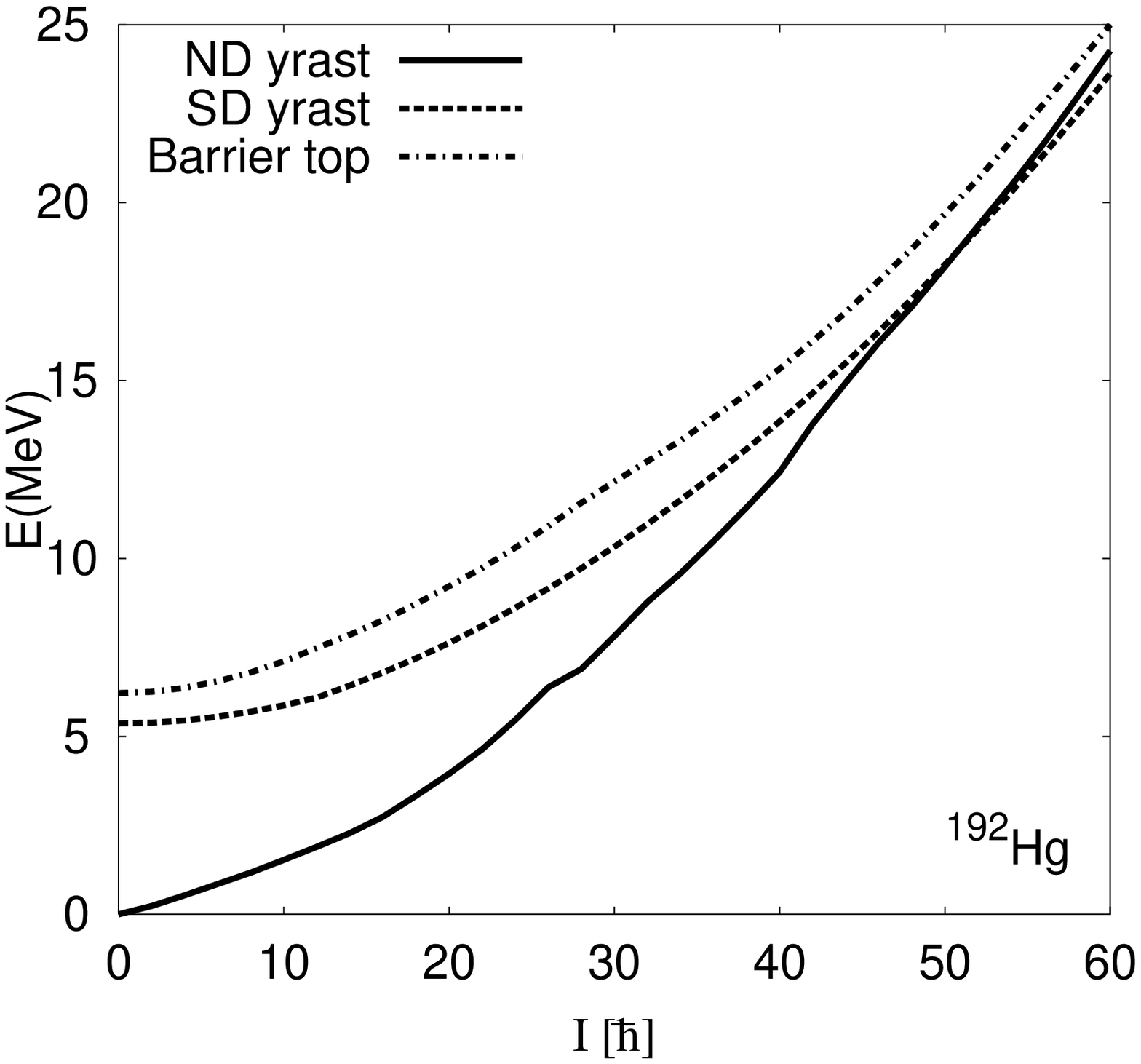}}
\caption{\label{Fig:EIP}
The energies of the ND and SD yrast states, $E_\NDyr$ and $E_\SDyr$,
and the saddle-point energy $\Vb$ as a function of angular
momentum for
$^{152}$Dy (left), $^{143}$Eu (middle) and $^{192}$Hg (right).
The signature and parity is taken the same as in Figs.~\ref{Fig:PES}
and~\ref{Fig:ACT}. 
}
\end{figure}

The energies of the yrast superdeformed and normal deformed
states, $E_\SDyr$ and $E_\NDyr$, and
the height of barrier (the saddle-point energy), $\Vb$,
are plotted in Fig.~\ref{Fig:EIP} 
as a function of spin for the parity and signature quantum numbers
of the yrast SD state. 
The correction for the zero-point oscillation energy 
is included in the ND yrast energy
$E_\NDyr$ in the same way as for the 
superdeformed yrast states: The zero-point energy is taken
the same as that of the superdeformed minimum.

Here difference of potential barrier in Fig.~\ref{Fig:EIP}
among these nuclei should be emphasized.
The barriers in $^{152}$Dy and $^{143}$Eu
increase monotonically
(more precisely, the barrier height of $^{143}$Eu almost saturates
at $I \approx 40 \hbar$, which is clearly seen from the action
in Fig.~\ref{Fig:ACT} for $E_\ex \gesim 1.5$ MeV),
while that in $^{192}$Hg increases
at low-spins but starts to gradually decrease at $I \approx 30 \hbar$.
This fact is seen directly in Table~\ref{Tab:TUN},
and the value of actions precisely reflects
the effect of this feature of barrier height in Fig.~\ref{Fig:ACT}.
Namely, the action of $^{192}$Hg does not increase linearly
as function of spin, while that of $^{152}$Dy does,
especially at higher excitation energies;
the action even changes to decrease around $I \approx 25 \hbar$
as the barrier height does.
This feature that the barrier height saturates and changes to decrease
at $I \approx 20-30 \hbar$ is commonly observed for nuclei
in the $A \approx 190$ region,
which can be seen in other calculations~\cite{WernerDudek,Satula}.
We will see in the next section (see \S\ref{Sec:Results})
that the decay-out properties in $^{192}$Hg are dramatically
different from those of $^{152}$Dy and $^{143}$Eu.

\subsection{Level density}
\label{Sec:LevDen}

The average level spacing $\Dn$ and $\Ds$ of 
the ND and SD states are calculated as
follows. Using the same cranked Nilsson single-particle 
orbits as employed in the shell model diagonalization, 
we construct excited many-particle many-hole configurations for each 
spin and parity. The Nilsson quadrupole deformation parameters
are fixed to those associated with the
normal deformed (or superdeformed) minimum.
The density of these excited configurations are calculated 
as a function of the excitation energy 
by explicitly counting the configurations in energy bins.
For normal deformed states 
the resultant level density as a function of the excitation energy
fluctuate but nicely follow in average the Fermi gas
formula appropriate for the cranking model
for fixed spin and parity~\cite{Aberg-lv},
\begin{equation}
   \rho_\FG (U) = \frac{\sqrt{\pi}}{48}a^{-\frac{1}{4}}
                  U^{-\frac{5}{4}}\exp{2\sqrt{aU}}.
\label{EQ:FGF}
\end{equation}
This allows us to use the Fermi gas expression to represent the 
average level density $\rhon$ of the normal deformed states.
The formula used in Ref.~\cite{ShimizuB} was the one
like Eq.~(\ref{EQ:FGF}) but with factor two being multiplied.
This was because the states with different parity mixes through
the statistical E1 transitions.
However, the level density in our model is used
through Eq.~(\ref{EQ:GamTun}), where the mixing is caused by
the tunneling process which conserves the spin and parity.
Therefore, we believe that Eq.~(\ref{EQ:FGF}) should be used.

As for the level density parameter $a$,
we use the one that is adjusted to fit the microscopic
calculation of the level density. Since spin dependence of the
calculated level density is found to be weak, we adopt a single value
of $a$ for all spin values by neglecting the spin dependence. 
The fitted value is
$a_{\rm n}=A/9.43,A/9.27$ and $A/10.15$ for normal deformed states 
in ${}^{152}$Dy, ${}^{143}$Eu and ${}^{192}$Hg,
respectively. 
The values of the level density parameter is different
from the traditional values $a=A$/8 or $A$/10. This is because of the
shell effect in the density of the single-particle orbits
near the Fermi surface. These values are close to a recent semi-empirical
estimate $a_{\rm n}$(emp.)$= A/8.57, A/9.46$ and $A/11.32$ 
for ${}^{152}$Dy, ${}^{143}$Eu and ${}^{192}$Hg
based on the neutron resonance data~\cite{Mughab}.  
The average spacing is given by $\Dn=1/\rho_\FGn (U',I)$ 
by using the fitted Fermi gas formula $\rho_\FGn (U',I)$, where
the excitation energy $U'$ is the energy measured from the 
yrast ND energy.  We take the lowest ND state
among the four set of the parity and signature quantum numbers
as the yrast configuration to define the yrast ND energy, for which
the zero-point energy is included. For
the other signature than that of the yrast configuration, we make
an interpolation to define the yrast ND energy.

We adopt a similar procedure for $\Ds$ of the superdeformed states. Since 
the excitation energy dependence is slightly different from Eq.~(\ref{EQ:FGF})
due to the strong shell effect, 
we make a fit at the excitation energy 
$E_\ex = 2.5$ MeV which is relevant for the quasi-continuum
spectra. The obtained level density parameter is $a_{\rm s} = A/18.9,A/18.5$
and $A/13.0$ for $^{152}$Dy, $^{143}$Eu and $^{192}$Hg~\cite{Yoshida1}. 

\subsection{Electromagnetic decay widths}
\label{Sec:EMwidth}

We evaluate the electromagnetic decay width
in a way similar to Ref.~\cite{ShimizuB}.
For the SD states, only collective rotational
E2 transitions are considered, and the decay width is given by
\begin{equation}
\Gammas =\mGamma_\Etwo^\SD= 3.0 \times 10^{-4} \, Q^2E_{\gamma}^5
\end{equation}
in the unit of MeV where the quadrupole moment $Q$ (in fm$^{2}$) 
is calculated by the liquid
drop model for the deformation parameters of the SD minimum.
The transition energy $E_\gamma$
is evaluated for each spin and parity as the energy difference 
between the SD potential minima at $I$ and $I-2$.
We use the same value for the excited SD states
by neglecting excitation energy dependence
which are expected to be small.
As for the effect of rotational damping, we note that
the rotational damping widths for the thermally excited SD bands
are less than 100 keV~\cite{Yoshida2},
while the E2 transition energy is typically $500-1000$ keV.
Therefore it is also expected to affect the electromagnetic width
only slightly and so its effect is neglected.

For the transitions associated with ND states,
the collective rotational E2 and the statistical E1 transitions
are taken into account. Namely the decay width is given by
$\Gamman = \mGamma_\Etwo^\ND
+\mGamma_\Eone^\ND$. Here $\mGamma_\Etwo^\ND$ is calculated
in the same way as in the SD case. In ND case, however, there
often exist more than one local minima with different quadrupole
deformations. Therefore, we take
average of the E2 width over the local minima. 
The weight factor for averaging is taken proportional to
the level density associated with each local minimum evaluated
with the fitted Fermi gas formula $\rho_\FGn$.
This averaging avoids possible
discontinuity of $\mGamma_\Etwo^\ND$ 
associated with crossing of different yrast configurations.
In addition, even if the lowest minimum is
spherical $\mGamma_\Etwo^\ND$ remains finite. 
The averaging, however, has little influence on the result except 
in the case in which the yrast minimum turns out spherical. Dependence of
$\mGamma_\Etwo^\ND$ on excitation energy is weak as far as the yrast state
is well deformed. The E1 decay width $\mGamma_\Eone^\ND$ is calculated 
by using the GDR strength function $f_\GDR$,
\begin{equation}
   \mGamma_\Eone^\ND=C_\Eone\int_0^{U'}\frac{\rho_\FGn (U'-E_\gamma)}
       {\rho_\FGn (U')}f_\GDR(E_\gamma)E_\gamma^3dE_\gamma,
\end{equation}
where $U'$ is the energy of the SD state relative to the ND yrast,
and $C_\Eone$ is the E1 hindrance factor.
Actually we use the analytic expression for
$\mGamma_\Eone^\ND$~\cite{DossVige}, which is a good approximation
at excitation energy $U'$ under consideration,
\begin{equation}
 \mGamma_\Eone^\ND=C_\Eone \, 2.3 \times 10^{-11} \, NZ A^{1/3}
      \biggl[ \frac{U'}{a} \biggr]^{5/2},
\end{equation}
where the Thomas-Reiche-Kuhn sum rule value is used for
the strength function $f_\GDR$ and the GDR parameters,
$\mGamma_\GDR=4$ MeV and $E_\GDR=80 A^{-1/3}$, are employed.
We use the E1 attenuation factor $C_\Eone=0.15$ taken from Ref.~\cite{Barth}.
The E1 width $\mGamma_\Eone^\ND$ increases with the fifth power of
the temperature, $T=\sqrt{U'/a}$ with increasing the
excitation energy $U'$.

\section{Results}
\label{Sec:Results}

\subsection{Decay-out of yrast superdeformed bands}
\label{Sec:yrast}

\begin{figure} 
\centerline{
\epsfxsize=40mm\epsffile{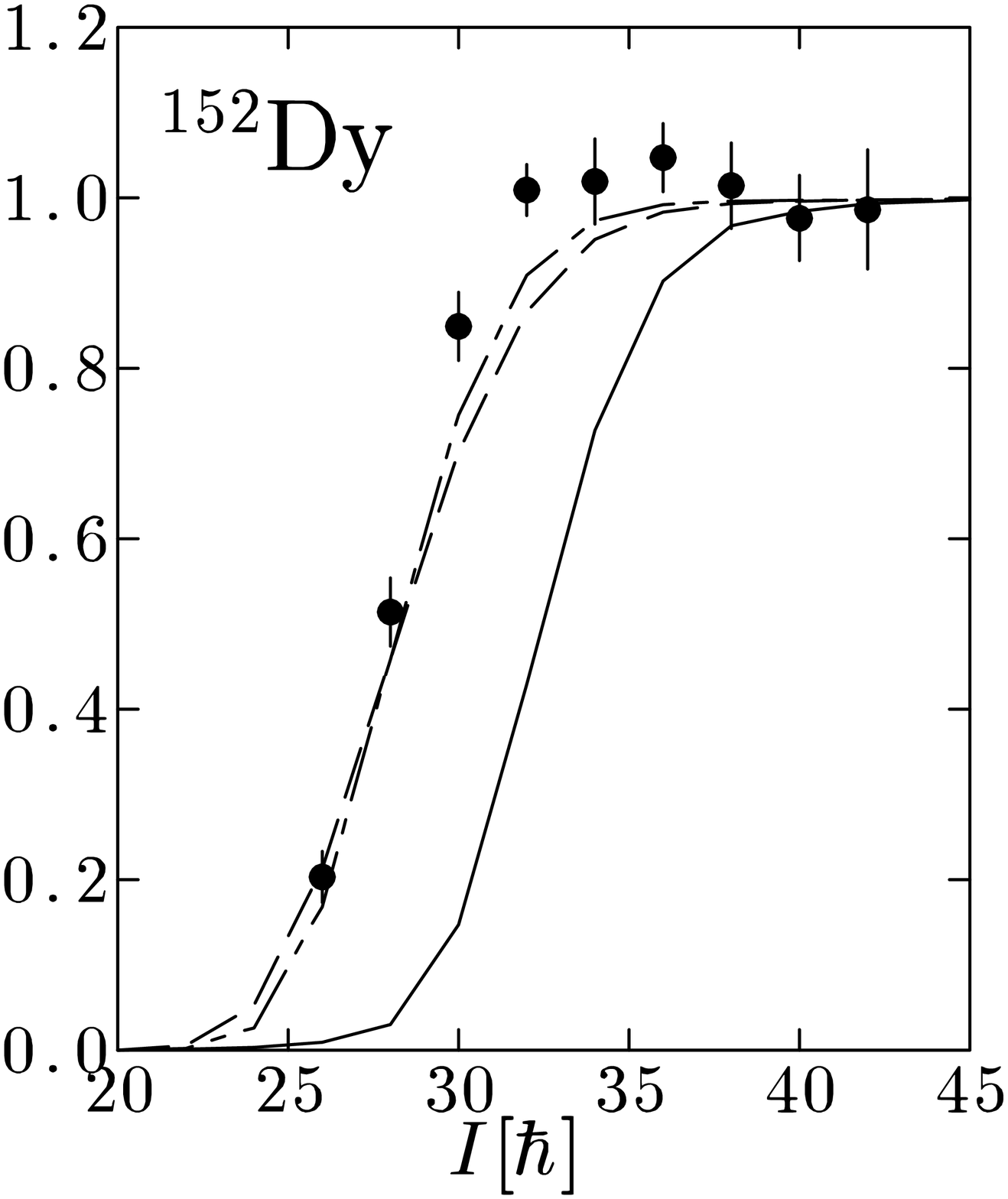}
\epsfxsize=40mm\epsffile{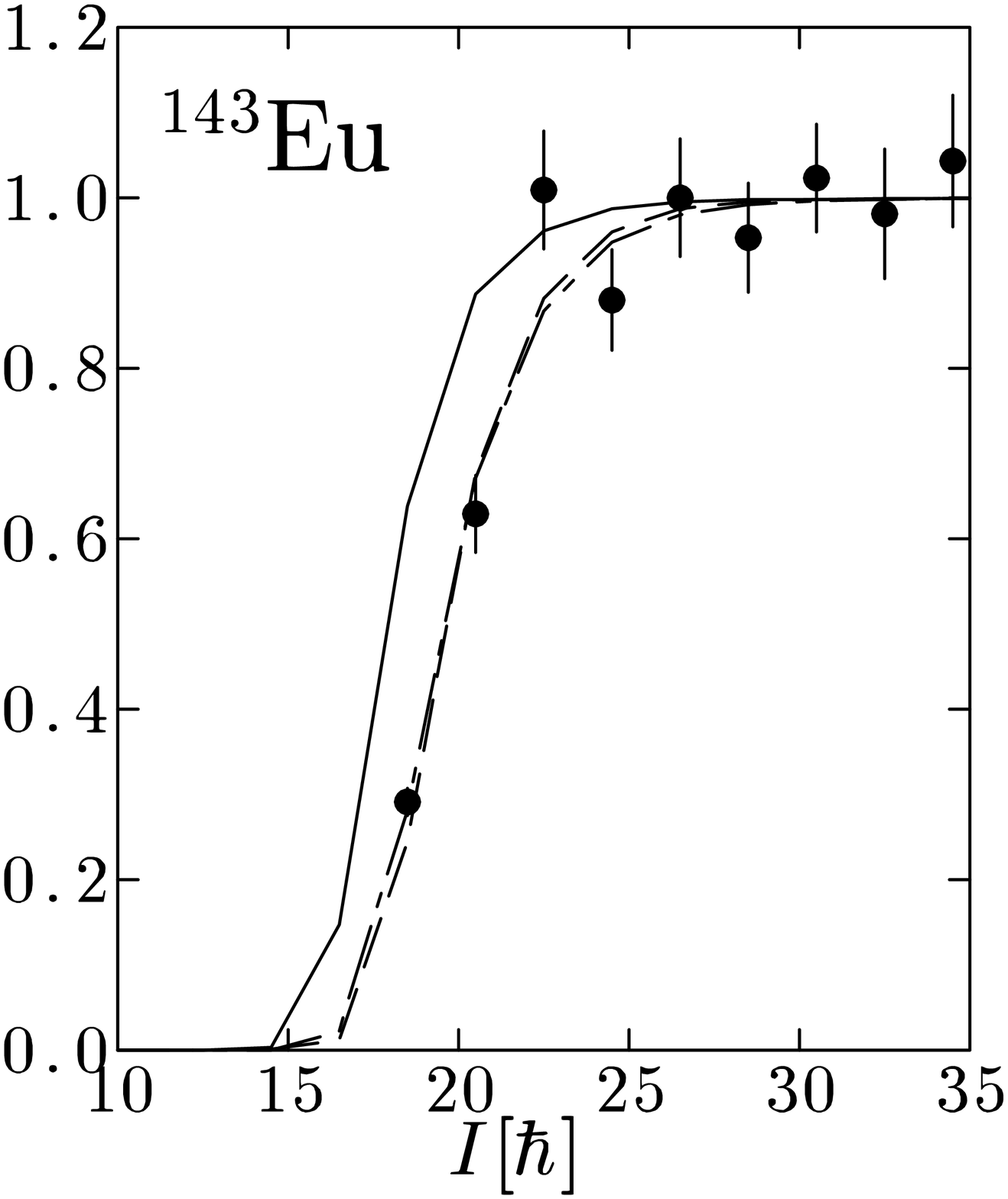}
\epsfxsize=40mm\epsffile{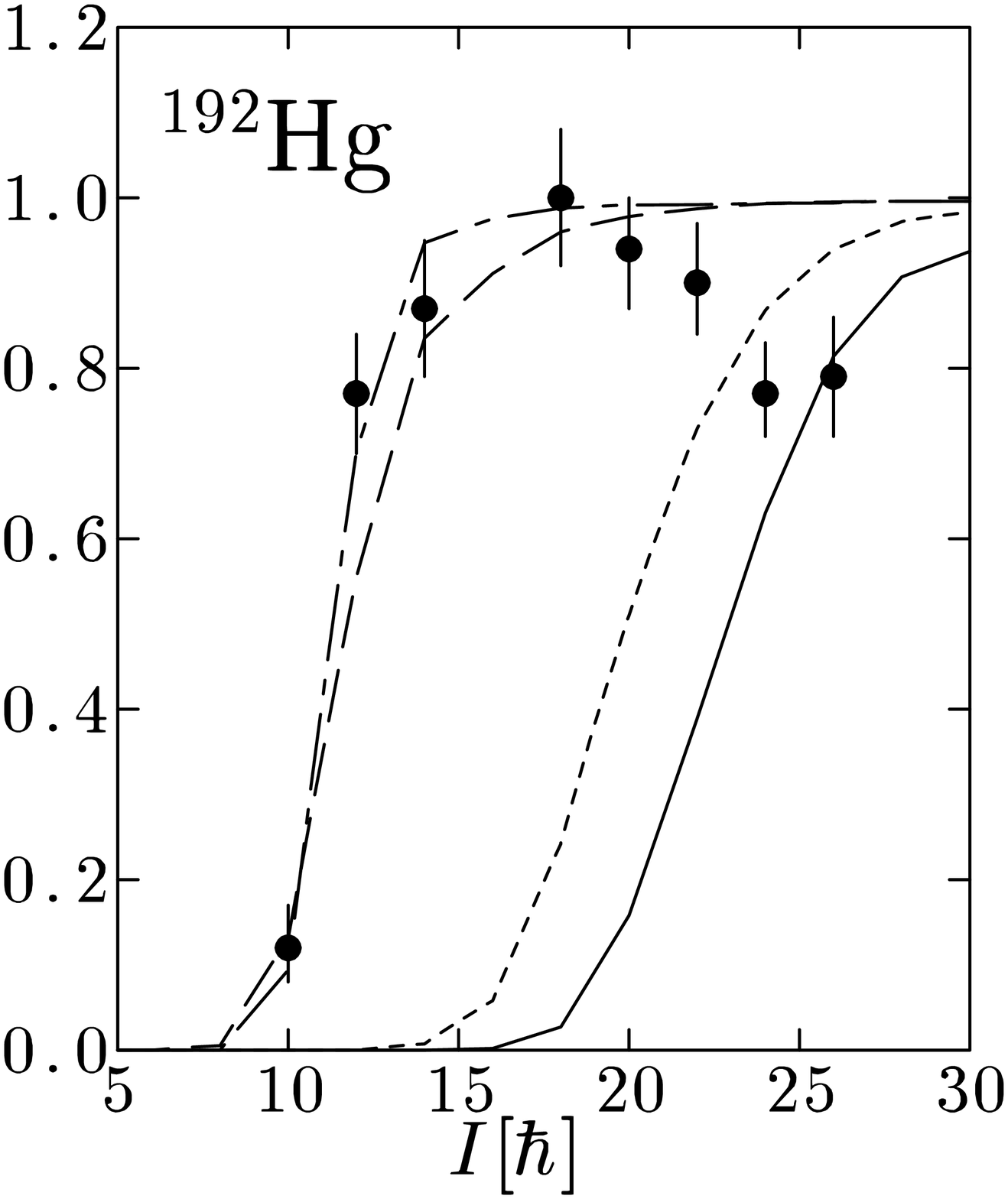}}
\caption{\label{Fig:GamInt}
The $\gamma$-ray intensity for decay-out of the yrast SD band
in $^{152}$Dy (left), $^{143}$Eu (middle) and $^{192}$Hg (right).
The solid, dashed and dot-dashed lines denote the results
without any adjustment, with modification by $C_\rho$,
with modification by $C_\mass$ (both are explained in the text),
respectively.
The result with using the ``back-shifted level density''
with $E_{\rm BS}=1.0$ MeV,
is also included as a dotted line for $^{192}$Hg.
}
\end{figure}

We have performed calculations for yrast SD bands in
both the $A \approx$ 150 and $A \approx$ 190 regions, 
in which the calculational procedure has been improved
over the previous calculations~\cite{ShimizuB},
as is outlined in the previous sections.
Here we use Eq.~(\ref{EQ:WKB1}) for the tunneling width
$\Gammat$, appropriate in describing the tunneling of the yrast SD bands.
The only definite observable for the decay-out of the yrast
SD band is the $\gamma$-ray intensity, $I_\gamma(I)$,
of the transitions $I \rightarrow I-2$ inside the band.
This quantity is calculated by using the average decay-out
probability $\Nout$ discussed above as
\begin{equation}
I_\gamma(I)=\prod_{I' \ge I+2}\Bigl(1-\Nout(U=0,I')\Bigr).
\end{equation}
We show in Fig.~\ref{Fig:GamInt},
the calculated $\gamma$-ray intensities by solid lines
for selected yrast SD bands in
$^{152}$Dy, $^{143}$Eu and $^{192}$Hg in comparison
with the data;
the intensity data are taken from~\cite{ExpDySys} for $^{152}$Dy,
from~\cite{Exp143Eu,Eu-newexp} for $^{143}$Eu
(see also~\cite{Eu-spinassign} for the most possible spin-assignment),
and from~\cite{Exp192Hg} for $^{192}$Hg.
In contrast to the previous calculations~\cite{ShimizuB},
where schematic rotational spectra were used for both
the SD and ND yrast bands, all the quantities entering
in the calculations are obtained theoretically; namely,
there is no adjustable parameters in the present calculations.
As a result, it is not so easy to reproduce the observed intensity.
The pattern of the $\gamma$-ray intensity is characterized
in two ways; one is the decay-out
spin $\Iout$, that is defined as a spin value at which
the $\gamma$-ray intensity is reduced to half, $I_\gamma(\Iout)=1/2$,
and the other is a sharpness of the intensity-drop.
The sharpness is well reproduced in all three SD bands
as was in the previous calculations~\cite{ShimizuB},
especially if the decay-out spin is adjusted in some way,
see the following discussions.
The problem lies in the calculated decay-out spin, which
is too large in $^{192}$Hg,
$I_{\rm out}^{(\rm cal)} -I_{\rm out}^{(\rm exp)} \approx 11 \hbar$,
while it is rather well reproduced in $^{152}$Dy,
$I_{\rm out}^{(\rm cal)} -I_{\rm out}^{(\rm exp)} \approx 3 \hbar$,
and in $^{143}$Eu, $\approx -2 \hbar$.
In order to set up a reliable model for the thermally
excited superdeformed states, it is very important to reproduce
the decay-out spins.  Therefore, let us examine possible origins
of the above discrepancies.

\begin{table}
\begin{center}
\begin{tabular}{|c|c |c| c| c |c |c |}
\hline
& \multicolumn{2}{c|}{$^{152}$Dy} 
& \multicolumn{2}{c|}{$^{143}$Eu} 
& \multicolumn{2}{c|}{$^{192}$Hg} \\
\hline
           & I=28 & I=32 & I=37/2& I=41/2& I=12 &I=22 \\
\hline
$\Gammat$  &$5.9\times10^{-5}$&$4.7\times10^{-6}$
           &$3.7\times10^{-5}$&$1.1\times10^{-5}$
           &$5.9\times10^{-5}$&$2.8\times10^{-6}$ \\
$\Dn$      &$1.1\times10^{-5}$&$1.2\times10^{-5}$
           &$5.9\times10^{-4}$&$1.1\times10^{-3}$
           &$5.7\times10^{-6}$&$2.2\times10^{-5}$ \\
$\Gammas$  &$1.8\times10^{-8}$&$2.9\times10^{-8}$
           &$1.5\times10^{-9}$&$3.2\times10^{-9}$
           &$5.8\times10^{-11}$&$2.9\times10^{-9}$\\
$\Gamman$  &$5.7\times10^{-9}$&$5.6\times10^{-9}$
           &$1.3\times10^{-9}$&$1.0\times10^{-9}$
           &$4.3\times10^{-9}$&$3.7\times10^{-9}$ \\
$U'_{\rm SDyr}$ & 4.68 & 4.61 & 2.57 & 2.25 & 4.24 & 3.53  \\
$S$            & 4.0 & 5.2 & 4.3 & 4.9 & 4.0 & 5.4  \\
$S(C_\mass)$   & 5.1 & $-$ & $-$ & 3.7 & 8.2 & $-$  \\
\hline
\end{tabular}
\end{center}
\caption{\label{Tab:Ydecay}
The calculated values of $\Gammat,\Dn,\Gammas,\Gamman$ and the action $S$
at $U=0$, $I\approx I_{\rm out}^{\rm (cal)}$ and $I_{\rm out}^{\rm (exp)}$, 
responsible for the
decay-out of the yrast SD band.
The energy $U'_{\rm SDyr}$of the SD yrast states
measured from the ND yrast is also listed.
The last line lists the
value of the action $S$ at $I \approx I_{\rm out}^{\rm (exp)}$
calculated with the collective mass
renormalization $C_\mass=1.4$, 0.7 and 3.0
for $^{152}$Dy, $^{143}$Eu and $^{192}$Hg, respectively. 
The unit for $\Gammat,\Dn,\Gammas,\Gamman$ and $U'_{\rm SDyr}$ is MeV.
}
\end{table}

As is discussed in the previous sections (see \S\ref{Sec:tunnel}),
the decay-out probability,
$\Nout$, and so the $\gamma$-ray intensity, $I_\gamma$,
is determined by four quantities;
the tunneling width, $\Gammat$, the electromagnetic decay widths
for the SD and ND bands, $\Gammas$ and $\Gamman$, and
the mean level distance of ND states, $\Dn$
(or equivalently the level density, $\rhon$).
Table~\ref{Tab:Ydecay} shows the values of
these quantities responsible for the decay-out,
evaluated at $U=0$,
$I\approx I_{\rm out}^{\rm (cal)}$ and $I_{\rm out}^{\rm (exp)}$.
As for the electromagnetic transition rates,
$\Gammas$ is that of the stretched E2 and $\Gamman$
is of the statistical E1 in the decaying spin region.
They are well determined by the equilibrium shape and
the excitation energy of the SD band from the ND yrast band,
namely by the minima of the calculated potential energy surface.
On the other hand, $\Gammat$ is calculated by the collective mass
tensor and the potential energy barrier, i.e.
the behaviour of the saddle point region.
Our potential energy surface gives correct equilibrium shape,
which is deduced from the measured life time of the SD band,
and the calculated barrier also seems to be consistent with
other calculations, see for example Refs.~\cite{WernerDudek,Satula}.
Therefore we do not try
any adjustment for quantities related to the potential
energy surface in the present work,
although there may exist a room for some improvement
if we use a different mean-field from
the Nilsson potential or a different type or strength
for the residual pairing interactions.
Remaining quantities in question are the level density
and the collective mass parameters.  We try to adjust these
quantities in order to obtain agreement of the decay-out spin
as in the following.

We remark that some additional effects on the level density
are not included in the present calculations. For example, the
pairing correlation that may be present at relatively low spins
may reduce the level density by a factor of up to $\sim 10^2$ 
while the effects of collective
vibrations such as beta and gamma modes may increase the level
density. To represent these effects and some other origins of ambiguity,
we shall allow a renormalization factor of the level density 
$C_\rho$ that multiply the level density $\rhon$ of normal deformed
states
as $\rhon \rightarrow C_\rho\rhon$.
Since we explicitly deal with the superdeformed states in the
shell model description, we do not apply such a renormalization
for the superdeformed states to avoid an inconsistency.
The results fitting the decay-out spin by using the factor $C_\rho$
are shown by the dashed lines in Fig.~\ref{Fig:GamInt};
the values of which are $C_\rho=0.1$, $15$ and $2 \times 10^{-4}$,
for $^{152}$Dy, $^{143}$Eu and $^{192}$Hg, respectively.
It should be mentioned that in the calculation of Ref.~\cite{ShimizuB},
the calculated $\gamma$-ray intensity was mistaken by two spin units,
i.e. it was $I_\gamma(I-2)$; this is the main reason why
the $C_\rho$ factor is about an order of magnitude
smaller in the present calculation than in~\cite{ShimizuB}.
The resultant $C_\rho$ factor in $^{152}$Dy and 
$^{143}$Eu may be acceptable. 
But it is too small for $^{192}$Hg since the ND level spacing
at the excitation energy $U'\approx 4$ MeV becomes $\Dn \approx 30-100$ keV,
which is unphysically large. The fluctuation analysis of the
quasi-continuum decay-out transitions in  $^{192}$Hg~\cite{Lopez-Dossing} 
suggests that
the ND level density $\rhon$ is smaller than the Fermi gas estimate due to
the pairing correlation, and that the reduction of $\rhon$ is the order of
$10^{-1} - 10^{-2}$.
As for the level density, the back-shifted Fermi gas formula,
namely $U$ is replaced by $(U-E_{\rm BS})$ in Eq.~(\ref{EQ:FGF}),
is often used to include the effect of the pairing correlation.
In the decay of the Hg region, the decay-out spin is small and
this effect may be more important than in the case of the Dy region.
Therefore, the results with using $E_{\rm BS}=1.0$ MeV is also included
as a dotted line for $^{192}$Hg in Fig.~\ref{Fig:GamInt}.
As seen in the figure, it helps to shift the calculated decay-spin
lower by about 3 $\hbar$ but the amount is not enough.

As an another origin of the discrepancy, we consider
a possibility that the collective mass could be affected
by some additional effects. The essential contents of the
pair hopping model is the number of single-particle level crossings
and the pairing gap. We have estimated the number of crossings
by using a lowest order semiclassical (Fermi-gas) approximation
in Eq.~(\ref{EQ:MassTens}),
and detailed properties of individual nucleus cannot be reflected.
However, if one calculate the number of crossing microscopically 
or take into account the shell effects in the single-particle
orbits, this quantity could be largely modified. This may
influence the collective mass $m_{ij}(q_1,q_2)$. To represent
such effects, we introduce a renormalization factor 
$C_\mass$ that modifies as $m_{ij}(q_1,q_2) \rightarrow 
C_\mass\,m_{ij}(q_1,q_2)$. 
It should be noticed that the scaling factor $C_\mass$ affects the
calculated action in two ways: one is a simple scaling of action
by a factor $\sqrt{C_\mass}$ as is clear from
Eqs.~(\ref{EQ:action}) and (\ref{EQ:s}), the other is a change
through the variation of the zero-point energy $\hbar\omegas/2$ 
in the action integral.
Both effects make the action larger (smaller) if $C_\mass > 1$ ($< 1$).
The results fitting the decay-out spin by using the factor $C_\mass$
are shown by the dot-dashed lines in Fig.~\ref{Fig:GamInt};
the values of which are $C_\mass=1.4$, $0.7$ and $3.0$,
for $^{152}$Dy, $^{143}$Eu and $^{192}$Hg, respectively.
As in the case of the introduction of $C_\rho$,
the factor seems large for Hg.

  From the above discussion, it is clear that we have large discrepancy
for the decay-out spin value in $^{192}$Hg.
According to the results of more systematic calculation~\cite{NS2000}
it is one of the worst cases in the $A \approx 190$ nuclei.
In this systematic calculation the empirical level density
parameters~\cite{Mughab} are used so that the results for
the intensity data are slightly different from those shown in this paper.
Also, the back-shifted level density with $E_{\rm BS}=1.0$ MeV is
used for comparison with the data for nuclei in the $A \approx 190$ region.
The results of Ref.~\cite{NS2000} show that
the basic characteristic features of intensity pattern
are reproduced in both the $A \approx 150$ and 190 regions;
especially the common rapid decrease of transitions at lower spins.
The observed decay-out spin is about 25$\hbar$ on average
in the $A \approx 150$ region, while it is about 10$\hbar$
in the $A \approx 190$ region.
On average calculated decay-out spins are comparable to the observed ones
in the $A \approx 150$ region, but those in the $A \approx 190$ region
are a little  larger than the observed ones systematically
even though the back-shift in the level density formula is employed.
The calculation  reproduces qualitatively the
empirical observation that
the average decay-out spins in the $A \approx 190$ nuclei are
systematically smaller than those in the $A \approx 150$ nuclei,
but the amount is not enough.  This trend is mainly due to
the facts that the calculated action as a function of spin is somewhat
different between nuclei in the two regions (see Fig.~\ref{Fig:ACT}),
which reflects the difference in the potential energy surfaces,
and that the level density parameter, $a_{\rm n}/A$, is systematically
smaller in the $A \approx 190$ region.
We have to admit that the calculation overestimates systematically
the decay-out spins in the $A \approx 190$ region,
which suggests that the calculated 
action values are too small in these nuclei.
It seems that some further mechanisms
which discriminate SD bands in these two regions are necessary.
Presently we do not have any clue to solve this problem.

\subsection{Decay-out of excited superdeformed states}
\label{Sec:DecOutExSD}

\subsubsection{Tunneling width}\label{Sec:TunWidthGen}

Let us first discuss the
tunneling width $\Gammat$ which is the most important quantity
in describing the decay-out of excited superdeformed states.
As discussed in \S\ref{Sec:TunWidth}, we evaluate the tunneling width 
with use of Eq.~(\ref{EQ:WKB2}), which is expected to give an
average behaviour of $\Gammat$.  In the analysis below, 
we shall also show results obtained by using Eq.~(\ref{EQ:WKB1}),
which provides an upper limit of $\Gammat$,
in order to
illustrate the difference in the two evaluations of $\Gammat$.
The use of Eq.~(\ref{EQ:WKB2}) is referred 
to as the case (a) in the following, while
the evaluation of the upper limit with Eq.~(\ref{EQ:WKB1}) 
is as the case (b).

\begin{figure} 
\centerline{
\epsfxsize=50mm\epsffile{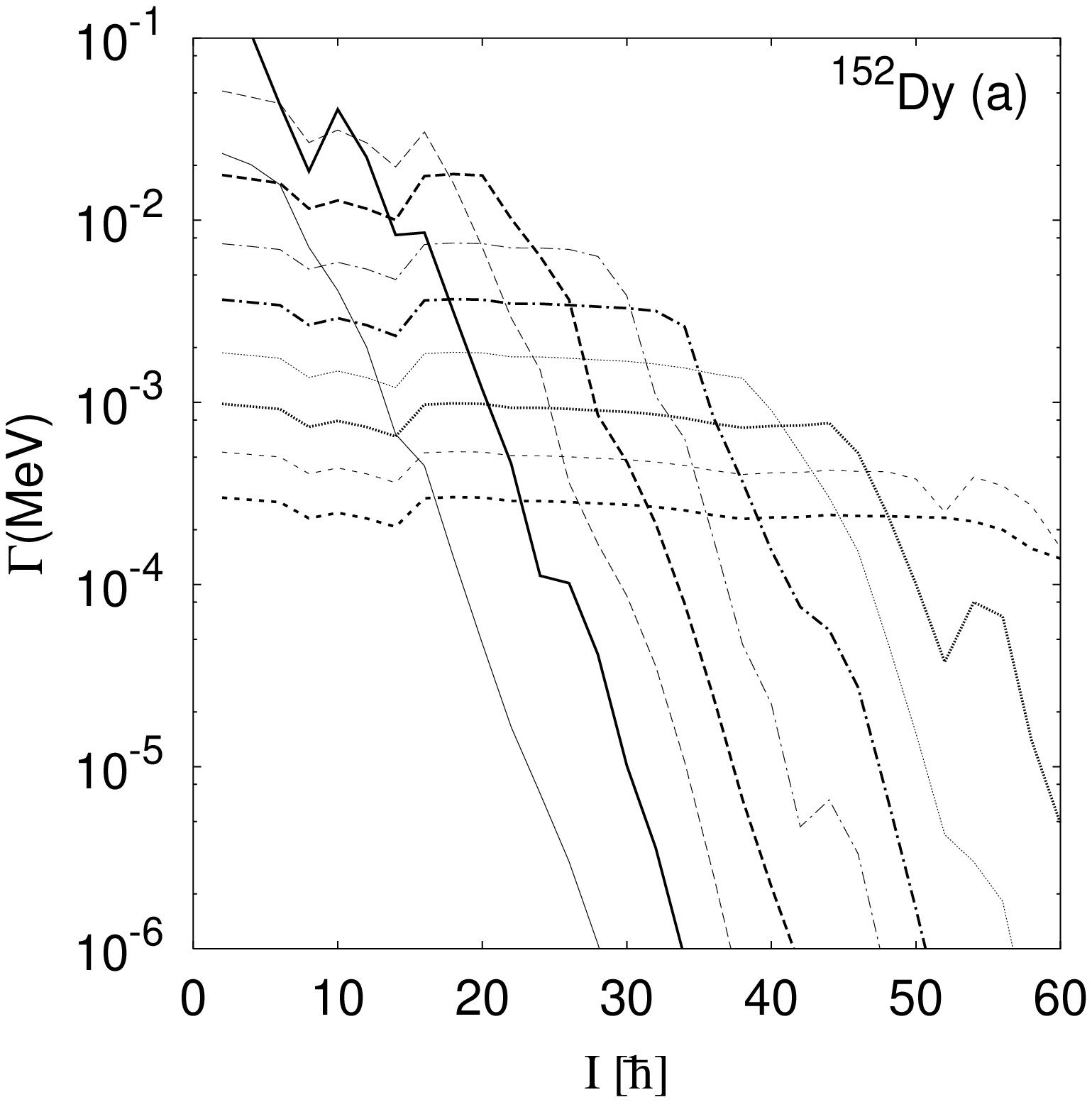}
\epsfxsize=50mm\epsffile{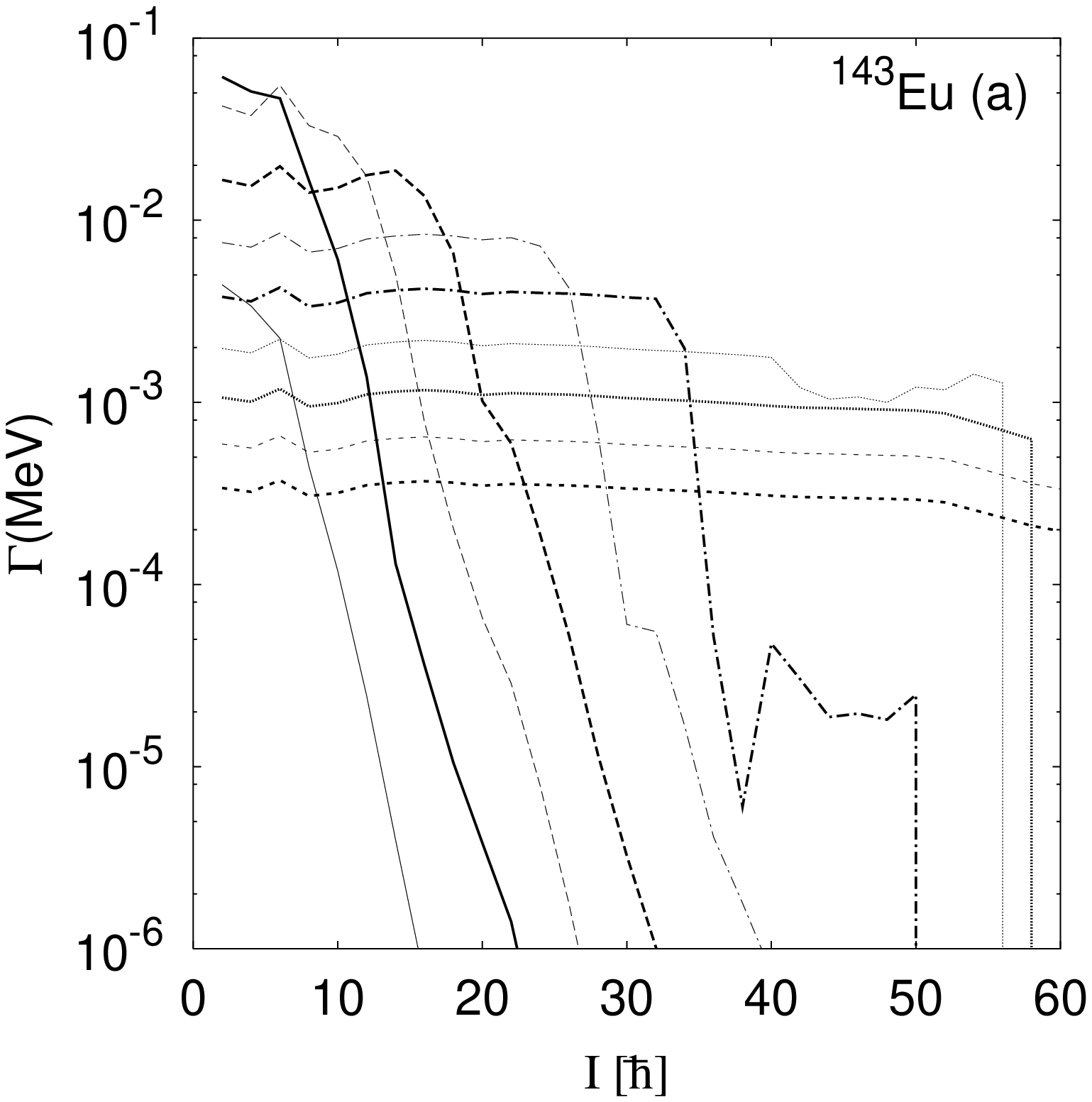}
\epsfxsize=50mm\epsffile{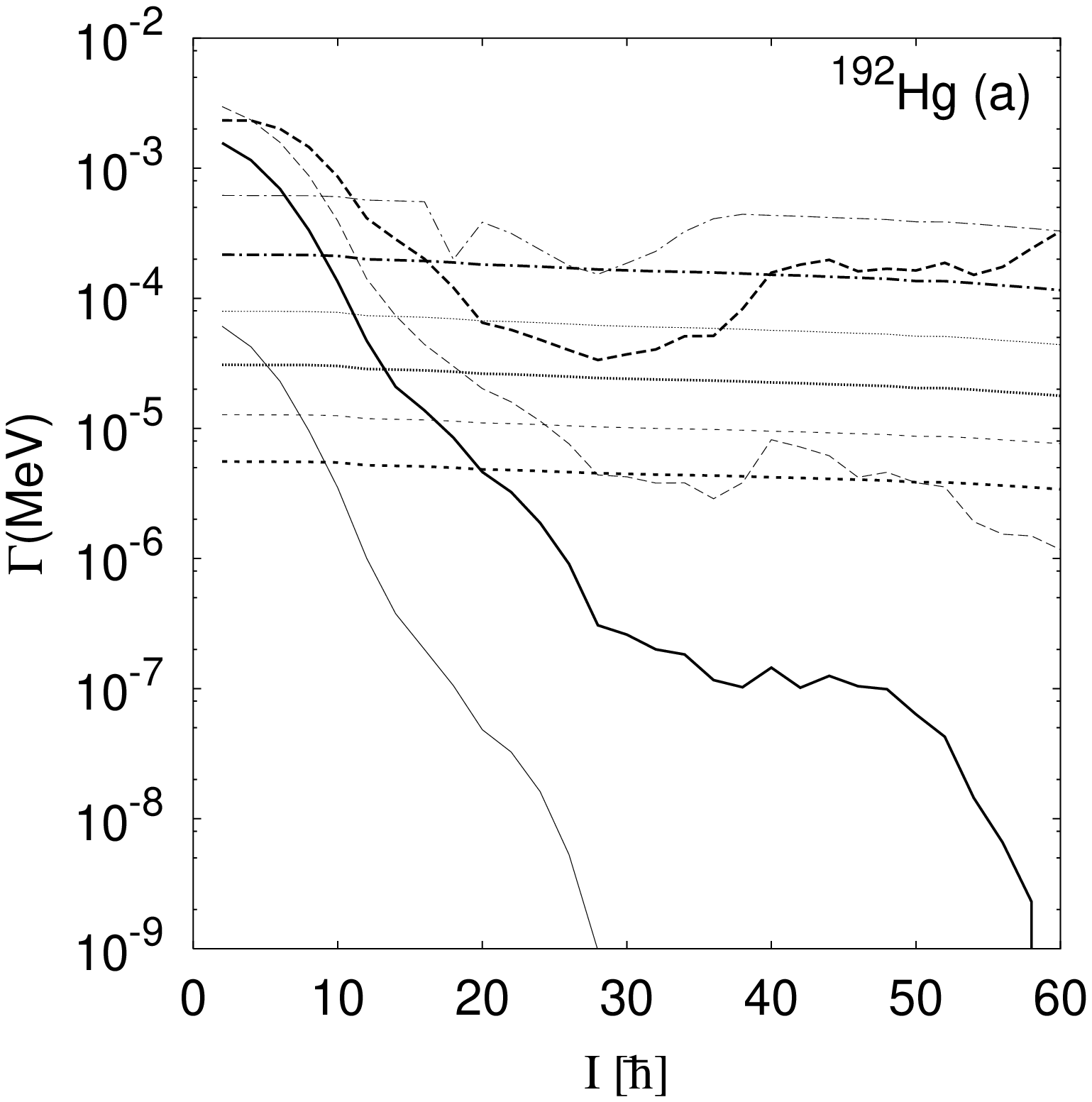}}
\centerline{
\epsfxsize=50mm\epsffile{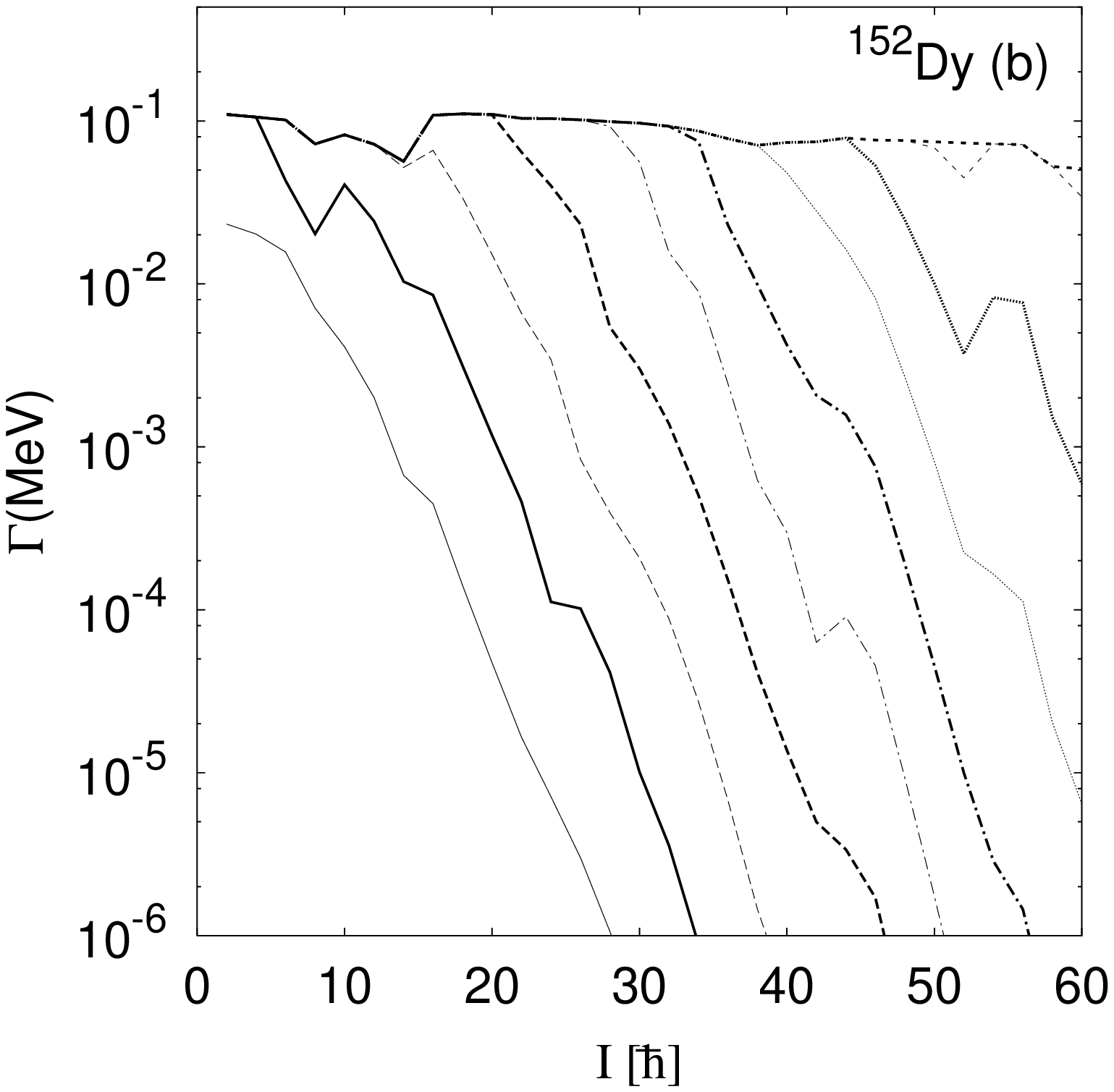}
\epsfxsize=50mm\epsffile{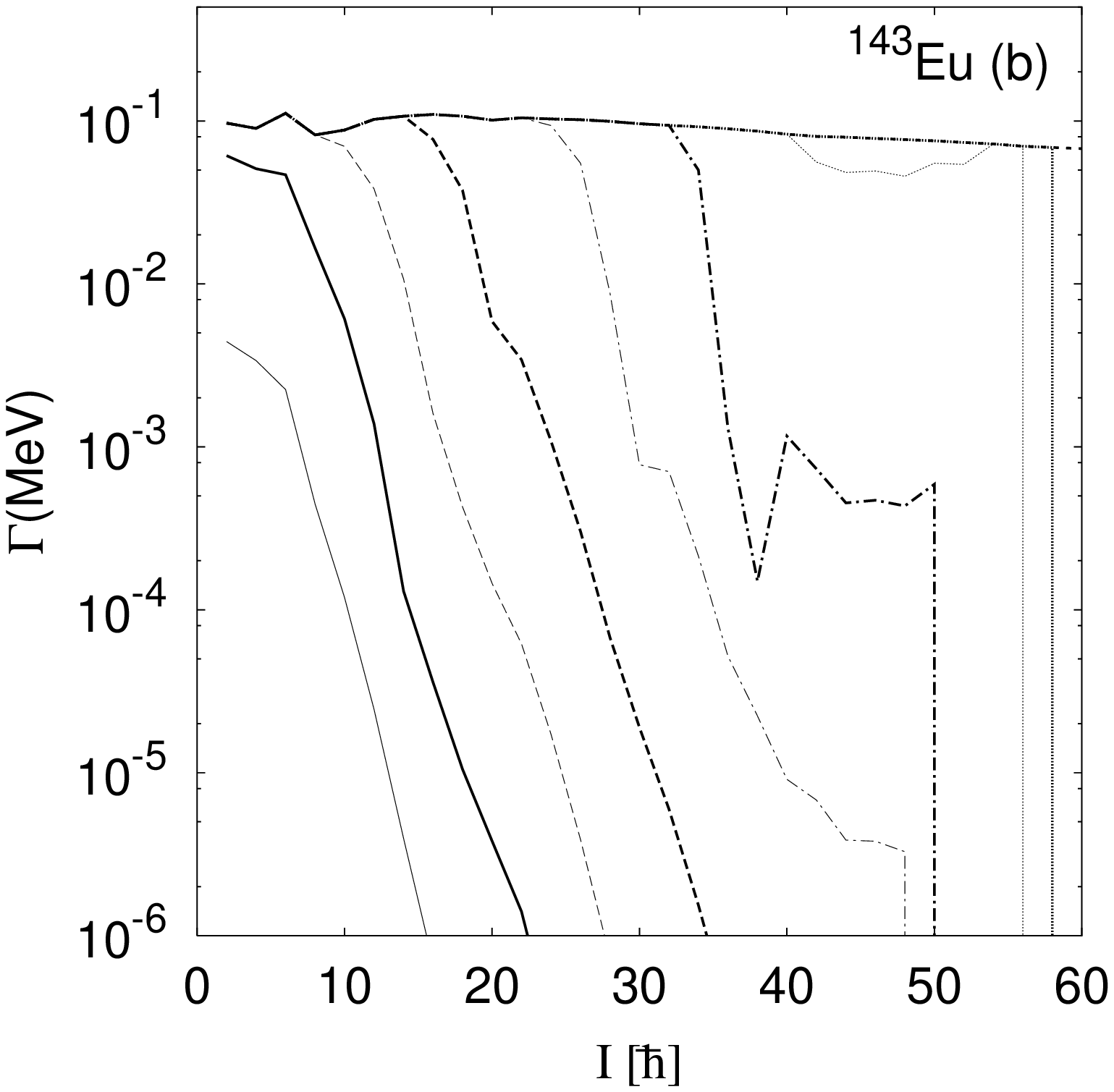}
\epsfxsize=50mm\epsffile{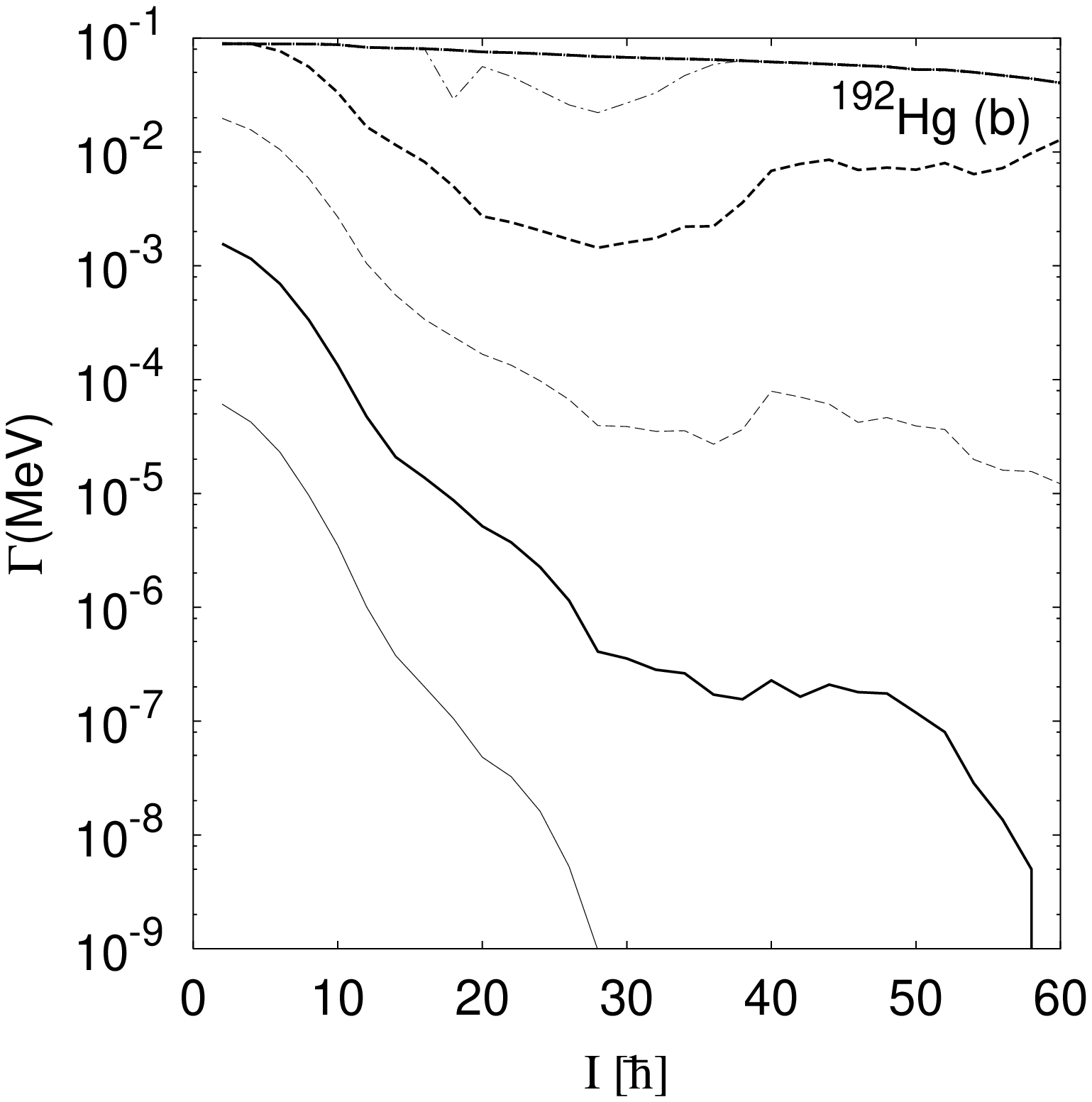}}
\caption{\label{Fig:SPW}
The tunneling width $\Gammat$ for $^{152}$Dy (left),
 $^{143}$Eu (middle) and $^{192}$Hg (right) as a function of spin
for different excitation energies. The line styles are
defined in the same manner as in Fig.~\ref{Fig:ACT}.
The upper three panels show the results in the case (a) using
 $\Gammat=(\Ds/2\pi) T$ whereas the lower panels show those in
the case (b) calculated with $\Gammat=(\hbar\omegas/2\pi)T$. 
The signature and parity is the same as in Fig.~\ref{Fig:ACT}. 
}
\end{figure}

In Fig.~\ref{Fig:SPW}, we
show dependence of the calculated tunneling
width on the spin and the excitation energy. The upper three
panels show results in the case (a), calculated
for the three nuclei, 
$^{152}$Dy(left), $^{143}$Eu(middle) and $^{192}$Hg(right).
The calculated tunneling width $\Gammat$ increases exponentially
with the excitation energy until the energy reaches
the barrier height.
Note that the tunneling transmission coefficient $T=1/(1+\exp 2S)$ 
increases exponentially with excitation energy 
since the action integral
$S$ decreases almost linearly with increasing the excitation energy
(cf. \S\ref{Sec:path} and Fig.~\ref{Fig:ACT}).
On the other hand, the knocking probability proportional
to the level spacing $\Ds$ 
exhibits an exponential decrease, but this decrease is weaker than
the increase in the transmission coefficient. 
A systematic spin-dependence is also seen in $\Gammat$
for the energy region below the barrier height.
Namely, the tunneling width
for the yrast SD states (corresponding to the thick curve in
Fig.~\ref{Fig:SPW}) decreases exponentially
with increasing spin except at high spins ($I\gesim 30$)
in $^{192}$Hg.  The spin-dependence originates mainly
from the tunneling transmission coefficient since the spin-dependence in the
level density of SD states is weak.
As the excitation energy exceeds
the barrier height, the calculated tunneling width
turns to decrease. This is because the transmission coefficient saturates 
to a constant value $\approx 1/2$ and 
the tunneling width $\Gammat \approx {\Ds/4\pi}$ then
decreases with excitation energy as it is governed by the level spacing 
$\Ds$ of the excited SD states.
In this regime, the tunneling width depends mostly on the
excitation energy measured from the SD yrast, but it is insensitive to 
the spin as the level density $\rhos$ of SD states is. 

The lower panels show the tunneling width in the case (b) 
where the knocking frequency is given by the collective 
vibrational frequency $\omegas$. Since  $\omegas$ depends little
on the excitation energy and spin, the energy dependence of the
tunneling width is solely governed by the
exponential increase of the transmission coefficient below the barrier, 
and it saturates at $\Gammat\approx {\hbar\omegas/4\pi}$ above the barrier
and becomes spin independent. 

Comparing the two cases quantitatively, it is seen that the
tunneling width in the case (a) is significantly smaller than
that in the case (b) by more than the order of $10^2$ for the
excitation energy $E_\ex=2 \sim 3$ MeV and the difference increases further
as  $E_\ex$ reaches near and above the barrier. 
(Note, however, that the value of $\Gammat$ in the case (b)
for the energy above the  barrier height is not meaningful
since Eq.~(\ref{EQ:WKB1}) does not give correct 
description of the statistical mixing that is expected above the barrier.)
We see also 
that the difference of $\Gammat$ between the two cases is not significant 
at the low excitations near the yrast SD states ($E_\ex \lesim 1$ MeV).
This is because $\Ds$ in this energy region is of the same order as
$\hbar\omegas\sim 1.0$ MeV.

\subsubsection{Decay-out properties}\label{Sec:DecOutGen}

\begin{figure} 
\centerline{
\epsfxsize=70mm\epsffile{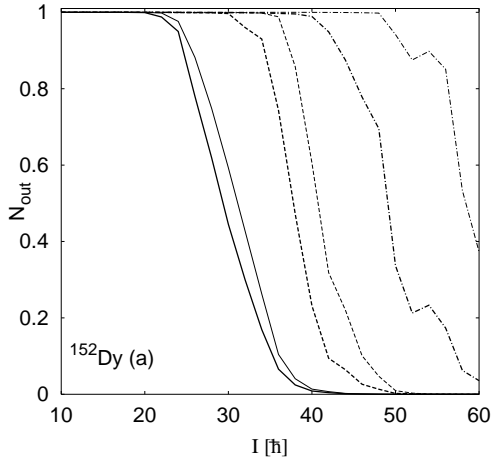}}
\caption{\label{Fig:NOUT}
The average decay-out probability $\Nout(E,I)$ plotted
as a function of spin, calculated for $^{152}$Dy $((\alpha,\pi)=(0,+))$
with $C_\rho=0.1$. The thick solid,
dashed and dot-dashed lines indicate the results in the case
(a) using  $\Gammat=(\Ds/2\pi) T$
for excitation energies $E_\ex =1.0$, 2.0 and 3.5 MeV, respectively.
Thin lines represent the results in the case (b) employing
$\Gammat=(\hbar\omegas/2\pi) T$. }
\end{figure}

Figure~\ref{Fig:NOUT} shows the calculated average decay-out branching
ratio $\Nout$ in $^{152}$Dy.
The value of $\Nout$ is plotted as a function of spin $I$ 
for different excitation energies $E_\ex=1.0,2.0$ and $3.5$ MeV 
measured from the SD potential minimum $E_\SD^0$.
A steep increase of $\Nout$ with decreasing $I$ around 
$\Nout\sim 1/2$ indicates a sharp decay-out, and this feature  
is commonly seen for all excitation energies. As the excitation
energy increases, the decay-out occurs at higher spins. 
This is because the tunneling probability increases.
In Fig.~\ref{Fig:NOUT}, we also show $\Nout$ calculated 
with use of Eq.~(\ref{EQ:WKB1}) (the case (b)) with thin  lines.
The difference
between the two cases (a) and (b) is significant for
high excitation energy $E_\ex \gesim 2$ MeV 
because of the
difference in $\Gammat$ becomes large as discussed above.
On the other hand, the two
evaluations give similar results for $E_\ex \lesim 1$ MeV.

\begin{figure} 
\epsfxsize=100mm\epsffile{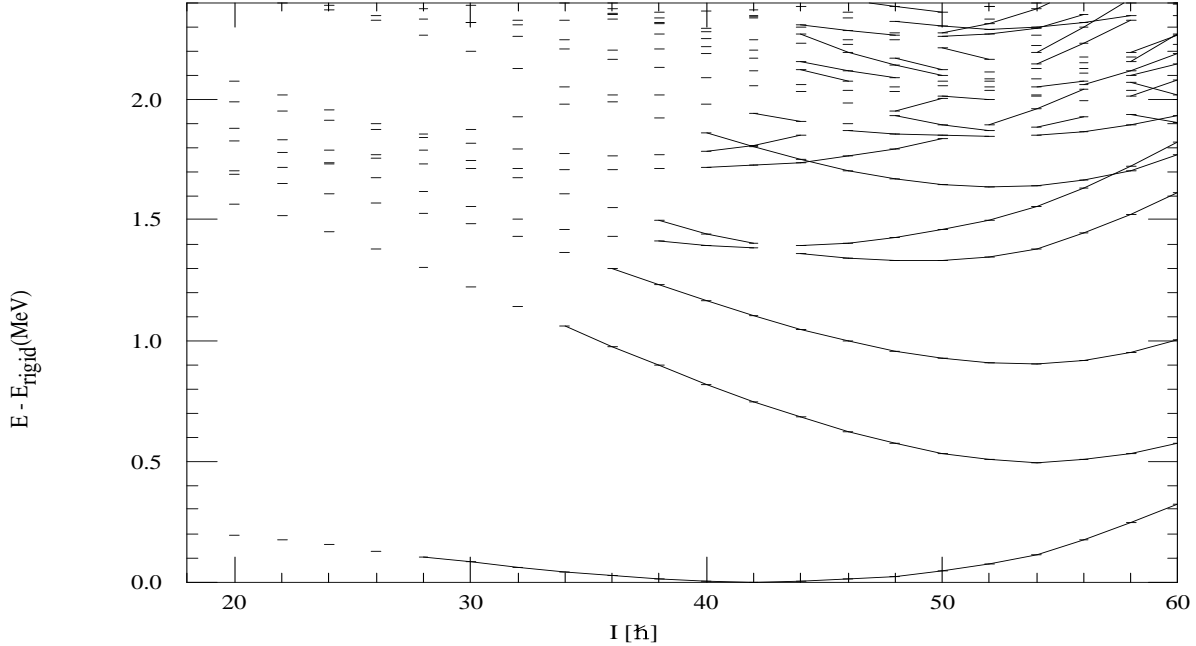}
\caption{\label{Fig:FLOW}
The energy levels of the yrast and excited SD states, shown with
horizontal bars, calculated for $^{152}$Dy ($(\alpha,\pi)=(0,+)$)
by means of the cranked shell model
diagonalization. A rigid body rotational energy 
$E_{\rm rigid}=I(I+1)/{\cal J}_{\rm rigid}$ is subtracted. 
Solid lines connecting the energy levels indicates 
strong E2 transitions satisfying $\tilde{S}_{\alpha I, \beta I-2}>1/\sqrt{2}$.
}
\end{figure}

The decay-out branching ratio $\Nout$  modifies the E2 transition
probability of exited superdeformed states
as described in \S\ref{Sec:TunWidth}. The energy levels $E_\alpha(I)$
of SD states and the E2 transition probabilities 
$\tilde{S}_{\alpha I,\beta I-2}$ calculated for
$^{152}$Dy are plotted in Fig.~\ref{Fig:FLOW}.  
In this figure shown are such strong E2 transitions that satisfy
$\tilde{S}_{\alpha I,\beta I-2}>1/\sqrt{2}$ 
by connecting the energy levels with solid lines.
Compared with the previous calculation which does not include
the decay-out effect (Fig.~6 in Ref.~\cite{Yoshida1}),
strong E2 transitions disappear at low spin and high
excitation energy due to the decay-out to normal deformed states. 
The sequence of strong E2 transitions associated with the yrast SD states
correspond to the yrast SD rotational band, and the figure shows that
the yrast SD band decays out around $I=28$ (See also Fig.~\ref{Fig:GamInt}). 
There exist several rotational bands above the yrast band
up to the excitation energy  $U\lesim 2$ MeV.
The decay-out spin of these bands is higher than
that of the yrast bands since the decay-out branching ratio $\Nout$ increases
with excitation energy.

Since the decay-out occurs at the spin where $\Nout$ increases
sharply from zero and reaches to $\Nout\approx 1/2$,
it is possible to define the
decay-out spin $\Iout(U)$ for a given excitation energy $U$ by means of 
a condition $\Nout=1/2$.
This definition of the decay-out spin is slightly different from
that used in \S\ref{Sec:yrast} to describe the decay-out of the yrast
SD band (It was defined there by $\prod_{I'}(1-\Nout(I'))=1/2$). 
However, the actual difference is less than $2\hbar$ for the 
value of the decay-out spin,  and in the following discussion we adopt 
the condition $\Nout=1/2$ as the definition of the decay-out
of excited superdeformed states.  Since $\Nout$ is a function of
the spin $I$ and the energy $E$, the criterion $\Nout=1/2$
gives a boundary in the $(E,I)$ plane. We can represent 
the decay-out boundary also in terms of the decay-out energy $\Eout(I)$
as a function of spin $I$ as well as  the decay-out spin  $\Iout(U)$.

\begin{figure} 
\centerline{
\epsfxsize=50mm\epsffile{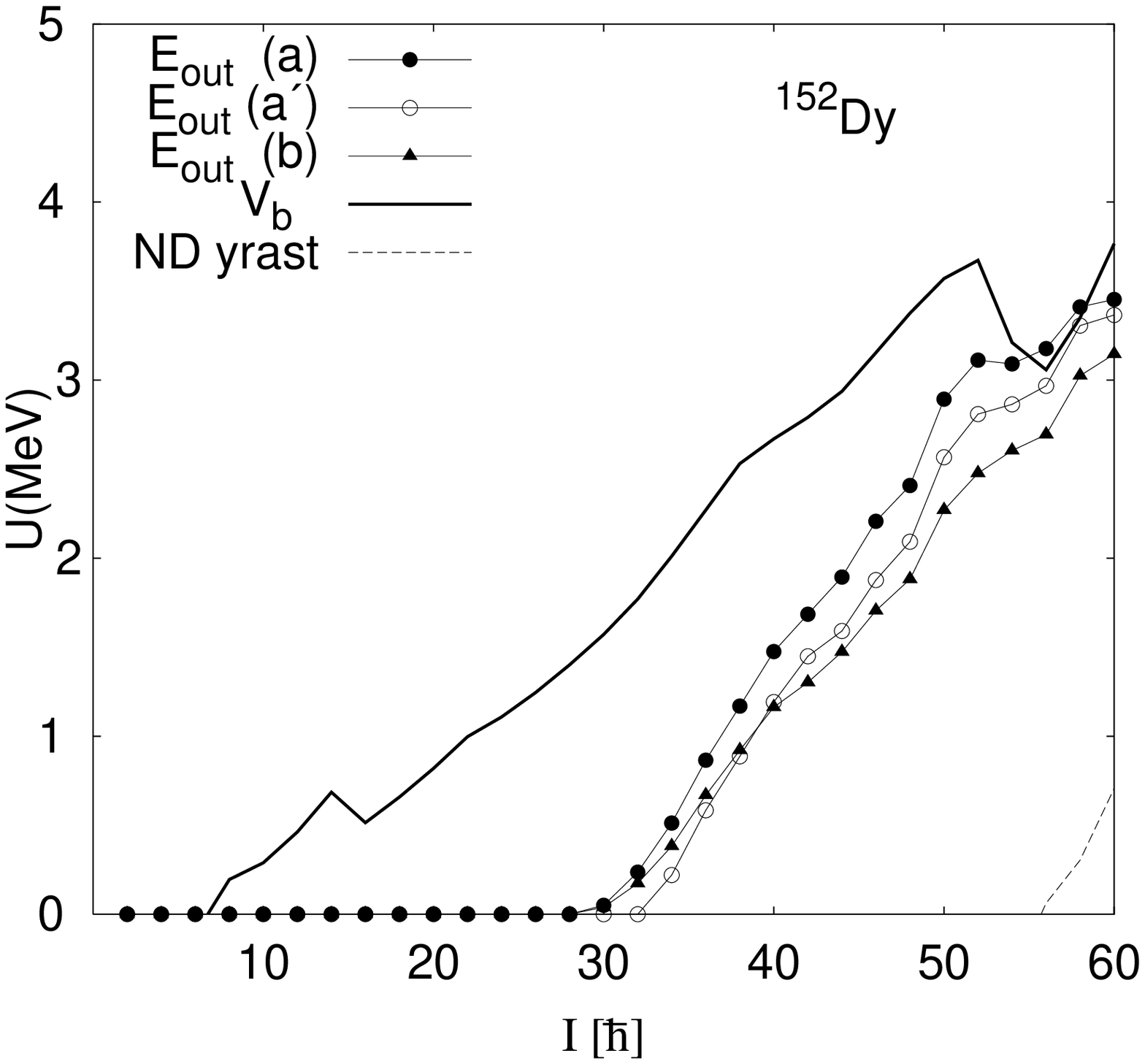}
\epsfxsize=50mm\epsffile{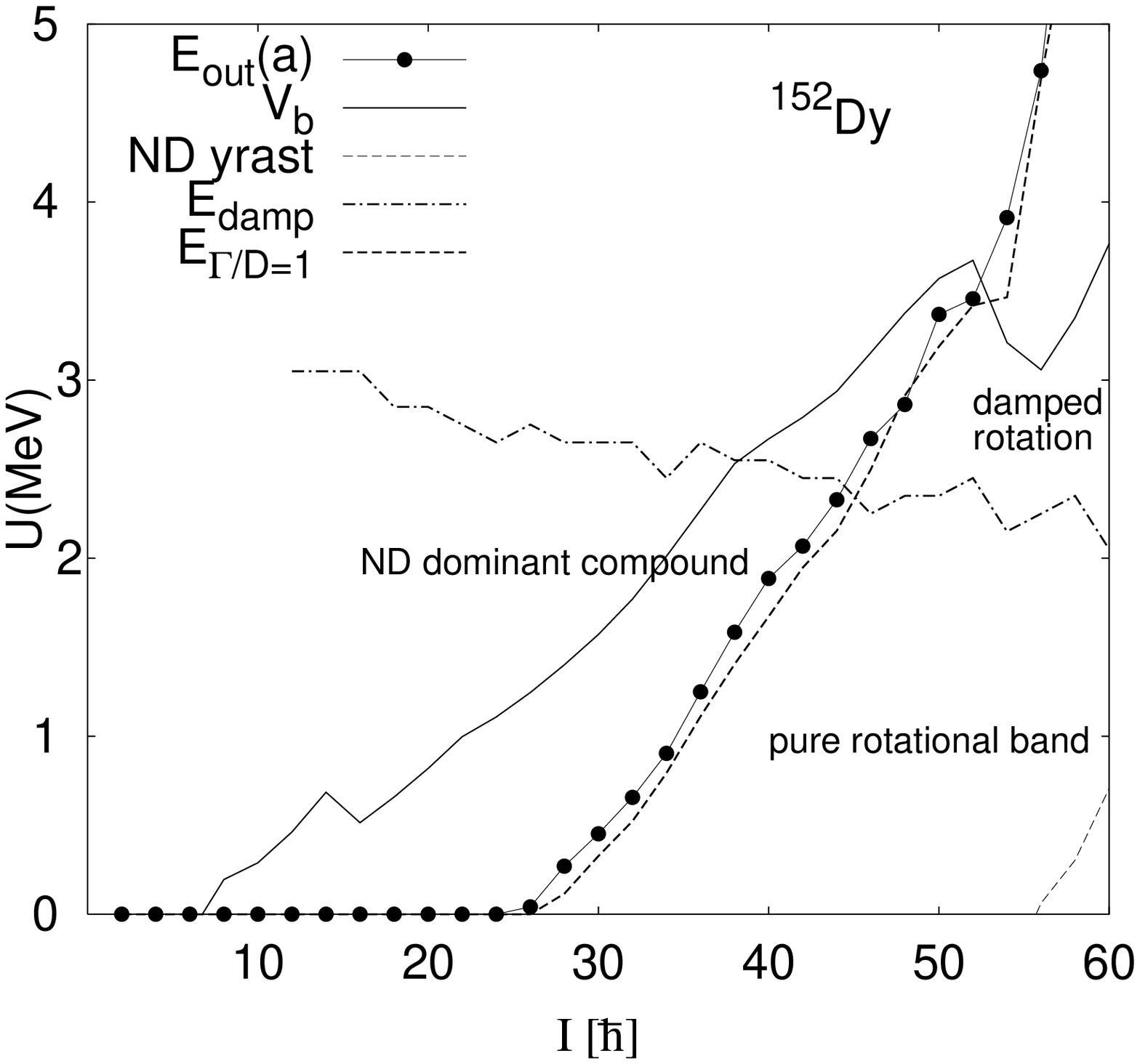}
\epsfxsize=50mm\epsffile{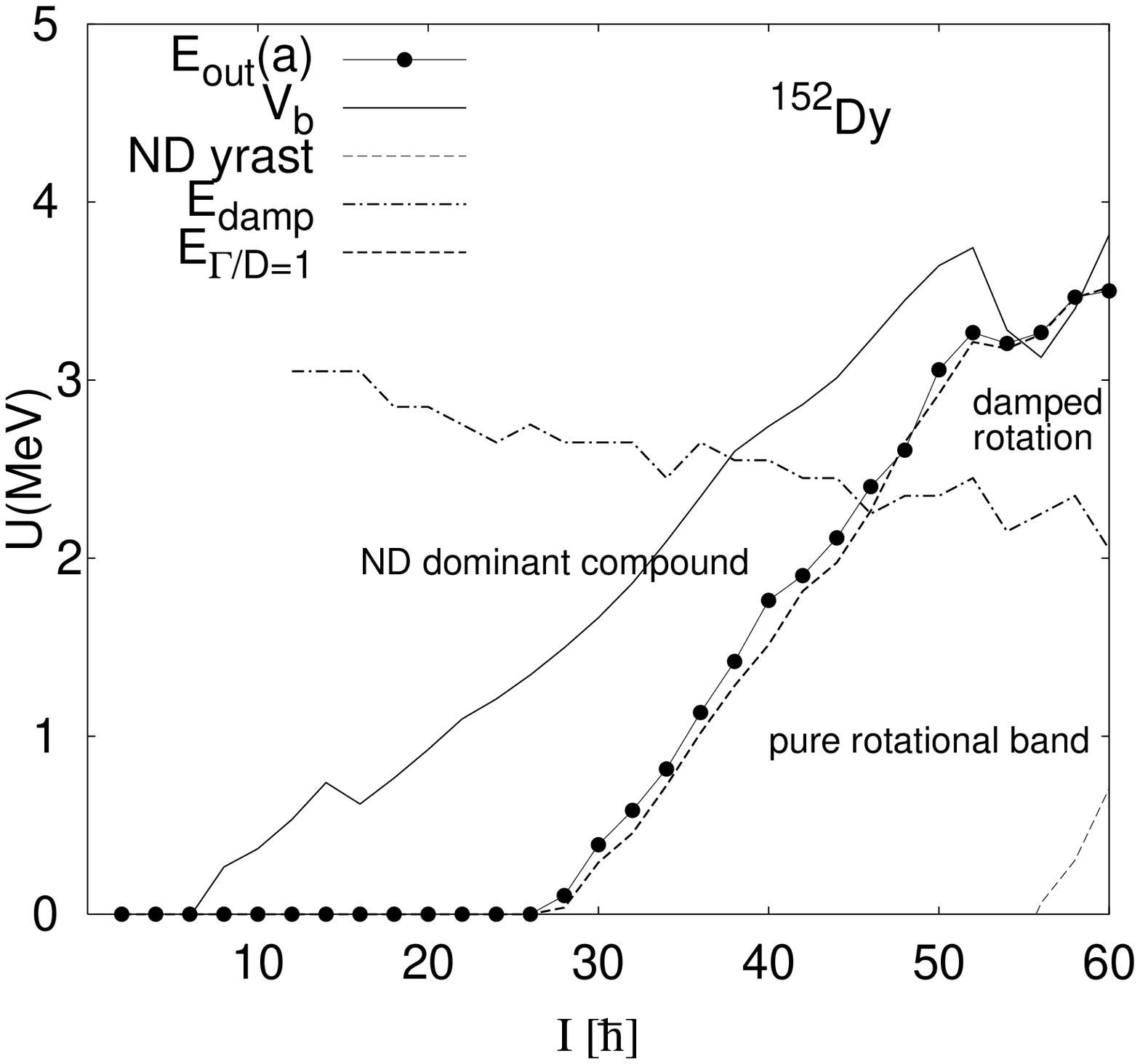}}
\caption{\label{Fig:BORDER1}
Left panel:
Decay-out energy $\Eout(I)$ for $^{152}$Dy ($(\alpha,\pi)=(0,+)$)
as a function of spin $I$,  together with the barrier height 
$\Vb$ and the ND yrast energy $E_\NDyr$. All these quantities
measured the SD yrast energy $E_\SDyr$ are plotted. 
The decay-out energy are shown in the three cases,
(a) $\Gammat=(\Ds/2\pi) T$, (a') $\Gammat=\Ds T$ and
(b) $\Gammat=(\hbar\omegas/2\pi) T$. The 
calculation is done without any adjustment
of the ND level density and the collective mass.
Middle panel: The onset energy of rotational damping
$\Edamp(I)$ is plotted together with the boundary energy for the SD-ND mixing
defined by $\Gammat/\Dn=1$ and the decay-out energy $\Eout$,
calculated in the case (a) with $C_\rho=0.1$.
The three regions  separated by the two boundaries $\Eout(I)$ and
 $\Edamp(I)$ are indicated as
``pure rotational band'',``damped rotation'',
and ``ND dominant compound'' (see text).
Right panel: Same as the middle except for
the adjustment of the collective mass with $C_{\rm mass}=1.4$.
}
\end{figure}

The decay-out energy $\Eout(I)$ calculated for 
$^{152}$Dy is shown 
in Fig.~\ref{Fig:BORDER1}.
Results in the left panel is obtained with no adjustment
of the level density nor of the mass parameter. 
In this figure, we show not only the 
results in the case (a), evaluated by using $\Gammat={(\Ds/2\pi)}T$,
Eq.~(\ref{EQ:WKB2}), which gives the average behaviour of the
decay-out, but also two other results in the cases (a') and (b). 
In the case (a') 
we assumed $\Gammat=\Ds T$ that gives the tunneling width different
from that in the case (a) by a factor of $2\pi$. Comparing the results
in the two cases (a) and (a'), we see that the 
ambiguity in the numerical factor of the tunneling width (difference
of the factor $2\pi$) does not cause large difference 
in the decay-out boundary. In the case (b), on the other hand, 
we evaluate with use of the other expression 
$\Gammat={(\hbar \omegas/2\pi)} T$, Eq.~(\ref{EQ:WKB1}), 
which is expected to give an upper limit of $\Gammat$.
The decay-out energies $\Eout(I)$ for the two cases
(a) and (b) differ by several hundred keV at maximum,
and the difference is larger than the ambiguity
of a numerical factor in the cases (a) and (a').
We can conclude from these comparisons that the reduction of the
knocking probability (the factor $\sim \Ds/\omegas$ in 
Eq.~(\ref{EQ:WKB2}) v.s. Eq.~(\ref{EQ:WKB1}))
plays important roles for the description of the excited SD states.

The decay-out boundary for the zero excitation energy $U=0$ corresponds
to the decay-out of the yrast SD band. As the excitation energy
increases, the decay-out boundary moves to higher spins.
In other
words, the decay-out of excited superdeformed states takes place at
higher spin than that of the yrast SD band. 
The calculation indicates that the decay-out spin $\Iout(U)$ increases by
about $8\hbar$ 
 for increasing excitation energy by 1 MeV.
Equivalently the decay-out energy $\Eout(I)$ increases 
monotonically with increasing spin. 
The slope of the decay-out
energy $\Eout(I)$ is larger than that of the barrier height $\Vb(I)$ 
since the decay-out probability $\Nout$ depends 
not only on the tunneling transmission coefficient 
but also on other quantities such as the level density of ND states, 
which vary strongly with spin.

The decay-out spin for $U=0$ is seen in the figure
to be $\Iout(U=0) \approx 30\hbar$, which is slightly larger than
the experimental decay-out spin of the yrast SD band. 
In order to have more reliable description,
we have discussed in \S\ref{Sec:yrast} 
two ways of adjustment by introducing the
renormalization factor $C_\rho$ of the ND level density  and
by the renormalization $C_\mass$ of the collective mass parameter.
In the middle (right) panel, we show the decay-out boundary for the
excited SD states, evaluated in the case (a), with use of the 
adjusted ND level density with $C_\rho=0.1$ (the adjusted collective
mass parameter $C_\mass=1.4$). 
Since the renormalization factors $C_\rho$ and $C_\mass$ are 
not very different from unity,
difference between the left and middle (right) panels is not large.
The difference between the two adjustments (middle vs. right panels)
is very small for the lower energy region $U \lesim 2$ MeV
since the decay-out spin of the yrast SD band is adjusted 
in both cases. The difference, however, is not negligible for higher
energy region, especially for the region above the barrier height.
The renormalization of the collective mass affects the
decay-out property only below the barrier, while the renormalization
of the ND level density influences both below and above the barrier.

In order to clarify the decay-out properties, it is useful to recall
that the average decay-out probability $\Nout$ is a function of
two ratios $\Gammat/\Dn$ and $\Gammas/\Gamman$~\cite{Vigezzi,ShimizuB}.
We found that the behavior of the decay-out energy $\Eout$ is 
mainly governed by the ratio  $\Gammat/\Dn$. To show this,
we plot in the middle and right panels of Fig.~\ref{Fig:BORDER1}
a boundary energy that is defined by the condition
${\Gammat/ \Dn}=1$, and compare with the decay-out energy $\Eout(I)$. 
If the ratio ${\Gammat/ \Dn}$ is more than 1,
it implies that a significant amount of components of ND states mix into
SD states. The condition  ${\Gammat/ \Dn}=1$ defines a
boundary for the strong mixing among SD and ND states. It is
seen from the figure that
the boundary of strong mixing follows rather well with the
decay-out energy $\Eout(I)$. 
The ratio of the electromagnetic
transition widths $\Gammas/\Gamman$ plays a relatively
minor role in determining the spin and excitation energy dependence 
of the decay-out, especially for the spin region much higher
than the decay-out spin of the yrast SD band. 
For the high spins, the E2 decay width of ND states
dominates over the E1 decay width, and 
the ratio of electromagnetic widths $\Gammas/\Gamman$ is
approximated by
$\left(Q_{\rm s}/Q_{\rm n}\right)^2 \times
\left(\Jmomn/\Jmoms \right)^5$,
where $\Qs$ ($\Qn$) and $\Jmoms$ ($\Jmomn$) are
the static quadrupole moment and moment of inertia
for the SD (ND) states.  The ratio then depends
neither on excitation energy nor spin, and takes 
a value which is a little more than 1 but does not
exceed $10^1$.  

The decay-out boundary $\Eout(I)$  
in the region above the barrier height 
shows behaviour different from that below the barrier.
As seen in the middle panel
(and also in Figs.~\ref{Fig:BORDER2} for $^{143}$Eu
and \ref{Fig:BORDER3} for $^{192}$Hg),
the boundary rises very sharply above the barrier.
In the region above the barrier the SD and ND states mix statistically
to form compound states as 
the ratio $\Gammat/\Dn$ reduces to the ratio of the level spacings 
$\Ds/\Dn$ of SD and ND states.
(Note that  the results of the cases (a) or (a') should be adopted
for the region above the barrier since the case (b) is not applicable 
in this region.)
As spin increases,  the
ND yrast energy relative to the SD yrast becomes higher and 
the ratio $\Gammat/\Dn \sim \Ds/\Dn=\rhon/\rhos$ decreases exponentially 
for a fixed value of $U$. In other words, 
superdeformed states become dominant
components of the statistically mixed compound states. 
On the other hand, the ratio $\Gammas/\Gamman$ increases with
spin since the E2 decay probability
$\Gammas$ associated with SD states becomes larger
due to the $E_\gamma^5$ factor 
while the ND decay width $\Gamman$ decreases due to the decrease
of the relative excitation energy.  Both two effects, the
decrease in  $\Gammat/\Dn \sim \Ds/\Dn$ and the increase in
$\Gammas/\Gamman$, favor dominance of 
the SD transitions at very high spins while the ND transitions
dominates at lower spins. The decay-out spin above the
barrier height represents a border where the relative dominance
of the SD and ND transitions is interchanged.

\subsubsection{Rotational damping effects}\label{Sec:RotDampGen}

Let us now discuss effects of the rotational damping on the
excited superdeformed states. As seen in Fig.~\ref{Fig:FLOW}, 
The strong E2 transitions with 
$\tilde{S}_{\alpha I,\beta I-2} > 1/\sqrt{2}$ gradually 
disappear in the region with excitation energy $U$ greater than 
about 2 MeV. 
This disappearance of the
strong E2 transitions is caused by the rotational damping. Although
the transition from the rotational band structures seen near the yrast line
to the rotational damping in the highly excited states is gradual, let us
characterize it by introducing a boundary energy for the onset 
of rotational damping. 
For this purpose, we calculate the branching number 
$\nbranch=(\sum_\beta S_{\alpha I,\beta I-2}^2)^{-1}$~%
\cite{Matsuo,Yoshida1,Yoshida2} that quantifies the fragmentation of 
E2 decay from a SD state $\alpha$ at spin $I$. The onset energy of damping
$\Edamp(I)$ is then defined as the energy where the average 
branching number calculated as a function of the
energy $E$ exceeds 2 for a given spin $I$. 
The calculation of $\Edamp(I)$ is the same as 
in Ref.~\cite{Yoshida1,Yoshida2}. 
The onset energy $\Edamp(I)$ of the rotational damping is plotted
in the middle and right panels of Fig.~\ref{Fig:BORDER1}. 

Using the decay-out energy $\Eout(I)$ and the
onset energy $\Edamp(I)$ of rotational damping, 
the excited superdeformed states are divided into three regions.
The E2 transitions associated with the 
superdeformed states can exist only in the low energy region 
below the decay-out boundary $\Eout(I)$ or at the high spin side of 
this boundary. This region is further
divided into two by the second boundary $\Edamp(I)$. Below $\Edamp(I)$,
the superdeformed states form the rotational band structures, i.e.,
sequences of levels connected by the strong E2 transitions. The superdeformed
rotational bands are confined in this region, which is labeled
in Fig.~\ref{Fig:BORDER1} with ``pure rotational band''. 
On the other hand, concerning the
excited SD states which are located above $\Edamp(I)$ and at the high spin 
side of the decay-out boundary $\Eout(I)$, they decay also by the
rotational E2 transitions characteristic to the SD shape, but the
the rotational E2 transitions are fragmented (i.e. the rotational
damping) due to the admixture of superdeformed
many-particle many-hole configurations caused by the residual
interaction.
This second region is labeled in Fig.~\ref{Fig:BORDER1}
with ``damped rotation''. 
Note that the damped E2 transitions can be found even
above the barrier. 
The third region that is situated at low-spin side 
(and high energy side) of the
decay-out boundary $\Eout(I)$ is labeled with ``ND dominant compound''
in Fig.~\ref{Fig:BORDER1}. 
The SD states in this region mix strongly with the ND states,
and the decay of these states are dominated by the E1 or E2 transitions
associated with the ND states. 

We may use the onset energy $\Edamp(I)$ as a guideline of whether
the excited superdeformed states have character of compound states.
At excitation energy higher than $\Edamp(I)$, the SD states become strong
admixture of different SD $n$p-$n$h configurations. Thus it may be
justified to use the average estimate of the tunneling
width, given by Eq.~(\ref{EQ:WKB2}) (the case (a)  with 
the knocking frequency $\propto \Ds$),
for the energy region sufficiently higher than
$\Edamp(I)$. But for the energy region below
$\Edamp(I)$, we can expect possible fluctuation around the average estimate.
As discussed above, an upper-limit estimate of the fluctuation is given
by Eq.~(\ref{EQ:WKB1}) (the case (b)).  
Note, however, that the ambiguity due to the fluctuation is not large 
in this energy region since 
the difference in the decay-out
boundary $\Eout(I)$ between the two cases
is small (see the left panel of Fig.~\ref{Fig:BORDER1}).  

\subsubsection{$^{143}$Eu}\label{Sec:ResEu143}

\begin{figure} 
\centerline{
\epsfxsize=50mm\epsffile{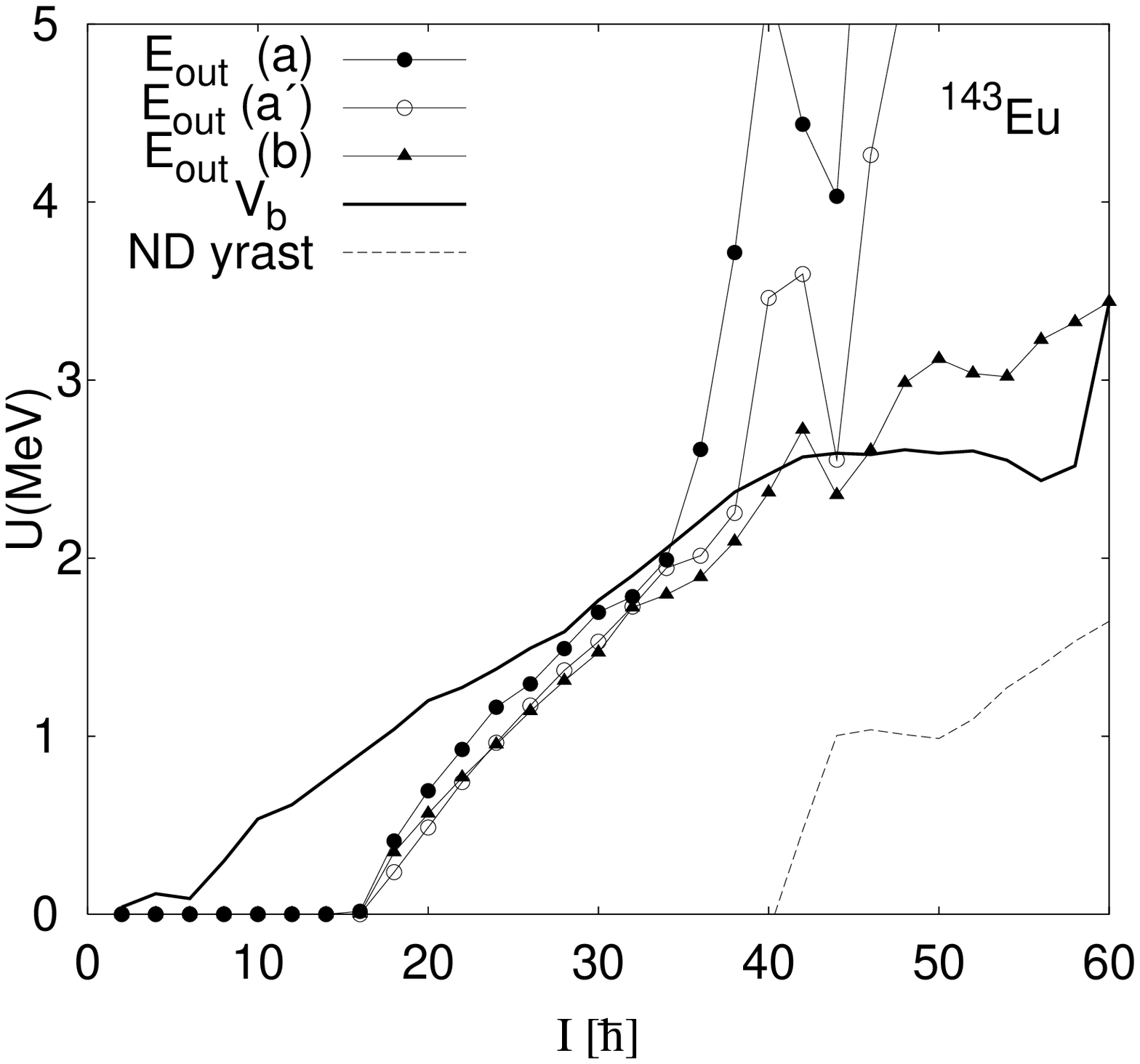}
\epsfxsize=50mm\epsffile{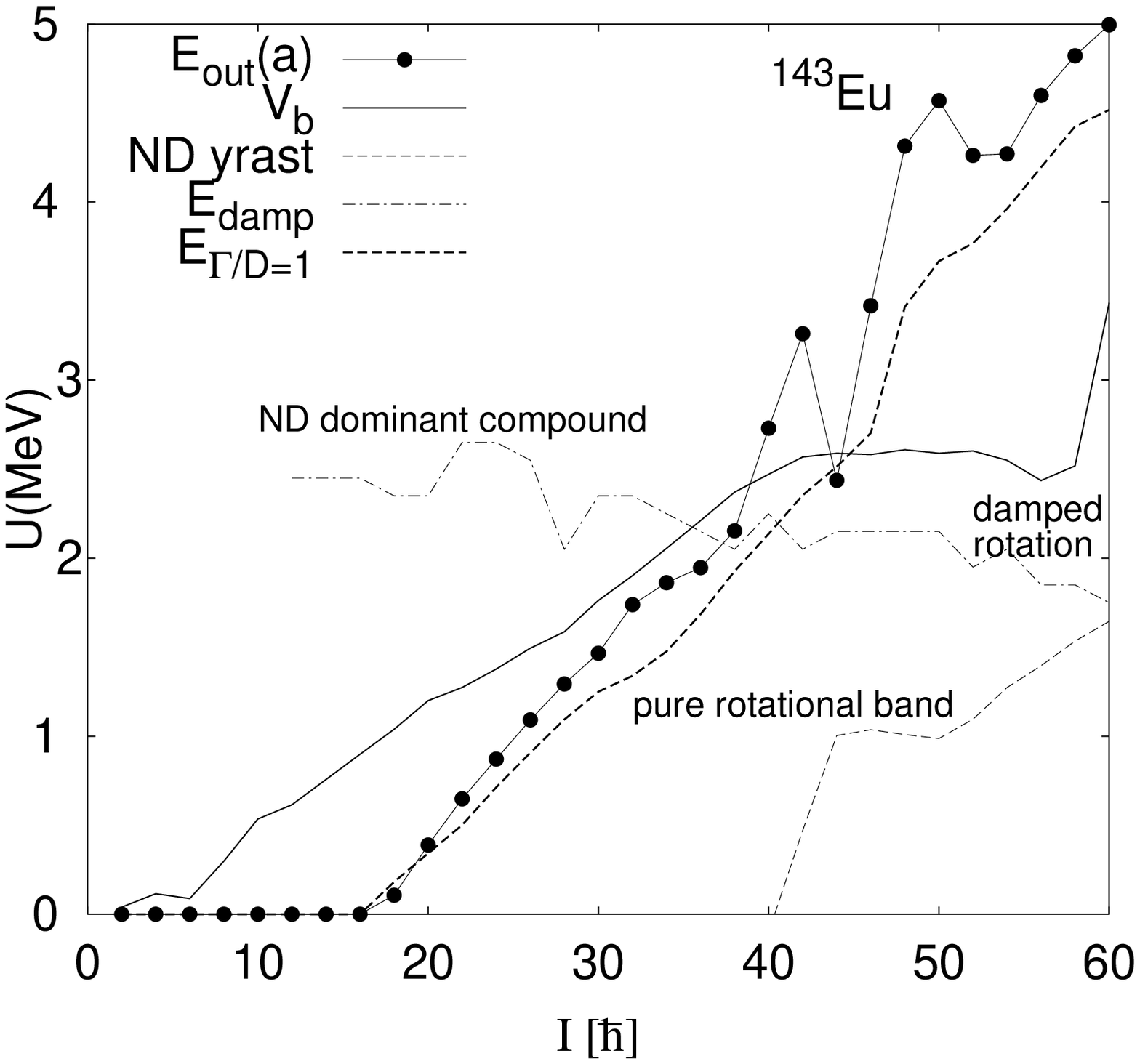}
\epsfxsize=50mm\epsffile{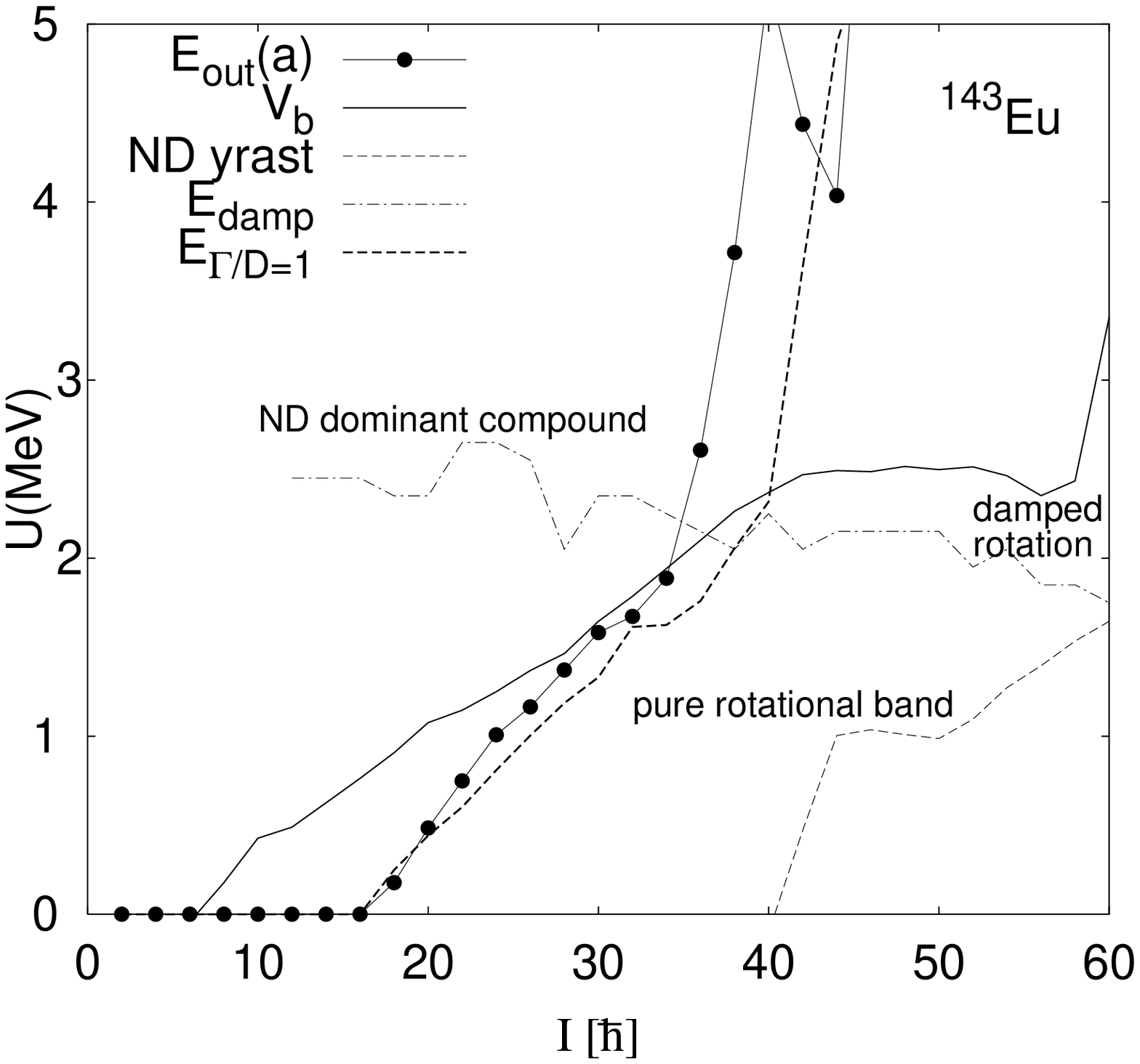}}
\caption{\label{Fig:BORDER2}
Same as Fig.~\ref{Fig:BORDER1} but for $^{143}$Eu 
($(\alpha,\pi)=(1/2,+)$) with  $C_\rho=15.0$ (middle panel) and with
 $C_{\rm mass}=0.7$ (right panel)}
\end{figure}

Figure \ref{Fig:BORDER2} shows the results for $^{143}$Eu.
Basic features are found similar to those in $^{152}$Dy.    
The influence of the adjustment of the ND level density (the middle
panel with $C_\rho=15.0)$ and of the collective mass parameter 
(the right panel with $C_\mass=0.7$) on the decay-out boundary is not large,
likewise in the case of $^{152}$Dy. 
There exits also in this nucleus 
the three regions classified in terms of 
the decay-out and the onset of the rotational damping, as shown in the 
middle and right panels. A remarkable difference from $^{152}$Dy
is that the decay-out boundary is located at lower spin by about
$10\hbar$. The low values of decay-out spin is seen not only for 
the yrast SD states $U=0$ (as discussed in \S\ref{Sec:yrast}) but also
for the highly excited SD states including the states above the barrier.
Consequently the region of ``damped rotation'' is much
larger than in $^{152}$Dy. The calculation predicts 
the decay-out spin $\Iout \sim 35-45\hbar$
for the highly excited SD states above
the barrier height. The SD states lying 
$I \gesim 35-45\hbar $ and  $U \gesim 2$ MeV emit 
characteristic rotational E2 gamma-rays with
fragmentation caused by the rotational damping. This feature
seems consistent, at least qualitatively, with the experimental
observation of intense quasi-continuum E2 transitions~\cite{Leoni-decay}.

\subsubsection{$^{192}$Hg}\label{Sec:ResHg192}

\begin{figure} 
\centerline{
\epsfxsize=50mm\epsffile{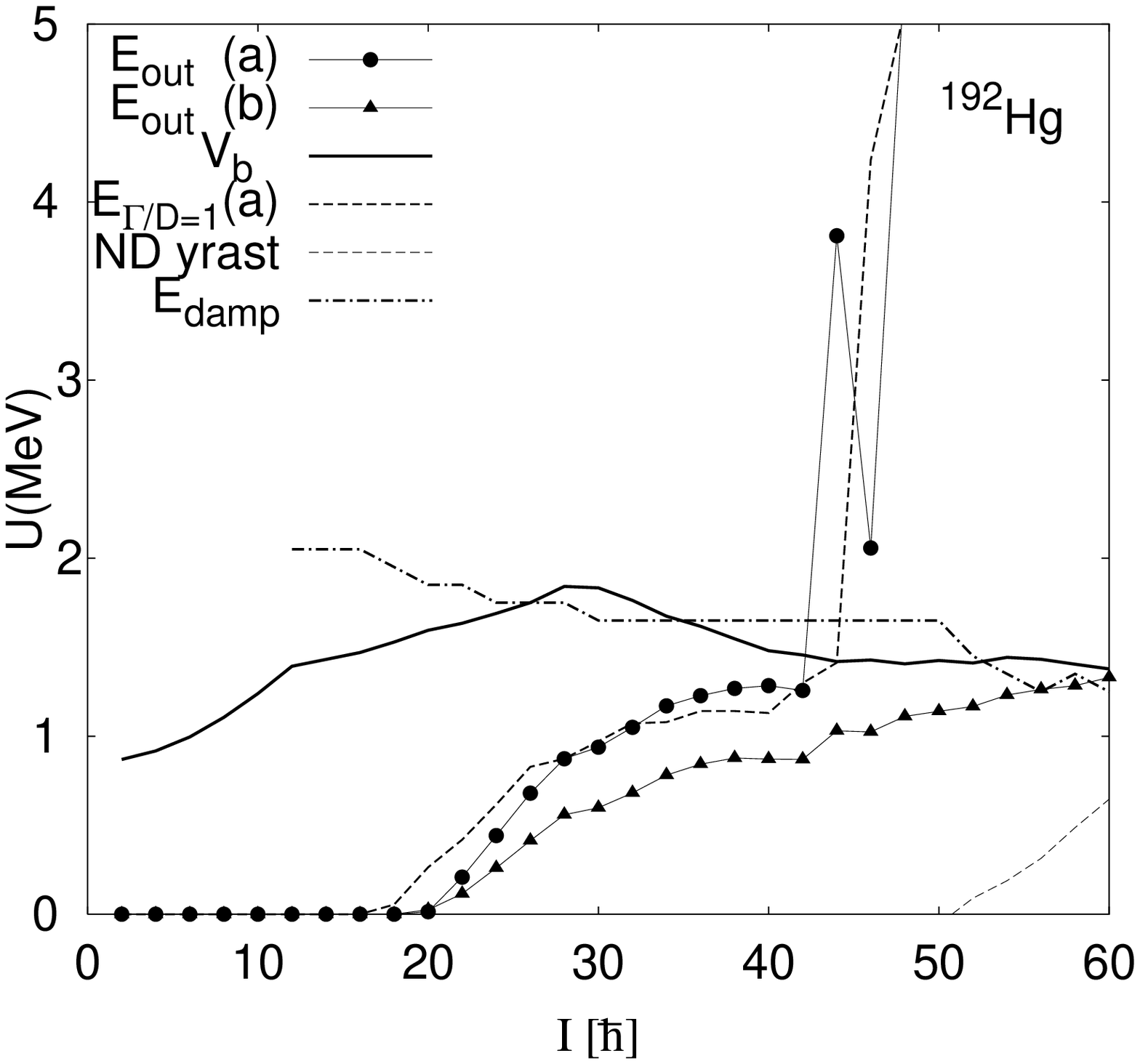}
\epsfxsize=50mm\epsffile{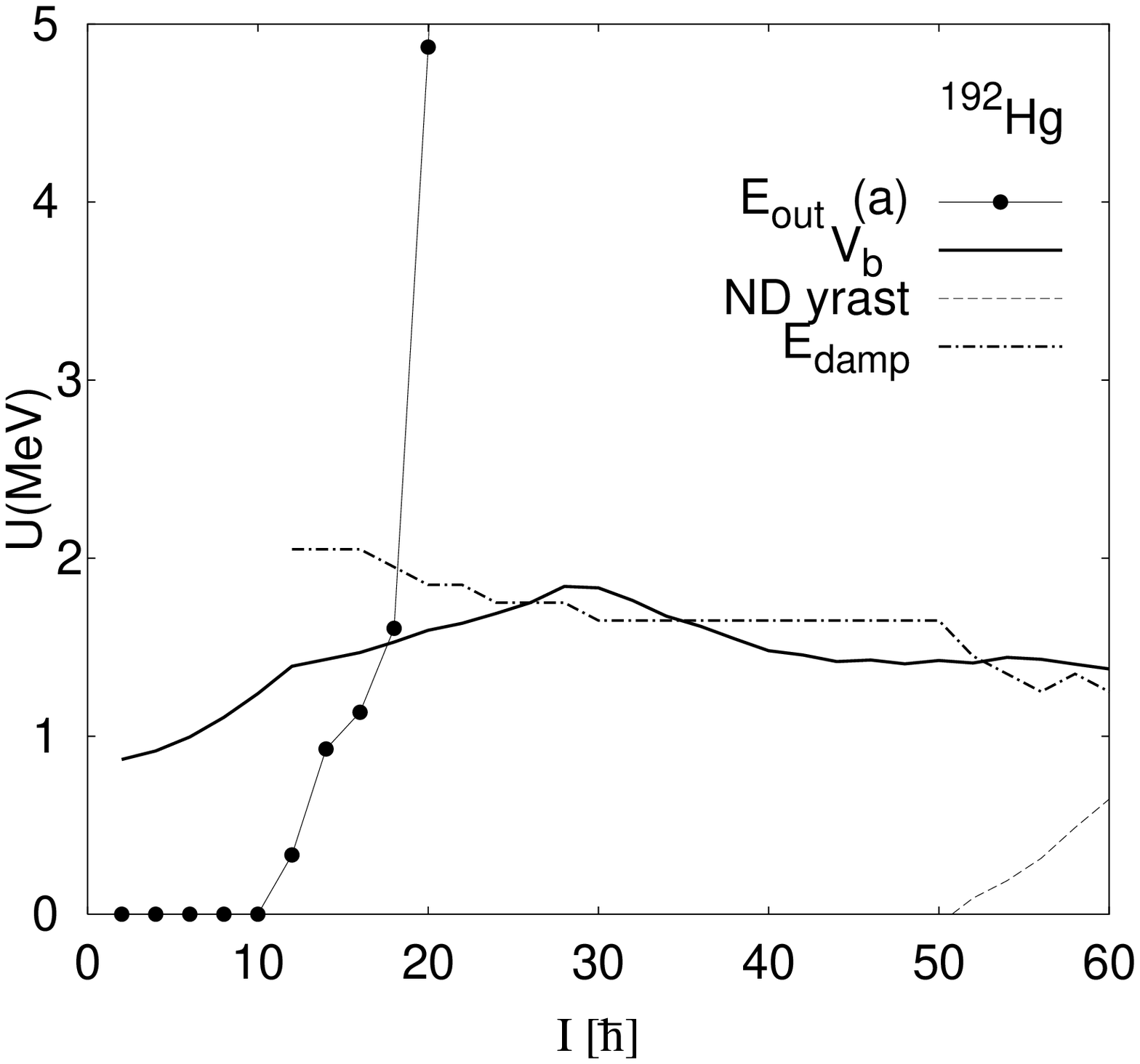}
\epsfxsize=50mm\epsffile{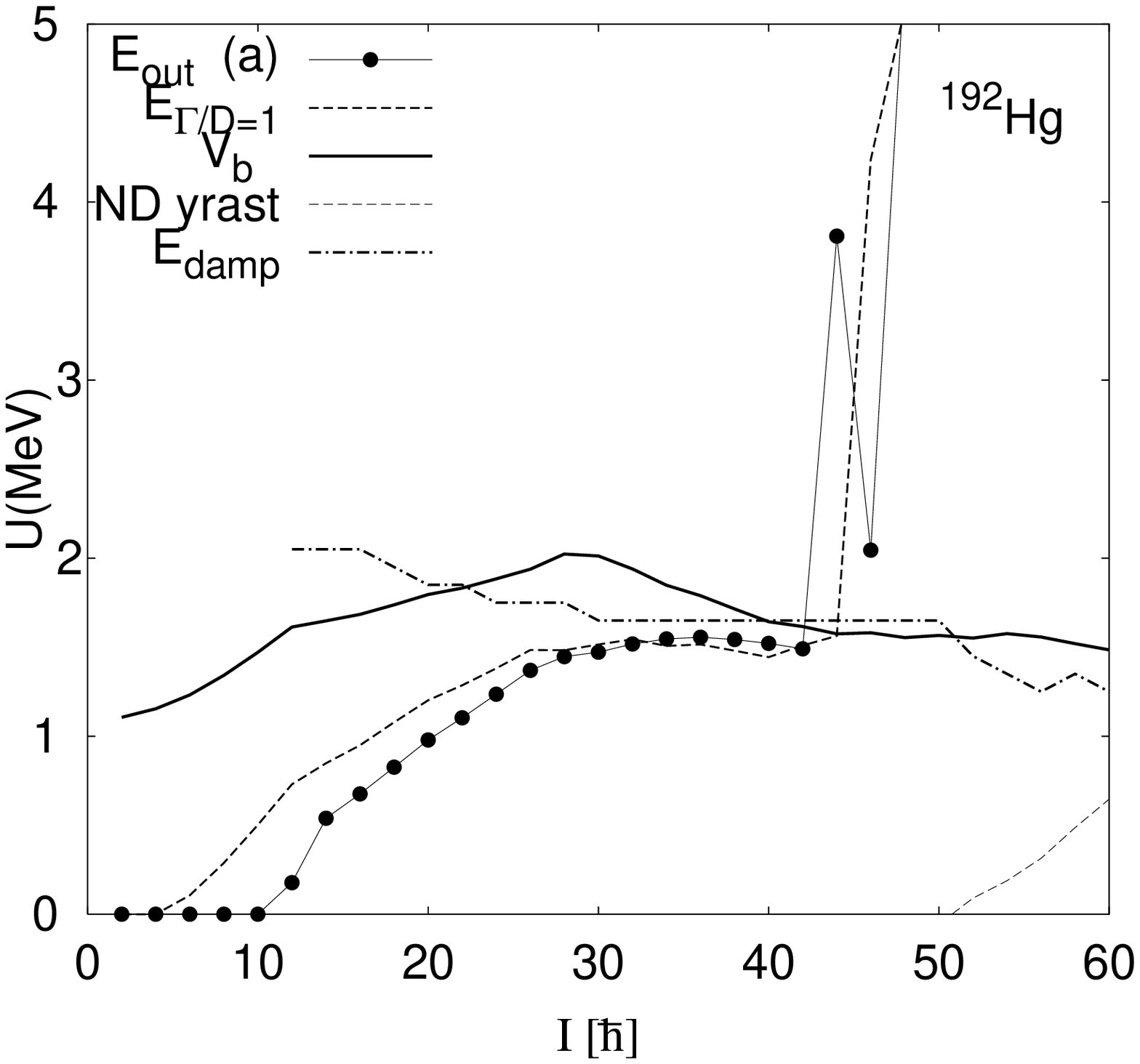}}
\caption{\label{Fig:BORDER3}
Left panel: Calculated decay-out energies $\Eout(I)$ in the
cases (a) and (b) (see text) for 
$^{192}$Hg ($(\alpha,\pi)=(0,+)$) calculated with
no adjustments ($C_\rho=1,C_\mass=1$), plotted together with the
onset energy of rotational damping $\Edamp(I)$,
the boundary energy for the SD-ND mixing
defined by $\Gammat/\Dn=1$ calculated in the case (a),
the barrier height $\Vb$ and the ND yrast energy $E_\NDyr$.
Middle panel:  
The decay-out energies $\Eout(I)$ in the case (a),
calculated with
the level density renormalization $C_\rho=2.0\times 10^{-4}$.
Right panel:  The decay-out energies $\Eout(I)$ in the case (a),
calculated with the collective mass renormalization $C_\mass=3.0$
and the boundary energy for the SD-ND mixing.
}
\end{figure}

Figure~\ref{Fig:BORDER3} shows the results for $^{192}$Hg.
Although there are some common features to those in $^{152}$Dy and
$^{143}$Eu, large differences are observed. The left panel shows
the results without any adjustment. 
It is seen that difference between
the case (a) and the case (b) is larger than in  $^{152}$Dy and
$^{143}$Eu. This is because the spin dependence of the tunneling
width $\Gammat$ is much weaker than in $^{152}$Dy  and $^{143}$Eu
as seen in Fig.~\ref{Fig:SPW}, 
hence the difference in
$\Gammat$ (the cases (a) vs. (b)) 
causes larger change in the decay-out boundary.
Weak spin-dependence is also seen in the barrier height.
These features are due to the qualitative difference
of the calculated potential energy surface between the Dy and Hg nuclei,
which is stressed at the end of \S\ref{Sec:path}.
It is also noted that
the decay-out boundary above the barrier is located
at rather low spin ($I\approx 40\hbar$).
This is caused by the relatively high level density of the superdeformed
states in this nucleus. (The corresponding level density parameter is
a=A/13.0 MeV$^{-1}$.) 
An unwanted feature in the
left panel is that the theoretical prediction of the decay-out spin
for the yrast SD band is significantly higher than that observed in
the experiment by
more than $10\hbar$ as already discussed in \S\ref{Sec:yrast}.

In \S\ref{Sec:yrast},
we introduced two ways of adjustment by renormalizing the level
density of ND states by a factor 
$C_\rho=2.0\times 10^{-4}$ and by renormalizing
the collective mass parameter by $C_\mass=3.0$.
The results of these calculations are shown in the middle and right panels.
Remarkably large difference between the two calculations is seen 
for the decay-out boundary of the excited SD sates although
both are adjusted to reproduce the observed decay-out of yrast SD band.
The change in the collective mass parameter
increases the action value of the tunneling path by a factor of 1.9$-$2.1
and reduces the tunneling transmission coefficient, but it does not
influence the behavior above the barrier where the tunneling action
becomes zero. On the other hand, the
change in the ND level density influences both below and above 
the barrier.  Since the renormalization factor for 
the adjusted ND level density ($C_\rho=2.0\times 10^{-4}$) is very
small, its effect is drastic. For example, 
the decay-out boundary above the barrier
is largely shifted to
$I \approx20\hbar$, which is very different from
the decay-out boundary calculated with the adjusted collective mass
($C_\mass=3.0$). Note, however, that the
modification of the ND level density with the factor of $10^{-4}$ 
may not be realistic, although it is also not certain 
whether the renormalization
of the collective mass parameter by a factor of 3 is justified.
We can say that the decay-out boundary has different
spin dependence and stays at considerably
lower excitation energy in $^{192}$Hg compared to $^{152}$Dy and $^{143}$Eu
if a realistic ND level density is used,
and this is mainly due to the fact that the barrier height
in the $A \approx 190$ nuclei saturates or even decrease
gradually at $I \gesim 20-30 \hbar$.

\subsection{Number of rotational bands}
\label{Sec:Nband}

As Fig.~\ref{Fig:FLOW} illustrates,
the number of superdeformed rotational bands 
existing in a nucleus is limited to a finite number since the 
rotational E2 transitions associated with highly excited levels
exhibit the rotational damping.
In addition, the decay-out transitions are present at low spins. 
Since the rotational band structures are not observed 
if the decay-out probability dominates over the E2 transitions,
the decay-out further reduces the number of rotational bands especially 
at low spins where the decay-out plays a role. 
An effective
number of decay paths, which approximately gives 
the number of superdeformed rotational bands, has been evaluated
experimentally from analysis of  the quasi-continuum 
E2 gamma-rays in superdeformed nuclei~\cite{FAM,Leoni-Npath,Leoni-new}.
In contrast to the intensity data for the decay-out of yrast SD bands,
where the tunneling probability in a narrow spin-range is relevant
because of the sharp decay-out, the number of rotational band
is sensitive to the decay-out process in a broad spin-range,
as is shown in the following, and is  effective to study
the barrier penetration problem.
We, therefore, adopt in this subsection
the number of rotational bands as a typical observable 
on the excited superdeformed states, and discuss in detail
effects of the decay-out on this quantity. 

The number of superdeformed rotational bands, $\Nband$, is evaluated as
follows. To make a distinction between the levels forming rotational bands 
and those exhibiting the rotational damping, 
we consider the branching number of E2 transitions
$\nbranch=\left( \sum_\beta S_{\alpha I,\beta I-2}^2 \right)^{-1}$
defined for each  SD state $\alpha$ at spin $I$. 
The effect of the
decay-out on this quantity is taken into account by
means of Eq.~(\ref{EQ:E2-renorm}). Namely, 
the branching number is now given by 
\begin{eqnarray}
\tilde{n}_{\rm branch}(\alpha,I)&=&
\left( \sum_\beta \tilde{S}_{\alpha I,\beta I-2}^2 \right)^{-1}\\
&=&(1-\Nout(E_\alpha,I))^{-2}\nbranch(\alpha,I)
\end{eqnarray}
by multiplying the factor $(1-\Nout(E_\alpha,I))^{-2}$ to the
original branching number $\nbranch$. 
The modified branching number takes a
very large value if the decay-out transition is dominant (i.e. 
$\Nout\approx 1$). 
The condition $\tilde{n}_{\rm branch}<2$ 
gives a definition of the states having strong E2
transitions forming superdeformed rotational bands. 
For fixed spin and parity, we
count the excited superdeformed levels 
that satisfy $\tilde{n}_{\rm branch}<2$ for consecutive
two steps of strongest E2 transitions~\cite{Matsuo,Bracco-simul,
Yoshida1,Yoshida2}. 
Summing over four combinations of parity and signature quantum numbers,
we evaluate the number of superdeformed rotational bands $\Nband$. 

\begin{figure} 
\centerline{
\epsfxsize=80mm\epsffile{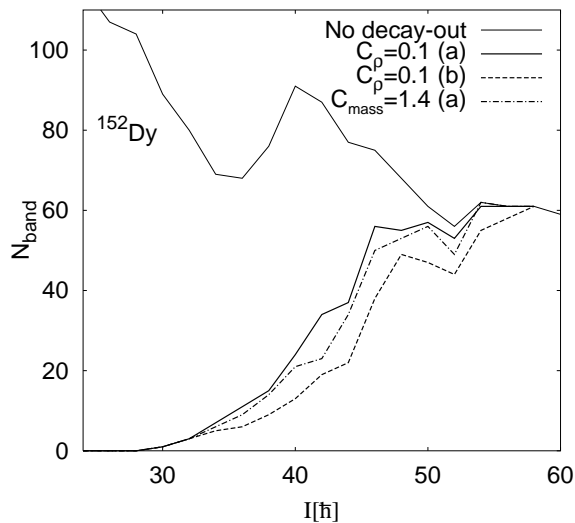}}
\caption{\label{Fig:NUMBER1}
The number of SD rotational bands, $\Nband$, for
$^{152}$Dy, calculated in the two cases (a) and (b) 
(see text) with $C_\rho=0.1$ and plotted as a function of spin $I$.
Also shown are the result with $C_\mass=1.4$ in the case (a) and 
the one without the decay-out effect.
}
\end{figure}

We show in Fig.~\ref{Fig:NUMBER1} the  number 
of superdeformed rotational bands calculated for $^{152}$Dy. 
Here we plot $\Nband$ obtained with the renormalization
of the ND level density ($C_\rho=0.1$) and of the
collective mass parameter ($C_\mass=1.4$) that are chosen to
reproduce the decay-out spin of the yrast SD band (cf. \S\ref{Sec:yrast}).
The average estimate 
of the tunneling width 
$\Gammat=(\Ds/2\pi) T$ (the case (a)) is adopted.
The result without the decay-out is also plotted for comparison,
which is the same as that obtained in the previous paper~\cite{Yoshida1}.
We find that the effect of decay-out  reduces 
 significantly the number of bands 
for the spin region $I\lesim 50\hbar$.
This behavior is easily understood from Figs.~\ref{Fig:FLOW} and 
\ref{Fig:BORDER1}.
Since the decay-out boundary cuts the region of pure rotational bands
at $I\sim26$ for $U=0$ (yrast states)
and at $I\sim50$ for $U\sim \Edamp$, the effect of the
decay-out on $\Nband$ become sizable in 
the spin region $I\approx 26-50\hbar$.
At high spins $I\gesim 50$ 
where $\Eout(I)$  exceeds $\Edamp(I)$,
the decay-out has  little influence on  $\Nband$.
There is only minor difference between the two ways of
renormalization. 
In the figure we plot also $\Nband$ obtained with 
the upper limit estimate of 
$\Gammat=(\hbar\omegas/2\pi) T$ (the case (b)).
The case (b) can be considered to provides a lower limit for 
the evaluated $\Nband$. The difference in $\Nband$
between the two cases (a) and (b) is up to about 
by $10\sim 15$.
The number of bands is as much as $50-60$ for long spin
interval  $I \sim 45-60$.
It is emphasized that
this number is larger than those for ND rare-earth nuclei 
($\Nband \sim 30$),
indicating that the shell effect leading to the large value of
$\Nband$~\cite{Yoshida1,Yoshida2} remains even with the presence of the
decay-out effect.

\begin{figure} 
\centerline{
\epsfxsize=80mm\epsffile{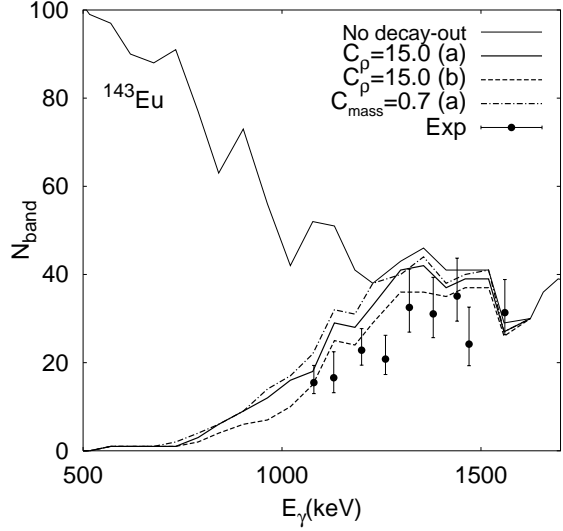}}
\caption{\label{Fig:NUMBER3}
Same as Fig.~\ref{Fig:NUMBER1} but for
$^{143}$Eu. Here $\Nband$ is calculated either
with the adjusted $C_\rho=15.0$ or with the mass
renormalization $C_\mass=0.7$. It is plotted as
a function of the average gamma-ray energy $E_\gamma$, 
and compared with the experimental
effective number of decay-path~\cite{Leoni-new}.
}
\end{figure}

Figure~\ref{Fig:NUMBER3} shows the results for 
$^{143}$Eu. Here the
experimental data~\cite{Leoni-new} are compared.
To make a direct comparison, we plot $\Nband$ as a function 
of the E2 gamma-ray energy. The transformation from the spin to
the gamma-ray energy is done by calculating the 
average gamma-ray energy calculated  over all E2 transitions as
a function of spin.
We show the calculated results for both cases (a) and (b).
The difference between (a) and (b) and also the difference
between the two renormalizations are
found smaller than that in $^{152}$Dy,
and it is not more than the size of experimental error bars.
The decay-out has little influence on the number of bands for
spin region $I\gesim 50$ ($E_\gamma\gesim1400$ keV). 
Gradual decrease of number of bands with decreasing spin 
in the spin region $I\lesim 50$ ($E_\gamma\lesim1400$ keV) is
the decay-out effect. Note that with the decay-out effect
we obtain overall agreements between
the calculated results and the experimental data.
The agreement is satisfactory although it is not perfect.
We may need more precise comparison using simulated gamma-ray
cascades if we want to evaluate the differences between
the theories and the experiment.

\begin{figure} 
\centerline{
\epsfxsize=80mm\epsffile{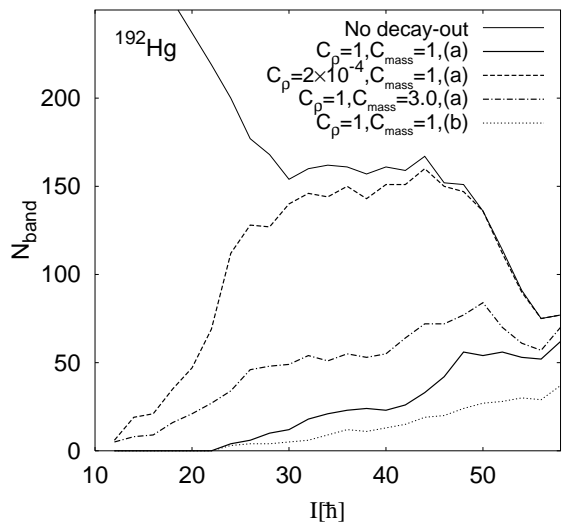}}
\caption{\label{Fig:NUMBER2}
The calculated $\Nband$ for $^{192}$Hg. The results with $C_\rho=C_\mass=1$
 using formula (a) and (b) are plotted by a thick solid line and a thin
 dotted one, respectively. Those using (a) with the fitted $C_\rho$ and
 the fitted $C_\mass$ are plotted by a thick dashed line and a thick
 dot-dashed one, respectively. They are compared with the number of
 bands without considering the decay-out effect(a thin solid line). 
}
\end{figure}

Finally we show in Fig.~\ref{Fig:NUMBER2}
the calculated number of bands for $^{192}$Hg.
We plot here four kinds of result; the 
calculations where no adjustment of the ND level density nor
of the collective mass is
introduced but we evaluate $\Nband$ in two ways 
(the case (b) as well as the case (a)), the calculation
where the renormalization of 
the ND level density with $C_\rho=2.0\times10^{-4}$ is taken into
account (in the case (a)), and the calculation
with the adjustment of the collective mass ($C_\mass=3.0$) 
instead of the ND level density.
In parallel to the results shown in Fig.~\ref{Fig:BORDER3}, 
significant differences in $\Nband$ are seen among these calculations.
Concerning the calculation without any adjustment ($C_\rho=C_\mass=1$),
the effect of decay-out is significant and $\Nband$ 
takes a very small number. However, 
since this calculation fails to reproduce the
decay-out of the yrast SD band, it should be an underestimate.
On the other hand, when we make the adjustments to
reproduce the yrast decay-out, the results depend very strongly
on the method of adjustment. When we adjust the level density
of ND states ($C_\rho=2.0\times10^{-4}$),
the resultant $\Nband$ takes the value of more than 100 for $I\gesim20$.
This is because the decay-out
occurs only at very low spins, $I\lesim 20$, even at high excitation energy
(Fig.~\ref{Fig:BORDER3}), and 
the large value of $\Nband$ originating
from the shell effect~\cite{Yoshida2} remains.
When we perform the adjustment with the 
collective mass ($C_\mass=3.0$), on the other hand, 
the decay-out effect influences
$\Nband$ in the whole interval of spin, and reduces
$\Nband$ significantly taking a value less than 70.

It has been found in Ref.~\cite{Yoshida2} that 
an extraordinary large $\Nband$ 
due to the shell effect on the rotational damping
has been predicted
(also shown in Fig.~\ref{Fig:NUMBER2}).
However, the present result with including the decay-out process
suggests the dramatic reduction if we use a realistic ND level density
and believe the behaviour of the potential energy surface
at higher-spin region.
This is mainly because the barrier height of $^{192}$Hg is lower
at higher spins and consequently the decay-out boundary is relatively low
in excitation energy as is discussed in \S\ref{Sec:ResHg192}.
This fact clearly shows that the effect of barrier penetration
is crucial in this nucleus, and suggests a possibility to study
a large-scale change of the potential
energy surface as a function of angular momentum. 
Experimental information on $\Nband$ would be of
crucial importance in this respect.

\section{Conclusions}

We have constructed a microscopic model of 
thermally excited superdeformed states that can describe
both the rotational damping caused by the residual two-body
interaction and the decay-out associated with the barrier penetration
to normal deformation. Combining the cranked Nilsson-Strutinsky 
model for the deformed rotating mean-field, the shell-model
description of individual excited superdeformed states, and the
the pair hopping mass for the quadrupole shape transition, 
we define the model microscopically without any phenomenological fitting.
The calculations have been done for representative 
superdeformed nuclei $^{152}$Dy and $^{143}$Eu in the $A \approx 150$ region,
and $^{192}$Hg in the $A \approx 190$ region.

The model thus constructed is applied to the decay-out
of the yrast superdeformed states to check its validity. 
The sharpness of the observed decay-out is described well, and 
a good quantitative account of the decay-out spin is achieved
for $^{152}$Dy and $^{143}$Eu.
However, the model fails to describe the
decay-out spin in$^{192}$Hg.
We examined possible origins of the discrepancy in connection with
the level density of normal deformed states and the collective
mass parameter.

We discussed in detail decay-out properties of the excited 
superdeformed states having the thermal excitation energy up
to several MeV above the yrast state.  For this purpose, we 
extended the decay-out theory of Vigezzi et al.~\cite{Vigezzi}
by using the prescription of Bj{\o}rnholm and Lynn~\cite{Bjornholm-Lynn}.
The calculation shows that the excited superdeformed states 
decay out at much higher spins than the decay-out spin of
the yrast SD bands.

In terms of the decay-out and the onset of the rotational damping,
the excited superdeformed states are classified into three groups
having different transitional properties; the levels
that decay with rotational E2 transitions and form the rotational 
band structure, the levels that decay with rotational E2 but exhibit 
the rotational damping, and the others that mix strongly with
ND compound levels and decays to ND levels. 
The model predicts that both the decay-out
and the rotational damping influence significantly
the effective number of superdeformed bands,
an observable for the excited superdeformed states
obtained recently by means of the quasi-continuum spectroscopy.
The decay-out causes a characteristic decrease in
the effective number of superdeformed band with decreasing spin.
The study of thermally excited superdeformed bands gives
us an opportunity to investigate a large-scale shape dynamics
as well as the damping of collective rotational motion.

\section*{Acknowledgments}

We thank T.~D{\o}ssing, E.~Vigezzi and B.~Herskind for valuable
comments and discussion, and S.~Leoni for useful discussion
and providing us with the experimental data prior to publication.
This work has been supported in part by the Grant-in-Aid for
Scientific Research from the Japan Ministry of Education,
Science and Culture (Nos. 10640267 and 12640281).


\begin{thebibliography}{99}

\bibitem{Lauritzen}
B.~Lauritzen, T.~D{\o}ssing, and R.~A.~Broglia,
  Nucl. Phys. {\bf A457} (1986) 61.

\bibitem{FAM}
B.~Herskind,
 A.~Bracco, R.~A.~Broglia, T.~D\o ssing, A.~Ikeda, S.~Leoni, J.~Lisle,
 M.~Matsuo, and E.~Vigezzi,
Phys. Rev. Lett. {\bf 68} (1992) 3008. \\
T.~D{\o}ssing,  
B.~Herskind, S.~Leoni, A.~Bracco, R.~A.~Broglia, M.~Matsuo, E.~Vigezzi, 
Phys. Rep. {\bf 268} (1996) 1.

\bibitem{Aberg}
S.~\AA berg, Phys. Rev. Lett. {\bf 64}(1990)3119;\\
 S.~\AA berg, Prog. Part. Nucl. Phys. vol.~28 (Pergamon, 1992) p.~11.

\bibitem{Matsuo}
M.~Matsuo, T.~D\o ssing, E.~Vigezzi, R.~A.~Broglia,
and K.~Yoshida, Nucl. Phys. {\bf A617} (1997) 1.

\bibitem{Bracco-simul}
A.~Bracco, P.~Bosseti, S.~Frattini, E.~Vigezzi,
S.~Leoni, T.~D{\o}ssing, B.~Herskind, M.~Matsuo,
Phys. Rev. Lett. {\bf 76} (1996) 4484.


\bibitem{Dy-cont}
P.~J.~Twin, Proc. Intern. Conf. on Nuclear
shapes (World Scientific, 1988)p152;\\
P.~J.~Twin,  Nucl. Phys. {\bf A520} (1990) 17c.

\bibitem{Khoo}
T.~Lauritsen, Ph.~Benet, T.~L.~Khoo, K.~B.~Beard,
I.~Ahmad, M.~P.~Carpenter, P.~J.~Daly, M.~W.~Drigert, U.~Grag,
P.~B.~Fernandez, 
R.~V.~F. Janssens, E.~F.Moore, F.~L.~H. Wolfs, and D. Ye,
Phys. Rev. Lett. {\bf 69} (1992) 2479; \\
T.~L.~Khoo, T.~Lauritsen, I.~Ahmad,  M.~P.~Carpenter,
P.~B.~Fernandez, R.~V.~F.~Janssens,  E.~F.~Moore,  F.~L.~H.~Wolfs, 
Ph.~Benet,  P.~J.~Daly,  K.~B.~Beard, U.~Grag, D.~Ye, and 
 M.~W.~Drigert, Nucl. Phys. A557 (1993) 83c.

\bibitem{Hg192-cont-decay} 
R.~G.~Henry, T.~Lauritsen, I.~Ahmad, M.~P.~Carpenter, B.~Crowell,
T.~D{\o}ssing, R.~V.~F.~Janssens, F.~Hannachi, A.~Korichi, C.~Schuck,
F.~Azaiez, C.~W.~Beausang, R.~Beraud, C.~Bourgeois, R.~M.~Clark,
I.~Delonche, J.~Duprat, B.~Gall, H.~Hubel, M.~J.~Joyce, M.Kaci,
Y.~Lecoz, M.~Meyer, E.~S.~Paul, N.~Perrin, N.~Poffe, M.~G.~Porquet,
N.~Redon, H.~Sergolle, J.~F.~Sharpey-Schafer, J.~Simpson, A.~G.~Smith,
R.~Wadsworth, and P.~Willsau, 
Phys. Rev. Lett. {\bf 73} (1994) 777.

\bibitem{Leoni-Npath}
S.~Leoni, B.~Herskind,
T.~D\o ssing, K.~Yoshida, M.~Matsuo,
A.~Ata\c{c}, G.~B.~Hagemann, F.~Ingebretsen, H.~J.~Jensen, R.~M.~Lieder,
G.V.~Marti, N.~Nica, J.~Nyberg, M.~Piiparinen, H.~Schnare,
G.~Sletten, K.~Str\"{a}hle, M.~Sugawara, P.~O.~Tj{\o}m,
and A.~Virtanen, Phys. Lett. {\bf B353} (1995) 179. 


\bibitem{Leoni-decay}
S.~Leoni, B.~Herskind,
T.~D\o ssing, A.~Ata\c{c}, M.~Piiparinen, 
Phys. Rev. Lett. {\bf 76} (1996) 3281; \\
S.~Leoni, B.~Herskind,
T.~D\o ssing, A.~Ata\c{c}, I.~G.~Bearden, M.~Bergstrom, C.~Fahlander,
G.~B.~Hagemann, A.~Holm, D.~T.~Joss, M.~Lipoglavsk,
A.~May, P.~J.~Nolan, J.~Nyberg, M.~Palacz, E.S.~Paul, 
J.~Peerson, M.~J.~Piiparinen, N.~Redon, A.~T.~Semple, 
G.~Sletten, J.~P.~Vivien, 
Phys. Lett. {\bf B409} (1997) 71. 


\bibitem{Vigezzi}
E.~Vigezzi, R.~A.~Broglia and T.~D{\o}ssing,
Phys. Lett. {\bf B249} (1990) 163.\\
E.~Vigezzi, R.~A.~Broglia~and T.~D{\o}ssing,
Nucl. Phys. {\bf A520} (1990) 179c.

\bibitem{ShimizuA} 
Y.~R.~Shimizu, F.~Barranco, R.~A.~Broglia, T.~D\o ssing, E.~Vigezzi,
Phys. Lett. {\bf B274} (1992) 253.

\bibitem{ShimizuB} 
Y.R.~Shimizu, E.~Vigezzi, T.~D{\o}ssing and R.~A.~Broglia,
Nucl. Phys. {\bf A557} (1993) 99c.

\bibitem{Schiffer}
K.~Schiffer, B.~Herskind, and J.~Gascon,
Z. Phys. {\bf A332} (1989) 17; \\
K.~Schiffer and B.~Herskind,
Phys.  Lett.  {\bf B255} (1991) 508; \\
K.~Schiffer and B.~Herskind,
Nucl. Phys. {\bf A520} (1990) 521c.

\bibitem{Exp-decayout} 
R.~Kr\"{u}cken, A.~Dewald, P.~Sala, C.~Meier,
H.~Tiesler J.~Altmann, K.~O.~Zell, P.~von Brentano, D.~Bazzacco,
C.~Rossi-Alvarez, R.~Burch, R.~Menegazzo, G.~de Angelis, G.~Maron,
and M.~de Poli,
Phys. Rev. Lett. {\bf 73} (1994) 3359; \\
R.~Kr\"{u}cken, A.~Dewald, P.~von Brentano,  D.~Bazzacco,
and C.~Rossi-Alvarez, 
Phys. Rev. {\bf C54} (1996) 1182; \\
R.~K\"{u}hn, A.~Dewald, R.~Kr\"{u}cken,  C.~Meier, R.~Peusquens,
H.~Tiesler O.~Vogel, S.~Kasemann, P.~von Brentano, 
D.~Bazzacco, C.~Rossi-Alvarez, S.~Lunardi, and J.~de Boer,
Phys. Rev. {\bf C55} (1997) R1002.

\bibitem{Gu-Weiden}
J.~-Z.~Gu, and H.~A.~Weidenm\"{u}ller,
 Nucl. Phys. {\bf A660} (1999) 197.

\bibitem{Te-cont}
S.~Frattini, A.~Bracco, S.~Leoni,
F.~Camera, B.~Million, N.~Blasi, G.~Lo Bianco, M.~Pignanelli, E.~Vigezzi,
B.~Herskind, T.~D{\o}ssing, P.~Varmette, S.~T\"{o}rm\"{a}nen,
A.~May, M.~Kmiecik, D.~R.~Napoli, M.~Matsuo, 
Phys. Rev. Lett. {\bf 83} (1999) 83; \\
A.~Bracco, S.~Frattini, S.~Leoni, F.~Camera, B.~Million,
N.~Blasi, G.~Falconi, G.~Lo~Bianco, M.~Pignanelli, 
E.~Vigezzi, B.~Herskind, M.~Bergstr\"{o}m, P.~Varmette,
S.~T\"{o}rm\"{a}nen, A.~Maj, M.~Kmiecik, D.~R.~Napoli, M.~Matsuo,
Nucl. Phys. {\bf A673} (2000) 64.



\bibitem{Yoshida1} 
K.~Yoshida and M.~Matsuo,
Nucl. Phys. {\bf A612} (1997) 26.

\bibitem{Yoshida2} 
K.~Yoshida and M.~Matsuo,
Nucl. Phys. {\bf A636} (1998) 169.

\bibitem{Bush}
B.~W.~Bush, G.~F.~Bertsch, and B.~A.~Brown,
Phys. Rev. {\bf C45} (1992) 1709.

\bibitem{Weidenmuller}
H.A. Weidenm\"{u}ller, P.von Brentano and
B.R. Barrett,Phys. Rev. Lett. {\bf 81} (1998) 3603.

\bibitem{Barrett}
C.~A.~Stafford and B.~R.~Barrett,
Phys. Rev. {\bf C60}(1999) 051305.

\bibitem{RMP89}
Y.~R.~Shimizu, J.~D.~Garrett, R.~A.~Broglia, M.~Gallardo and E.~Vigezzi,
Rev. Mod. Phys. {\bf 61} (1989), 131.

\bibitem{ShimBrog}
Y.~R.~Shimizu and R.~A.~Broglia,
Nucl. Phys. {\bf A515} (1990) 38.

\bibitem{BenRag}
T.~Bengtsson and I.~Ragnarsson,
Nucl. Phys. {\bf A436} (1985) 14.

\bibitem{Brack}
M.~Brack, J.~Damgaard, A.~S.~Jensen, H.~C.~Pauli, 
V.~M.~Strutinsky, C.~Y.~Wong, Rev. Mod. Phys. {\bf 44} (1972) 320.

\bibitem{Bertsch}
G.~Bertsch, in Proc. of CIV ``E.Fermi'' International
School of Physics, Frontiers and borderlines in many-particle physics,
ed. R.~A.~Broglia and J.~R.~Schrieffer (North-Holland, 1988) p.~41.

\bibitem{Barranco}
F.~Barranco, G.~F.~Bertsch, R.~A.~Broglia and E.~Vigezzi,
Nucl. Phys. {\bf A512} (1990) 253.

\bibitem{MassTensor}
Y.~R.~Shimizu, F.~Barranco, E.~Vigezzi and R.~A.~Broglia,
in Proceedings of the 4th International Symposium on
{\it Foundations of Quantum Mechanics --- In the Light of New Technology ---}
(ISQM-Tokyo '92),  August 24-27, 1992, Tokyo,
Japanese Journal of Applied Physics Series 9 (1993) pp.~164-167.

\bibitem{Kisomer}
K.~Narimatsu, Y.~R.~Shimizu and T.~Shizuma,
Nucl. Phys. {\bf A601} (1996) 69.

\bibitem{Bjornholm-Lynn}
S.~Bj{\o}rnholm and J.~E.~Lynn,
Rev. Mod. Phys. {\bf 52} (1980) 725.

\bibitem{Creagh}
S.~C.~Creagh, Phys. Rev. Lett. {\bf 77} (1996) 4975.

\bibitem{Aberg-Chaos}
S.~\AA berg, Phys. Rev. Lett. {\bf 82}(1999) 229.

\bibitem{Schmid}
A.~Schmid, Ann. Phys. {\bf 170} (1986) 333.

\bibitem{WernerDudek}
T.~R.~Werner and J.~Dudek,
``Super- and Hyper-Deformed Nuclei for 58 $\le$ Z $\le$ 92'',
AIP Conf. Proc. {\bf 259} (1992) 683; \hfill\break
T.~R.~Werner and J.~Dudek, Atom. Dat. Nucl. Dat. Tab. 
{\bf 50} (1992) 179; {\bf 59} (1995) 1. 

\bibitem{Satula}
W.~Satula, S.~Cwiok, W.~Nazarevicz, R.~Wyss and A.~Johnson,
Nucl. Phys. {\bf A529} (1991) 289.

\bibitem{Aberg-lv}
S.~{\AA}berg, Nucl. Phys. {\bf A477} (1988) 18.

\bibitem{Mughab}
S.~F.~Mughabghab and C.~Dunford, 
Phys. Rev. Lett. {\bf 81} (1998) 4083.

\bibitem{DossVige}
T.~D\o ssing and E.~Vigezzi,
Nucl. Phys. {\bf A587} (1995) 13.

\bibitem{Barth}
G.~A.~Bartholomew, E.~D.~Earle, A.~J.~Ferguson,
J.~W.~Knowles, and M.~A.~Lone,
Adv. Nucl. Phys. {\bf 7} (1973) 229.

\bibitem{ExpDySys}
D.~Curien, G.~de~France, C.~W.~Beausang, F.~A.~Beck, T.~Byrski,
S.~Clarke, P.~Dagnall, G.~Duch$\hat{\mbox{e}}$ne, S.~Flibotte,
P.~D.~Forsyth, B.~Hass, M.~A.~Joyce, B.~Kharraja,
B.~M.~Nyak\'o, C.~Sch\"uck, J.~Simpson, C.~Theisen,
P.~J.~Twin, J.~P.~Vivien, and L.~Zolnai,
Phys. Rev. Lett. {\bf 71} (1993), 2559.

\bibitem{Exp143Eu}
A.~Ata\c{c}, M.~Piiparinen, B.~Herskind, J.~Nyberg, G.~Sletten, 
G.~de Angelis, R.~M.~Clark, S.~A.~Forbes, N.~Gj{\o}rup, G.~B.~Hagemann,
F.~Ingebretsen, H.~J.~Jensen, D.~Jerrestam, H.~Kusakari, R.~M.~Lieder,
G.~V.~Marti, S.~Mullins, P.~J.~Nolan, E.~S.~Paul, P.~H.~Regan,
D.~Santonocito, H.~Schnare, K.~Str\"ahle, M.~Sugawara, P.~O.~Tj{\o}m,
A.~Virtanen and R.~Wadsworth, Nucl. Phys. {\bf A557} (1993), 109c.

\bibitem{Eu-newexp}
A.~Axelsson, J.~Nyberg, A.~Ata\c{c}, 
M.~H.~Bergstr\"{o}m, B.~Herskind, G.~de Angelis. 
F.~B\"{a}ck, D.~Bazzacco, A.~Bracco, F.~Camera, B.~Cederwall, 
C.~Fahlander, J.~H.~Huijnen, S.~Lunardi, B.~Million, C.~R.~Napoli,
J.~Persson, M.~Piiparinen, C.~Rossi Alvarez, G.~Sletten, 
P.~G.~Varmette, M.~Weiszflog, Eur. Phys. J. {\bf A6} (1999) 175.

\bibitem{Eu-spinassign}
S.~Lunardi, L.~H.~Zhu,  C.~M.~Petrache,
D.~Bazzacco, N.~H.~Medina, M.~A.~Rizzuto, C.~Rossi Alvarez, 
G.~de Angelis, G.~Maron, C.~R.~Napoli, S.~Utzelmann, W.~Gast,
R.~M.~Lieder, A.~Geogiev, F.~Xu, R.~Wyss, 
Nucl. Phys. {\bf A618} (1997) 238.


\bibitem{Exp192Hg}
P.~Fallon, T.~Lauritsen, I.~Ahmad, M.~P.~Carpenter, B.~Cederwall,
R.~M.~Clark, B.~Crowell, M.~A.~Deleplanque, R.~M.~Diamond,
B.~Gall, F.~Hannachi, R.~G.~Henry, R.~V.~F.~Janssens,
T.~L.~Khoo, A.~Korichi, I.~Y.~Lee, A.~O.~Macchiavelli,
C.~Sch\"uck, and F.~S.~Stephens,
Phys. Rev. {\bf C51} (1995), R1609.



\bibitem{Lopez-Dossing}
A.~Lopez-Martens, F.~Hannachi, T.~D{\o}ssing, C.~Sch\"{u}ck, R.~Collatz,
E.~Gueorguieva, Ch.~Vieu, S.~Leoni, B.~Herskind, T.~L.~Khoo, T.~Lauritsen,
I.~Ahmad, D.~J.~Blumenthal, M.~P.~Carpenter, G.~Gassmann, R.~V.~F. Janssens,
D.~Nisius, A.~Korichi, C.~Bourgeois, A.~Astier, L.~Ducroux, Y.~Le~Coz,
M.~Meyer, N.~Redon, J.~F.~Sharpey-Schafer, A.~N.~ Wilson, W.~Korten,
A.~Bracco, and R.~Lucas, Phys. Rev. Lett. {\bf 77} (1996) 1707; \hfill\break
T.~D{\o}ssing, T.~L.~Khoo, T.~Lauritsen, I.~Ahmad, D.~J.~Blumenthal, 
M.~P.~Carpenter, B.~Crowell, D.~Gassmann, R.~G.~Henry, R.~V.~F. Janssens,
and D.~Nisius, Phys. Rev. Lett. {\bf 75} (1995) 1276.

\bibitem{NS2000}
Y.~R.~Shimizu, M.~Matsuo, and K.~Yoshida,
in Proceedings of the International Conference
{\it Nuclear Structure 2000},
15-19 August 2000, East Lansing, USA, to be published
(preprint nucl-th/0008060). 

\bibitem{Leoni-new}
S.~Leoni, private communication, to be published.


\end{thebibliography}
\end{document}